\documentclass[twocolumn,trackchanges]{aastex62}
\usepackage{amsmath}
\usepackage{epsf}
\usepackage{color}
\usepackage{color}
\usepackage{enumitem}
\usepackage{mathtools}

\usepackage[mathscr]{euscript}
 \let\mathscr\relax
\usepackage[scr]{rsfso}
\newcommand{\powerset}{\raisebox{.15\baselineskip}{\Large\ensuremath{\wp}}}

\shorttitle{Census of the Bright $z=$9--11 Universe}
\shortauthors{Finkelstein et al.}

\newcommand{\sol}{$_{\odot}$}

\newcommand{\hb}{$H_{160}$}

\def\arcs{\hbox{$^{\prime\prime}$}}

\turnoffedit 

\begin{document}
\title{A Census of the Bright $z\!\!=$8.5--11 Universe with the {\it
    Hubble} and {\it Spitzer} Space Telescopes in the CANDELS Fields}
\author[0000-0001-8519-1130]{Steven L. Finkelstein}
\affiliation{Department of Astronomy, The University of Texas at Austin, Austin, TX, USA}
\email{stevenf@astro.as.utexas.edu}

\author[0000-0002-9921-9218]{Micaela Bagley}
\affiliation{Department of Astronomy, The University of Texas at
  Austin, Austin, TX, USA}
\author{Mimi Song}
\affiliation{University of Massachusetts, Amherst, MA, USA}
\author{Rebecca Larson}
\altaffiliation{NSF Graduate Fellow}
\affiliation{Department of Astronomy, The University of Texas at Austin, Austin, TX, USA}
\author{Casey Papovich}
\affiliation{George P. and Cynthia Woods Mitchell Institute for Fundamental Physics and Astronomy, Department of Physics and Astronomy, Texas A\&M University, College Station, TX, USA}
\author{Mark Dickinson}
\affiliation{National Optical-Infrared Astronomy Research Laboratory, Tucson, AZ, USA}
\author{Keely Finkelstein}
\affiliation{Department of Astronomy, The University of Texas at Austin, Austin, TX, USA}
\author[0000-0002-6610-2048]{Anton M. Koekemoer}
\affiliation{Space Telescope Science Institute, Baltimore, MD 21218, USA}
\author{Norbert Pirzkal}
\affiliation{Space Telescope Science Institute, Baltimore, MD, USA}
\author{Rachel S. Somerville}
\affiliation{Center for Computational Astrophysics, Flatiron Institute, NY, USA}
\author[0000-0003-3466-035X]{L. Y. Aaron Yung}
\affiliation{NASA/Goddard Space Flight Center, Greenbelt, MD, USA}
\author{Peter Behroozi}
\affiliation{University of Arizona, Tucson, AZ, USA}
\author{Harry Ferguson}
\affiliation{Space Telescope Science Institute, Baltimore, MD, USA}
\author{Mauro Giavalisco}
\affiliation{University of Massachusetts, Amherst, MA, USA}
\author{Norman Grogin}
\affiliation{Space Telescope Science Institute, Baltimore, MD, USA}
\author{Nimish Hathi}
\affiliation{Space Telescope Science Institute, Baltimore, MD, USA}
\author[0000-0001-6251-4988]{Taylor A. Hutchison}
\altaffiliation{NSF Graduate Fellow}
\affiliation{George P. and Cynthia Woods Mitchell Institute for Fundamental Physics and Astronomy, Department of Physics and Astronomy, Texas A\&M University, College Station, TX, USA}
\author[0000-0003-1187-4240]{Intae Jung}
\affil{Astrophysics Science Division, Goddard Space Flight Center, Greenbelt, MD 20771, USA}
\affil{Department of Physics, The Catholic University of America, Washington, DC 20064, USA }
\author{Dale Kocevski}
\affiliation{Department of Physics and Astronomy, Colby College, Waterville, ME, USA}
\author{Lalitwadee Kawinwanichakij}
\affiliation{Kavli Institute for the Physics and Mathematics of the Universe, The University of Tokyo, Kashiwa 277-8583 (Kavli IPMU, WPI), Japan}
\author[0000-0003-2349-9310]{Sof\'ia Rojas-Ruiz}
\altaffiliation{Fellow
  of the International Max Planck Research School for\\ Astronomy and
  Cosmic Physics at the University of \\ Heidelberg (IMPRS--HD)}
\affiliation{Max-Planck-Institut f\"{u}r Astronomie, K\"{o}nigstuhl 17, D-69117, Heidelberg, Germany}
\author{Russell Ryan Jr.}
\affiliation{Space Telescope Science Institute, Baltimore, MD, USA}
\author{Gregory F. Snyder}
\affiliation{Space Telescope Science Institute, Baltimore, MD, USA}
\author[0000-0002-8224-4505]{Sandro Tacchella}
\affiliation{Center for Astrophysics $\vert$ Harvard \& Smithsonian, Cambridge, MA, USA}

\begin{abstract}
We present the results from a new search for candidate galaxies at $z
\approx$ 8.5--11 discovered over the 850 arcmin$^2$ area probed by the
Cosmic Assembly Near-Infrared Deep Extragalactic Legacy Survey
(CANDELS).  We use a photometric redshift selection including both
{\it Hubble} and {\it Spitzer Space Telescope} photometry to robustly
identify galaxies in this epoch at \hb\ $<$ 26.6. We use a detailed vetting procedure,
including screening against persistence and stellar contamination, and the
inclusion of ground-based imaging and followup {\it Hubble Space Telescope} imaging
to build a robust sample of 11 candidate galaxies, three presented
here for the first time.  The inclusion of {\it Spitzer}/IRAC
photometry in the selection process reduces contamination, and yields
more robust redshift estimates than {\it Hubble} alone.  We constrain
the evolution of the rest-frame ultraviolet luminosity function via a
new method of calculating the observed number densities without
choosing a prior magnitude bin size.  We find that the abundance at
our brightest probed luminosities ($M_{UV}=-$22.3) is consistent
with predictions from simulations which assume that galaxies in this
epoch have gas depletion times at least as short as those in nearby
starburst galaxies. Due to large Poisson and cosmic variance uncertainties we
cannot conclusively rule out either a  smooth evolution of the
luminosity function continued from $z\!\!=$4--8, or an accelerated
decline at $z >$ 8.  
We calculate that the presence of seven galaxies in a
single field (EGS) is an outlier at the 2$\sigma$ significance level,
implying the discovery of a significant overdensity.  These scenarios
will be imminently testable to high confidence within the first year
of observations of the {\it James Webb Space Telescope}.
\end{abstract}


\section{Introduction}\label{sec:intro}

The epoch of reionization marks the time when the first luminous
sources in the Universe, presumably massive stars in nascent galaxies, began to impact the
universe around them.  Ionizing ultraviolet (UV) photons from these
early objects leaked into the intergalactic medium (IGM), beginning
the process of ionizing the pervasive hydrogen (and singly-ionizing
helium) gas.  Significant effort has gone into studying the end of
this process, with a general consensus that reionization is complete
by $z \sim$ 6 \citep[e.g.,][]{fan06,pentericci18}, with some islands of neutral gas
remaining to $z \sim$ 5.5 \citep[e.g.][]{becker15, kulkarni18,becker21}.

Studies of galaxies have concluded that if the ionizing photon escape
fraction is high enough ($\sim$10--20\%) then the observed galaxy
population, extrapolated to a reasonable level beyond the limits of
modern observations, can supply the necessary ionizing photons
\citep[e.g.,][]{finkelstein10,finkelstein12b,finkelstein15,robertson13,robertson15,bouwens15b,naidu20}.
The reionization history implied by these moderately-leaking galaxies
is one with a very slow beginning and a rapid rise in the ionized
fraction at $z <$ 8 for a ``late reionization'' scenario.  However, theoretical models predict that
it is likely the tiniest dwarf galaxies which have the highest escape
fractions due to feedback processes more easily clearing channels for
escape \citep[e.g.][]{paardekooper15,xu16}, with the massive,
presently observable galaxies, contributing little \citep[e.g.,][]{yung20b}.  \citet{finkelstein19}
explored reionization scenarios adopting these simulation-based escape
fractions, and found that reionization can still be completed by $z
\sim$ 5.5 with population-averaged escape fractions of $<$5\% (and a
small contribution by active galactic nuclei at $z <$ 6).  The
reionization history in this scenario evolves more slowly than previous models
as the dependence on very faint galaxies yields an earlier start to
reionization and a quasi-linear evolution in the ionized fraction
through the end of reionization.  Both scenarios are consistent with
{\it Planck} measurements of the electron scattering optical depth to
the cosmic microwave background \citep[CMB, ][]{planck16}.

These scenarios differ the most at $z \sim$ 9, where the former scenario
implies a low ionized fraction of $\sim$20\%, and the latter implies a larger
$\sim$50\% ionized fraction.  Moving robust studies of the galaxy populations
to $z >$ 8 thus has the power to observationally constrain these
models, helping improve our understanding of reionization.  Observing
galaxies at $z >$ 8 also pushes us closer to the epoch when we expect
the first galaxies to be forming ($z \sim$ 10--15), constraining the
formation of the first massive galaxies at these epochs, measuring
their chemical enrichment, placing constraints on their
star-formation histories, and exploring whether the
physics which regulate star formation are evolving to the earliest
times \citep[][]{behroozi15,finkelstein15b,stefanon17b,tacchella18,behroozi20,yung20b}.

Understandably, after the first big near-infrared surveys with Wide
Field Camera 3 (WFC3) on the {\it Hubble Space Telescope} ({\it HST})
robustly constrained the $z =$ 4--8 galaxy population
\citep[e.g.][]{mclure13,finkelstein15,bouwens15}, attention turned to
the $z >$ 8.5 universe.  This is more difficult, as at $z \sim$ 9--10,
the Ly$\alpha$ break is passing through the WFC3 F125W filter, with
most deeply covered regions of the sky being covered by only F160W imaging
fully redward of the break (though a few small fields include F140W to
a commensurate depth).

The first large sample of $z >$ 8.5 candidate galaxies was published
by \citet{oesch13}, who found nine candidate $z \geq$ 9 candidate
galaxies in the Hubble Ultra Deep Field (HUDF, \citealt{beckwith06}) and Great Observatories
Origins Deep Survey \citep[GOODS][]{giavalisco04b} South field using Lyman break color-color selection.  This was followed up by
\citet{oesch14}, who did a similar search in the GOODS-North field,
finding four surprisingly bright candidates at $z \sim$ 9--10,
including a source with a grism spectroscopic redshift of $z =$ 11.1 \citep{oesch16}.  At a
similar time, \citet{bouwens15} also used color selection to identify six candidate $z \sim$ 10 galaxies in the two
GOODS fields (including two in the HUDF), followed
up by \citet{bouwens16} who used additional imaging to find a full
sample of 15 $z \sim$ 9--10 galaxies over the five fields from the
Cosmic Assembly Near-infrared Deep Extragalactic Legacy Survey
(CANDELS; \citealt{grogin11,koekemoer11}) fields (inclusive of the bright Oesch et
al.\ objects).  \citet{oesch18} added a few faint candidate $z \sim$
10 galaxies from the Hubble Frontier Fields survey \citep{lotz17},
while recently \citet{bouwens19} added a few additional bright sources
from a re-analysis of the CANDELS fields, this time using a
more inclusive photometric-redshift selection (updated again in \citealt{bouwens21}).  Each of these studies
examined the implications of their sample on the evolution of the
rest-UV luminosity function, and they each concluded that the
evolution from $z =$ 8 to 10 is more rapid (per unit redshift) than at
$z =$ 4--8, where the decline with increasing redshift is observed to be smooth
\citep[e.g.,][]{finkelstein15,bouwens15,finkelstein16}.

Other groups have also explored this epoch.  \citet{mcleod15} and
\citet{mcleod16} used photometric redshift selection to select $z
\sim$ 9--10 galaxies from the narrow yet deep HUDF and HFF fields.  They
concluded that the integrated luminosity density at these redshifts
is consistent with a continued smooth decline, and not an accelerated
drop-off, when integrating down to similar magnitudes as previous
studies ($M_{UV} =-$17.7).
\citet{ishigaki18} also used data from the HFF survey, making the
interesting point that they too find an accelerated decline at $z >$ 8
in the UV luminosity density when integrating to $M_{UV} =$ $-$17 (as previous
studies did), but when they integrate to $-$15, the decline is
smooth.  This should be unsurprising, as at earlier times, one may
expect more of the luminosity density to come from less luminous
systems (e.g., due to the evolution of the halo mass function).
\citet{bowler20} used ground-based imaging from the UltraVISTA survey
finding seven even brighter galaxies at $z >$ 8.  Their calculated luminosity
function was also in excess of that predicted from a smoothly evolving
Schechter function, potentially indicating a double power-law shape as
has been found at $z \sim$ 6--7 \citep{bowler14,bowler15}.
Other studies searching for bright galaxies in this epoch from {\it HST}
pure parallel surveys such as the Brightest of Reionizing Galaxies
(BoRG) survey have explored a new parameter space, probing several
randomly-placed fields across the sky, with results less susceptible to cosmic variance.  Interesting, several studies from
different teams with these data have found more bright galaxies than
expected, more consistent with a non-accelerated decline in the
luminosity density at $z >$ 8
\citep{trenti11,calvi16,bernard16,morishita18,rojasruiz20,roberts-borsani21}.  While these data are
less deep than the legacy fields such as CANDELS, and thus potentially
more prone to contamination, the $\sim$100 independent pointings also provide a more robust probe of
the underlying density field.

There is thus considerable uncertainty about the evolution of the UV
luminosity density, and hence the star-formation rate (SFR) density to
the highest redshifts presently accessible.  The
underlying evolution will soon be revealed in high fidelity with the launch
of the {\it James Webb Space Telescope} ({\it JWST}) and its
associated early-release science and guaranteed time programs, such as the Cosmic Evolution Early Release
Science (CEERS) survey \citep{finkelstein17}, the JWST Advanced Deep
Extragalactic Survey \citep[JADES,][]{williams18}, and the lensing
program ``Through the Looking Glass'' \citep{treu17}, as well as
several other recently approved large Cycle 1 General Observer programs.  However, more
insight can be gleaned from a re-examination of the existing data,
focusing on three aspects.  

First, the majority of previous studies, in
particular those which concluded that there is an accelerated decline
in the luminosity density at $z >$ 8, used simple color-selection to
identify their candidate galaxies.  While this is a robust way to
identify galaxy candidates, it requires galaxies to satisfy strict
color criteria, which means that photometric scatter can remove
otherwise valid candidates from a galaxy sample.  Photometric
redshift-based selection makes no such color cuts, though it is
necessary to implement ``quality control'' using the redshift probability
distribution function measured during this process.  A number of
previous studies have noted the inclusivity of photometric redshift
selection when opting to use this technique
\citep[e.g.][]{mclure10,finkelstein10,ellis13,finkelstein15,mcleod16},
which also has the advantage in that it provides information on
plausible low-redshift alternatives.  Second, previous studies were
heterogeneous in their usage of available {\it Spitzer}/IRAC imaging,
with some studies using it in the original selection, while others
first select galaxies with {\it HST} imaging only, and then examine
the IRAC imaging in a separate step.  As we discuss below, when used
in the sample selection process IRAC imaging has tremendous
leveraging power to probe likely contaminants and increase the
robustness of the photometric redshifts at $z >$ 8.  \edit1{Finally, three of
the CANDELS wide fields lack $Y$-band (1 $\mu$) imaging, which is just
below the Ly$\alpha$ break at $z \sim$ 9.  In this work we include new
followup {\it HST} imaging for all selected candidates below the
Ly$\alpha$ break (F098M in most cases) to further
increase our confidence in a $z >$ 8.5 solution.}

In this study, we aim to make a robust census of the $z =$ 9--11 universe
in the {\it HST} CANDELS fields, which have the most complete set of
moderately-deep yet wide imaging on the sky with both {\it HST} and
{\it Spitzer}/IRAC.  We make use of the IRAC photometry in the initial
sample selection, utilizing a deblending technique to measure accurate
IRAC photometry for all but the most crowded sources, which we then
use to select a galaxy sample via photometric redshifts.  In this
paper, we restrict ourselves to the bright end of the luminosity
function (H $<$ 26.6), as this is the luminosity regime where {\it
  Spitzer} has the greatest constraining power.  In \S 2 we discuss
the imaging data we used, and lay out our photometric measurement
process, including simulations we ran to justify our detection
algorithms.  In \S 3 we describe our sample selection, including our
contaminant rejection process.  We also discuss our screening for
persistence, which can mimic single-band-selected high-redshift
galaxies.  In \S 4 we describe the vetting process for our initial
sample, making use of all available ground-based photometry and
followup {\it Hubble} imaging, while in
\S 5 we compare to previous galaxy samples in this field.  In \S 6 we
present our luminosity function and discuss plausible physical
interpretations, and in \S 7 we discuss a potential overdensity in the
EGS field.  Our conclusions are
presented in \S 8.  In the appendix we show all objects manually
removed from our sample for a variety of reasons following visual
examination.  All magnitudes are presented in the AB system, and
we assume a cosmological model with $H_{0} =$ 67.74 km s$^{-1}$
Mpc$^{-1}$, $\Omega_m =$ 0.309 and $\Omega_{\Lambda} =$ 0.691,
consistent with the latest results from {\it Planck} \citep{planck20}.

\begin{figure*}[!t]
\epsscale{0.54}
\plotone{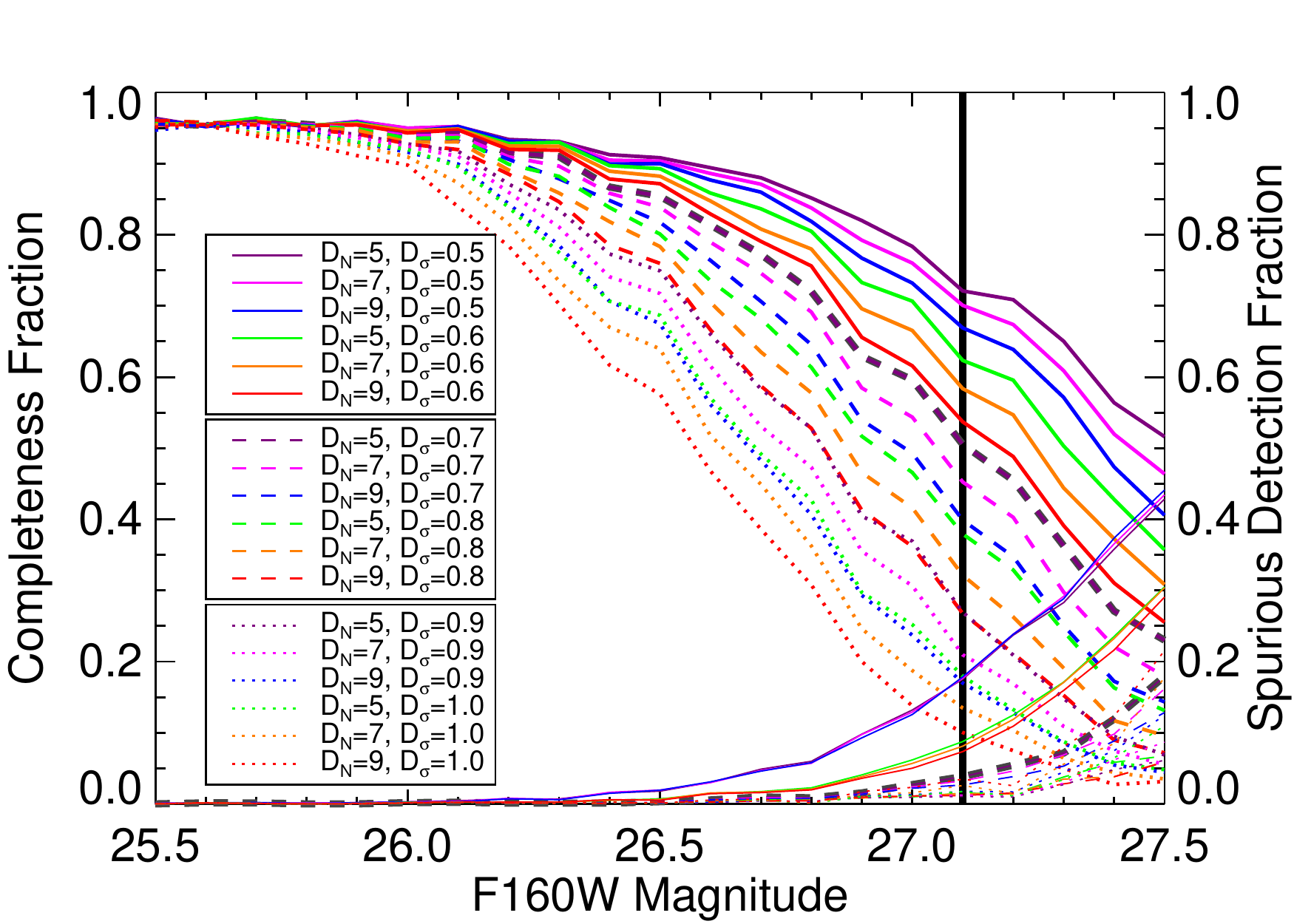}
\hspace{3mm}
\epsscale{0.58}
\plotone{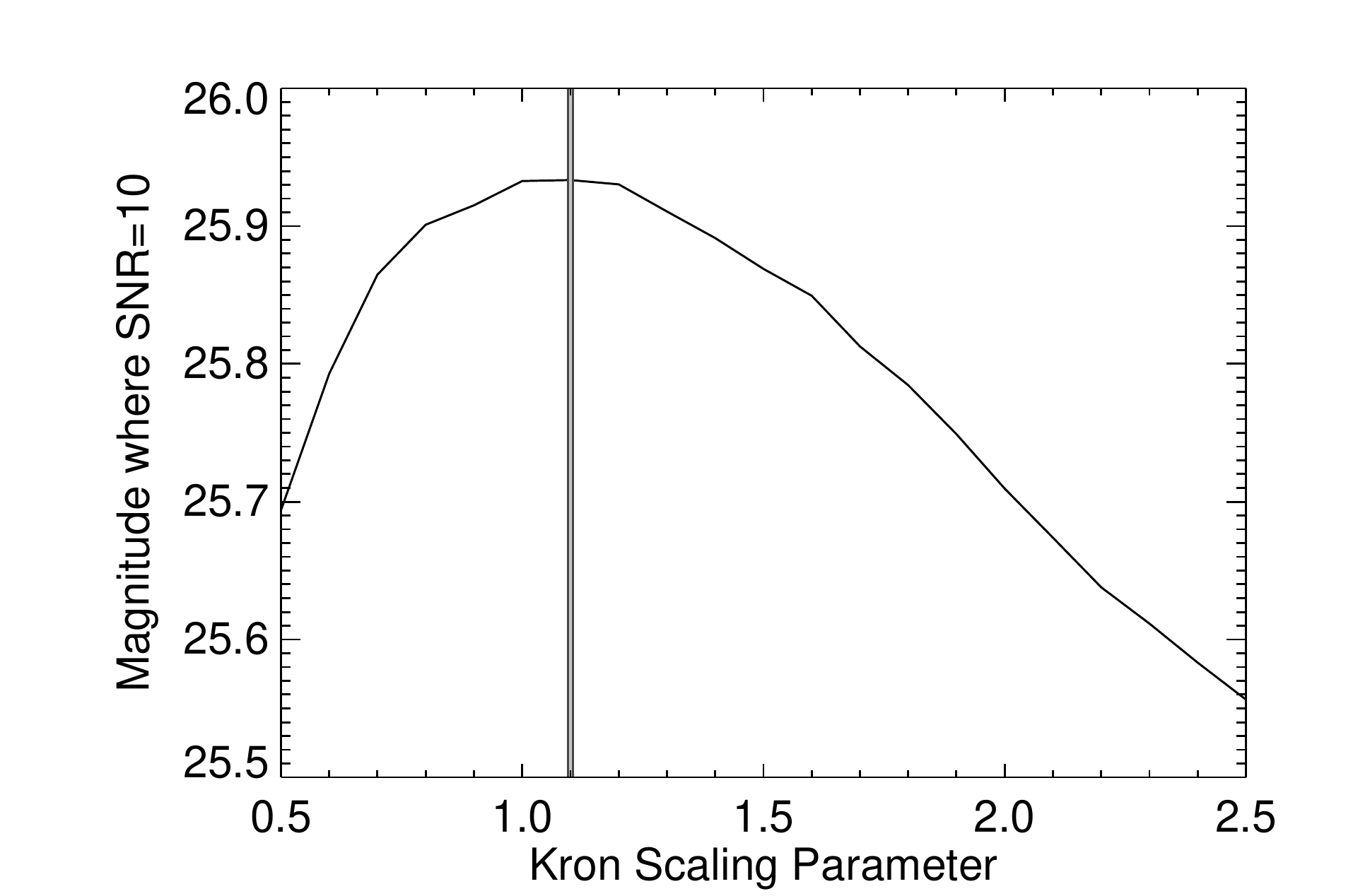}
\caption{The results from the optimization of our Source Extractor
  catalogs in the CANDELS EGS field.  \textit{Left}) The
  completeness (thick lines = fraction of recovered simulated
  sources recovered) and spurious fraction (thin lines = ratio of
  sources recovered from a negative image to those in the original
  image) as a function of total F160W magnitude.  We used this
  analysis to select fiducial values of DETECT\_THRESH (D$_{\sigma}$)
  $=$ 0.7 and DETECT\_MINAREA (D$_N$) $=$ 5 pixels, shown as the thickest
  dashed purple line.  This combination maximizes completeness while
  minimizing contamination.  A 20--40\% higher completeness is
  possible, but the catalogs become much less pure, with spurious
  fractions more than doubling.  \textit{Right}) The magnitude where the
  signal-to-noise ratio is equal to 10 as a function of the Kron
  scaling parameter $k$ (also for the EGS field).  This shows that Kron apertures with a value
  of $k=$1.1 results in a catalog which is nearly 0.4 mag deeper than the
  default value of $k=$2.5. }
\label{fig:sextractor}
\end{figure*}
\section{Observations and Catalogs}\label{sec:totalnion}

\subsection{Imaging}
The CANDELS survey \citep{grogin11,koekemoer11} imaged five fields in the sky: GOODS-North,
GOODS-South, COSMOS, the UKIDSS Deep Survey (UDS) and the Extended Groth
Strip (EGS).  The central 50\% of each of the GOODS fields was
observed longer and represents the CANDELS ``deep'' survey, while the
remaining regions of GOODS, and the other three fields, comprise the
CANDELS ``wide'' survey.  One exception is the northern 25\% of
GOODS-S, which was observed by the WFC3 Early Release Science (ERS)
team, and has a depth in between CANDELS wide and deep \citep{windhorst11}.
Each of these fields had previous {\it HST} imaging with the optical
Advanced Camera for Surveys (ACS), which was also utilized by CANDELS
in parallel.  The mosaics used here include all such previously
available data.  The data reduction and mosaic procedure is outlined
in \citet{koekemoer11}.  We use the latest available CANDELS
internal-team mosaics for each of our fields. 

\subsection{Catalog Detection Parameters}
We use v2.19.5 of the Source Extractor Software \citep{bertin96} to
measure photometry of our sources, using two-image mode with the F160W image as the
detection image.  \edit1{We set the weight type to MAP\_RMS, with the
CANDELS-provided rms image as the weight image.}  This software has a number of
user-definable choices which can impact the completeness and purity of a catalog,
and the accuracy of the photometry.  The primary factors affecting the
completeness and purity are DETECT\_THRESH ($D_{\sigma}$) and DETECT\_MINAREA ($D_N$).  The
former is a significance threshold, while the latter is a number of
pixels.  Source Extractor will extract a source if it has $D_N$ connected pixels
with a significance greater than $D_{\sigma}$.  In our previous work
\citep{finkelstein15} we assumed $D_{\sigma}=$ 0.6 and $D_{N}=$ 7
(with the same 60 milli-arcsecond pixel scale as we use here).  We
  arrived at these values based on visually inspecting the results of
  several parameter combinations, attempting to maximize completeness
  while minimizing inclusion of spurious sources.  As our goal in this
  paper is to push {\it Hubble} to its limits by probing the $z >$ 9
  universe, it is crucial to validate these choices via simulations.

To optimize these parameters we performed simulations in each of
the five fields, inserting 30,000 simulated sources in the F160W
science image.  We assumed the sources followed a Sersic light profile, with a
half-light radius in arcseconds drawn from a log-normal distribution \citep[e.g.,][]{shibuya16}
of the form: $r=e^{A+B\times\mathcal{R}}$, where $A=-0.1$ is the
central value, $B=1.8$ is the width, and $\mathcal{R}$ denotes a
Gaussian random deviate.  This functional form produces a distribution with a peak at
0.2 physical kpc and a relatively shallow decline to a maximum of 1
physical kpc ($\sim$0.25\arcs), expected for moderately bright $z >$
8 galaxies \citep[e.g.,][]{kawamata18}.  The Sersic index also assumed
a log-normal distribution, with a peak at the minimum of $n=1$, and a tail to $n=$5;
this distribution was much tighter than that of the radius, with the
majority of simulated objects having $n<$2.  The axis ratio also had a
lognormal distribution, peaked at 0.8, and the position angle was
random.  These simulated objects were created with the {\sc Galfit} package
\citep{peng02}, and added to the image at random positions.

We then ran Source Extractor on this image in two-image mode, using
this modified F160W-band image as both the detection and the
measurement image.  We ran 18 independent catalogs, testing values of
$D_{\sigma} \in$ \{0.5,0.6,0.7,0.8,0.9,1.0\} and $D_{N} \in$
\{5,7,9\}.  For this test, we used the default Kron aperture for
measuring photometry, as this approximately reproduces the total
flux.  As we know the true flux for each of our sources, we tested
this assumption by comparing the input and recovered fluxes for our
simulated sources.
We found that the ratio of our input flux to that measured in the
default Kron aperture was consistently $\sim$1.20 $\pm$ 0.01 in all five
CANDELS fields.
We thus applied this aperture correction to all sources for
the purpose of these simulations.
We matched each of these catalogs to our simulated source list, and
calculated the completeness as the fraction of sources recovered
divided by the total number input into the image in magnitude bins of $\Delta$m$=$0.1.

To calculate the fraction of spurious sources,
we performed another run of Source Extractor for each detection
parameter combination, now using a ``negative'' version of the
original science image (created by multiplying each pixel in the
original image
by $-$1).  \edit1{Assuming that the background is zero (see
\citealt{koekemoer11} for details on the reduction of these images)}, any significant ``source'' in this image is spurious, as it
is made up of a random clumping of negative pixels due to noise in the
image.  We calculated the spurious fraction as the number of sources
detected in this negative image in a given magnitude bin divided by the
number of sources detected in the original positive image (based on an
independent Source Extractor run) in that same magnitude bin.

The result of this exercise is an estimate of both the completness and
spurious detection fraction as a function of magnitude for each of our
Source Extractor parameter combinations.  In
Figure~\ref{fig:sextractor} we show the results for the EGS field,
comparing both the completeness and the spurious detection fraction to
the total 5$\sigma$ limiting magnitude in this field (where the
1$\sigma$ noise was estimated as the
standard deviation of fluxes measured in randomly placed
0.4\arcs-diameter apertures, corrected to total by dividing by the
fraction of a point-spread function [PSF] enclosed in a 0.4\arcs\
aperture [0.64 for F160W]).  This
figure shows that the completeness at this 5$\sigma$ limit falls
monotonically with increasing values of both $D_{\sigma}$ and $D_{N}$.  The
spurious fraction on the other hand shows little dependence on $D_N$,
but a strong dependence on $D_{\sigma}$.  We elected to use values of
$D_{\sigma} =$ 0.7 and $D_N =$ 5 pixels, shown by the dashed purple
line.  This produces a completeness fraction of 50.5\% at the 5$\sigma$
limiting magnitude in this field, and a spurious fraction (or catalog
impurity) of only 3.8\% at this limiting magnitude.  

\begin{deluxetable}{cccc}
\vspace{2mm}
\tabletypesize{\small}
\tablecaption{Quality of the CANDELS Photometric Catalogs}
\tablewidth{\textwidth}
\tablehead{
\colhead{Field} & \colhead{F160W 5$\sigma$} & \colhead{Completeness} & \colhead{Spurious}\\
\colhead{Field$ $} & \colhead{Limiting Mag} & \colhead{Fraction} & \colhead{Fraction}}
\startdata
EGS&27.10&0.505&0.038\\
COSMOS&26.87&0.486&0.061\\
UDS&26.92&0.526&0.066\\
GOODS-N Deep&27.64&0.634&0.029\\
GOODS-N Wide&26.98&0.586&0.094\\
GOODS-S Deep&27.47&0.615&0.025\\
GOODS-S Wide&26.79&0.426&0.068
\enddata
\tablecomments{The completeness fraction and spurious fraction (i.e.,
  catalog impurity) at the 5$\sigma$ limiting magnitude in each of the
  five CANDELS fields for our chosen detection parameters of
  DETECT\_THRESH$=$0.7 and DETECT\_MINAREA$=$5.  The limiting magnitude was measured in a
  0.4\arcs-diameter aperture, and corrected to total under the
  assumption of a point-source flux distribution.}
\label{tab:tab1}
\end{deluxetable}

As shown in Figure~\ref{fig:sextractor},
smaller values of $D_{\sigma}$ (solid lines) do result in higher
completeness fractions (55--70\%), but at the cost of significantly higher
spurious fractions (8--18\%).  We decided that a potential 40\% gain
in completness was not worth a 2--4$\times$ increase in the spurious fraction.
We emphasize that, though low, our fiducial spurious fraction is
non-zero, and we must take
care to ensure selected candidate galaxies are not spurious.  As
discussed below, this is done by imposing a detection significance
(i.e., minimal signal to noise) further tuned to eliminate such
spurious sources.  We note that the \citet{finkelstein15} parameters of $D_{\sigma}
=$ 0.6 and $D_N =$ 7 pixels result in a completeness fraction 15\%
higher than our fiducial value, at the cost of a spurious fraction
2.1$\times$ larger.

We performed this same analysis in all five CANDELS fields (including
splitting the GOODS fields into their different depth components).  We
find that our fiducial choice of detection parameters appears to be
preferred in all fields regardless of depth, thus we adopt them for
all fields.  In Table~\ref{tab:tab1}, we give the completeness and spurious
fraction at the typical 5$\sigma$ depth for each (sub) field.

\begin{figure*}[!t]
\epsscale{0.56}
\plotone{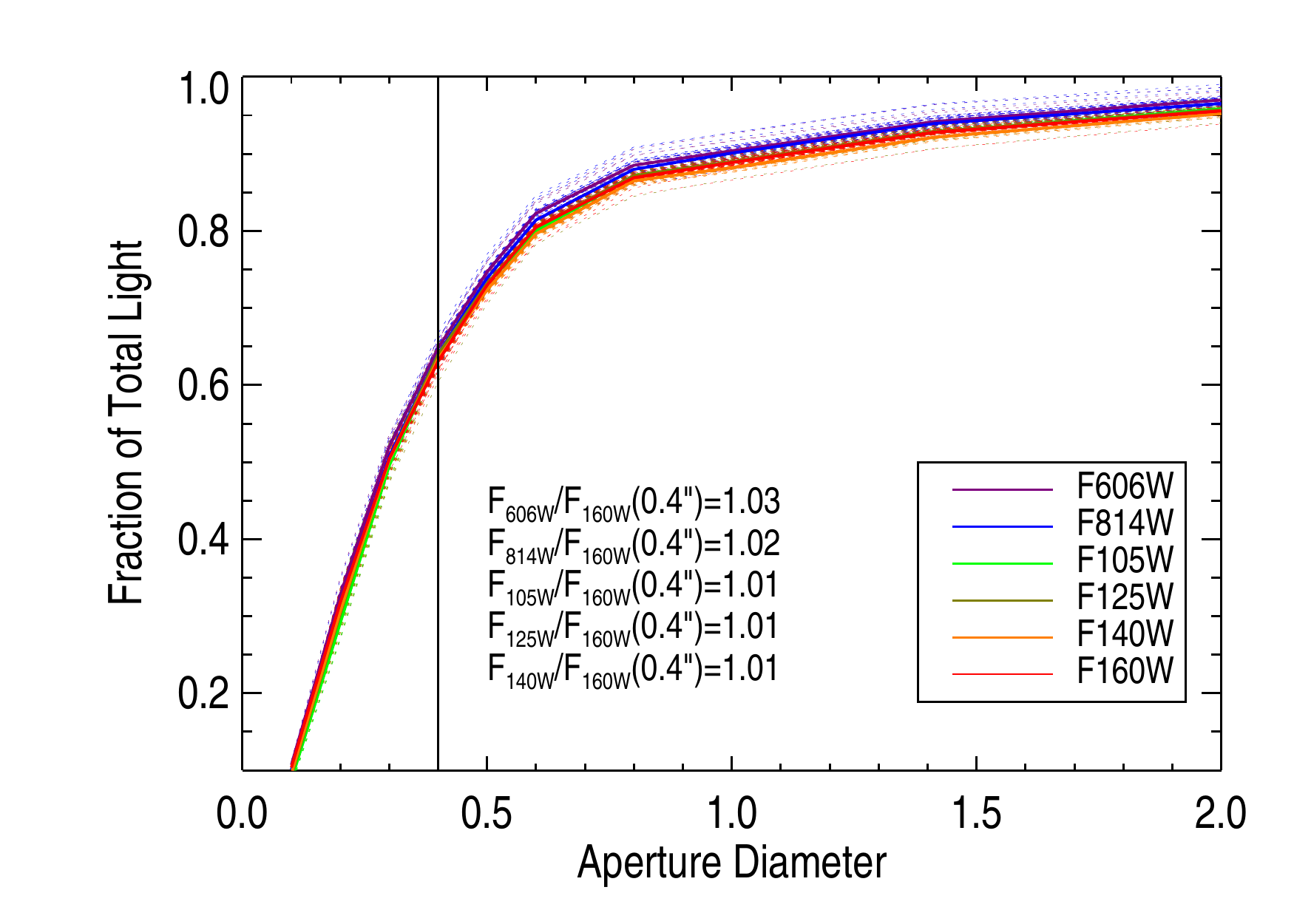}
\hspace{10mm}
\epsscale{0.52}
\plotone{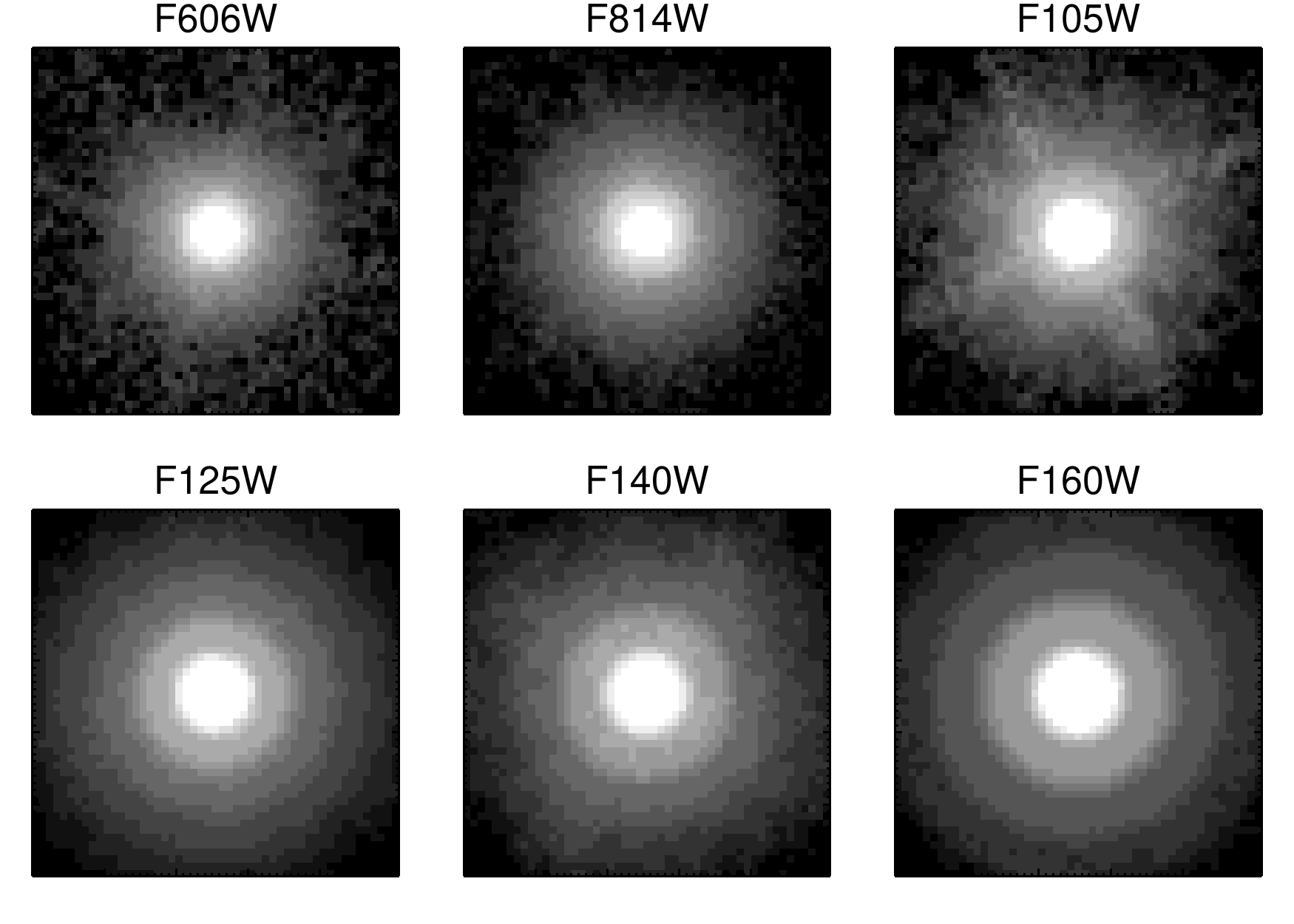}
\caption{\emph{Left}) The curve-of-growth measured for stars in the
  EGS field after the implementation of PSF matching.  The thin dotted
lines show individual stars, while the thick solid lines show the
average.  The vertical line indicates an aperture diameter of
0.4\arcs, while the inset text highlights that all images have an
enclosed flux within 3\% of the F160W image at this diameter.
\emph{Right}) Empirical stacked-star point-spread-functions for the
six images in the EGS field.  These cutout images show the central 51$\times$51 pixels
($\sim$3\arcs), and are displayed with logarithmic scaling.  The
measured full-width at half-maximum for these PSFs are 0.142\arcs,
0.147\arcs, 0.208\arcs, 0.222\arcs, 0.223\arcs, and 0.234\arcs\ for
F606W, F814W, F105W, F125W, F140W and F160W.}
\label{fig:psfs}
\end{figure*}   

\subsection{Photometric Aperture}
We follow our previous works
\citet{finkelstein10,finkelstein12a,finkelstein12b,finkelstein15} and
make use of small Kron elliptical
apertures to measure the colors of objects in our catalog \citep[see
also, e.g.,][]{bouwens07,bouwens15}.  These
elliptical apertures are tuned to match the shapes of the measured
objects, and typically achieve higher signal-to-noise than similarly sized circular
or isophotal apertures \citep[e.g.,][]{finkelstein10}.  Source Extractor parameterizes this radius
with a ``Kron factor'' $k$ and a minimum radius $R_{min}$.  As shown
in the Source Extractor manual, the default parameters of $k\!=$2.5,
$R_{min}\!=$3.5 contain $>$ 90\% of an object's flux \edit1{(see further
discussion in \citealt{graham05})}.  However, this
aperture is typically significantly larger than the region defined by
the detection threshold for most high-redshift galaxies, and so does
not result in the optimal signal-to-noise.  It is thus typical to use a
smaller Kron factor when measuring colors, with $k=$1.2 commonly used.  We explored a range of Kron
factors, running several iterations of Source Extractor with Kron factors
ranging from 0.5 to 2.5 (keeping the ratio of R$_{min}$/$k$ fixed to
3.5/2.5). using the same simulated sources from the preceding subsection.  The results are shown in the right panel of
Figure~\ref{fig:sextractor} for the EGS field (and are similar for
the other CANDELS fields).  This figure shows the magnitude where a
typical source has a signal-to-noise of 10, as a function of Kron
parameter $k$.  This distribution shows a clear peak at 1.0 $< k <$
1.2, with a steep decline to higher and lower values. Notably, at the
default value of $k\!=$2.5 a catalog would be nearly 0.4 mag shallower
than one using $k\!=$1.1.  We thus adopt $k\!=$1.1, R$_{min}\!=$1.6
for color measurements in our
analysis.

\subsection{Point-Spread Function Matching} \label{section:psfmatching}
To measure accurate colors in constant-sized apertures, it is
necessary to match the PSFs of the imaging used.
This is relevant as the diffraction limit of {\it Hubble} varies
by $>$2$\times$ across the wavelength range used here.  While this is
slightly mitigated as the pixel scale used in our drizzled images (60
mas) results in the bluer-band PSFs being somewhat larger than the
diffration limit in those bands, we find that the effective PSFs in our
WFC3/IR images are still significantly larger than that in our ACS
images.  We therefore match the PSF in all images to that in our F160W
images.

The first step is to construct the PSF.  As our main concern is
accurate colors, we elected to measure empirical PSFs by stacking
stars in our images.  We selected stars by identifying the stellar locus in a plot of the half-light
radius versus magnitude, constructed using a preliminary run of Source
Extractor.  The exact location of the stellar locus varies from
band-to-band (and a bit from field-to-field), thus the stellar
selection box was manually derived for each band/field combination. To mitigate
the effect of crowding, we only accepted a star as a PSF star if it had
no object within 200\arcs\ brighter than one magnitude fainter than
the star in question.  This preliminary list of PSF stars was then
visually inspected, to remove remaining crowded sources, objects near
the egde of an image, and non-stellar-like interlopers.  The stars
which passed this cut were then stacked together, in 101$\times$101
pixel cutouts.  During this process, the stars were centered in the
cutout by upsampling by a factor of 10, recentering, then sampling
back down to the original 60 mas pixel scale.  Each star was also rotated by a
random amount, such that the final PSF did not contain significant
diffraction spikes (this is relevant as some fields were obtained at
multiple orients, so the final PSF had several diffraction spikes; the
fraction of flux in these spikes is negligible, so we elected to
remove them from the PSF in this manner).  Each stacked-star PSF then
had any remaining residual background subtracted, and was normalized to
unity.  Example PSFs from the EGS field are shown in Figure~\ref{fig:psfs}.

For each field, we created kernels to match the individual PSFs to that in the F160W
band in each field using the \texttt{pypher} Python
routine\footnote{https://pypher.readthedocs.io}.  Each image was then
convolved with its respective kernel.  To examine the accuracy of this
PSF homogenization process, we measured the curves-of-growth of the
identified stars in these images, and compared those from each band to
the F160W band.  In nearly all cases, the enclosed flux at a
radius of 0.4\arcs\ matched that in the F160W band to within 1-3\%.
We show the results from the EGS field in Figure~\ref{fig:psfs}.

\subsection{Catalog Construction}

Using the conclusions from the above tests, we proceeded to
construct our photometric catalog in each of our fields.  For each
catalog, we used the F160W image as the detection image, and cycled
through every available image as the measurement image.  For the
remainder of this paper, we will refer to filters by their
characteristic letter, with their three-number identifier in a
subscript (e.g., F160W $\rightarrow$ $H_{160}$).  The EGS, COSMOS
and UDS fields are fully covered by the $V_{606}$, $I_{814}$,
$J_{125}$, $JH_{140}$ and $H_{160}$ filters.  We note that the $JH_{140}$ was
obtained as direct imaging for the 3D-HST slitless spectroscopy program \citep{momcheva16}, and
is thus quite shallow.  We also used available $Y_{105}$ data in
these fields, which consists of a few pointings obtained to followup potential
high-redshift sources (PID 13792, PI Bouwens).  The two GOODS fields
contain these same filters (including full-field $Y_{105}$ coverage),
as well as full-field coverage in the $B_{435}$, $i_{775}$, and
$z_{850}$ filters.  The exception is the ERS region of GOODS-S, which
has $Y_{098}$ rather than $Y_{105}$.  We collate the Source Extractor
results into a single catalog per field, including several measures of
the photometry (e.g, Kron, isophotal, and circular apertures).  Fluxes
and errors in all filters were corrected for Galactic extinction
assuming a \citet{cardelli89} Milky Way attenuation curve, with E(B-V)
$=$ 0.006, 0.016, 0.019, 0.010 and 0.008 for the EGS, COSMOS, UDS,
GOODS-N and GOODS-S fields, respectively.

\subsubsection{Aperture Correction}

Our default photometry consists of colors measured in the small Kron
aperture, corrected to total using an aperture correction.  To
calculate the appropriate aperture correction, we performed one
additional run of Source Extractor per field with the $H_{160}$-band
as the measurement image, this time with the default larger Kron
parameters, which as mentioned above approximately recover the total
flux.  The aperture correction is thus the ratio of the flux measured
in this larger Kron aperture to that in our tuned smaller Kron
aperture.  As we discovered in our simulations in \S 2.2, even
this larger Kron aperture fails to recover the flux in the wings of
the PSF.  Our simulations found that an additional 20\% correction was
needed to recover the total flux of simulated sources.  Our final
aperture correction thus includes this additional multiplicative factor.
As all images have been PSF-matched to the $H_{160}$-band PSF, we apply this
$H_{160}$-measured aperture correction to all filters, though we again
reiterate that colors are measured in small Kron apertures.

\subsubsection{Empirical Noise Estimation}
While Source Extractor provides a noise estimate based on the provided
rms map, we have elected to calculate our noise empirically.  We
follow the procedure outlined in \citet{papovich16}, omitting the
Poisson correction, as our sources of interest will be faint and this
noise component will be negligible (see also \citealt{wold19}, \citealt{rojasruiz20} and
\citealt{stevans21} for implementations of this procedure).  This
calculation relies on measuring the noise fluctuations as a function of
the number of pixels in an aperture.  We do this by measuring the flux
at 10$^{4}$ random postions in each science image.  We do this with
Source Extractor, making a detection image which contains zeroes at
all locations, except at these random positions, where we set the
pixel values to 1.0.  At each of these
positions, we measure the flux in 12 circular apertures, ranging from
1--35 pixels (0.06 -- 2.1\arcs) in diameter.  For each of these
aperture diameters, we measure the noise as the spread in the 10$^{4}$
measured values.  We do this by fitting a Gaussian to the negative
side of the measured flux distribution, to omit any positive flux from
real sources.  The measured $\sigma$ value from this Gaussian fit is
thus the 1$\sigma$ noise value in an aperture of that size.

\begin{figure}[!t]
\epsscale{1.2}
\plotone{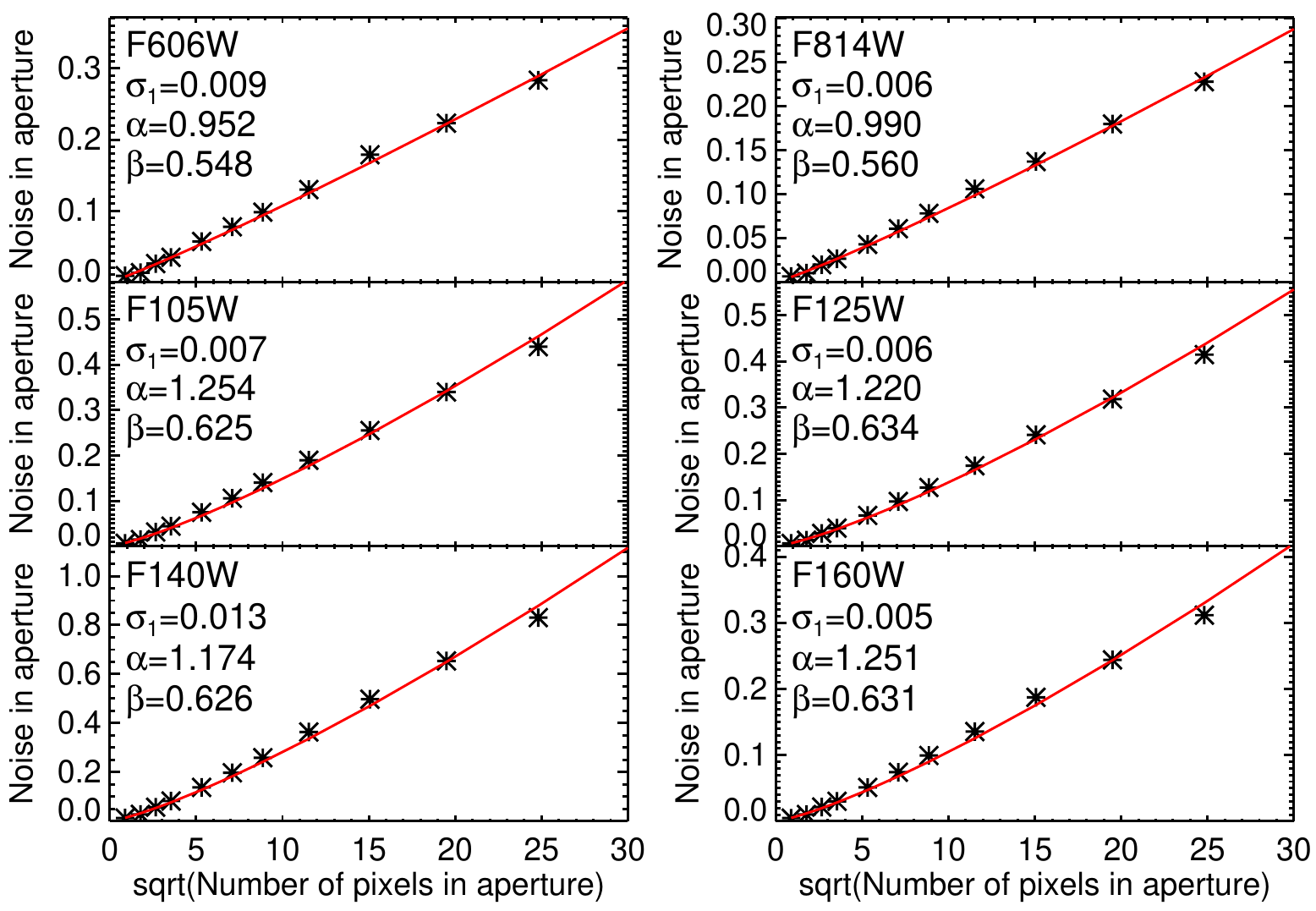}
\vspace{-3mm}
\caption{The results of our empirical noise simulations for each
  filter in the COSMOS field.  The data points show the measured
  $\sigma$ from a Gaussian fit to the negative side of the flux
  distribution in randomly placed apertures of different sizes
  (plotted versus the square root of the number of pixels in the
  aperture).  The red line shows the fit to the data using Equation 1,
with the parameters of the fit inset in the panels.  This fits were
used to calculate the noise for each object given the number of pixels
in its custom Kron aperture.}
\label{fig:noise}
\end{figure}

\begin{deluxetable*}{ccccccccccc}
\vspace{2mm}
\tabletypesize{\small}
\tablecaption{Limiting Magnitudes in {\it HST} Imaging}
\tablewidth{\textwidth}
\tablehead{
\colhead{Field} & \colhead{$B_{435}$} & \colhead{$V_{606}$} &
\colhead{$i_{775}$} & \colhead{$I_{814}$} & \colhead{z$_{850}$} &
\colhead{$Y_{098}^{\ast}$} & \colhead{$Y_{105}$} & \colhead{J$_{125}$} &
\colhead{$JH_{140}$} & \colhead{$H_{160}$}}
\startdata
EGS&---&27.95&---&27.60&---&---&26.79&27.05&26.36&27.10\\
COSMOS&---&27.63&---&27.31&---&---&26.93&26.94&26.33&26.87\\
UDS&---&27.63&---&27.53&---&---&27.01&26.96&26.23&26.92\\
GOODS-S Wide&28.04&28.43&27.71&27.80&27.46&27.32&27.02&27.14&26.24&26.79\\
GOODS-S Deep&28.20&28.59&27.96&28.13&27.52&---&27.74&27.79&26.34&27.47\\
GOODS-N Wide&27.99&28.06&27.82&27.92&27.40&---&27.01&27.13&26.59&26.98\\
GOODS-N Deep&28.01&28.23&28.14&28.46&27.59&---&27.66&27.94&26.69&27.64
\enddata
\tablecomments{The listed values are 5$\sigma$ limiting magnitudes
measured in 0.4\arcs-diameter apertures, and corrected to total
assuming a point source flux distribution.  $^{\ast}$$Y_{098}$ data is
only avaiable in the northern section of GOODS-S (in the WFC3 ``Early
Release Science" field).}
\label{tab:tab2}
\end{deluxetable*}

Inspecting these measured $\sigma$ values, it is apparent that, as expected, there
is a clear dependence of the noise on the number of pixels in an
aperture.  We were able to robustly fit a curve to this
distribution using the functional form of:
\begin{equation}
\sigma_N = \sigma_1 \alpha N^{\beta}
\end{equation}
where $\sigma_{N}$ is the noise in an aperture containing $N$ pixels, and
$\sigma_1$ is the pixel-to-pixel noise (measured in each image as the
sigma-clipped standard deviation of all pixels, omitting pixels
belonging to real objects using the Source Extractor segmentation
map), following several previous works \citep[e.g.,][]{labbe03,whitaker11,papovich16}.  The parameters $\alpha$ and $\beta$ are free parameters, which
we fit for using the IDL implementation of MPFIT.  We note that
\citet{papovich16} had an extra term in their noise equation
($\sigma_1 \gamma N^{\delta}$), however, in our exploration, we were
able to achieve good fits with just the two free parameters.  We show
an example of these fits for the COSMOS field in
Figure~\ref{fig:noise}.

To calculate the noise for a given source, we calculated $N$ for our
Kron aperture using the Source Extractor measured ellipse
properties (KRON\_RADIUS, A\_IMAGE and B\_IMAGE), and using a given
aperture diameter for the circular apertures.  Lastly, these noise
estimates were multiplied by the ratio of the rms image value at the
central postion of a given source to the median rms value of the whole
map, thereby allowing the noise to be representative of the local noise level.
In Table~\ref{tab:tab2} we list the implied 5$\sigma$ point-source depths for each
filter and field using these noise functions.  \edit1{As a sanity
  check, we calculated the ratio between our adopted empirical errors to the Source
  Extractor calculated errors, finding a median ratio of
  $\sim$0.8--0.9 in the catalogs for all five fields.}

\subsubsection{IRAC Photometry with TPHOT}

We performed IRAC photometry on the final S-CANDELS 3.6 and 4.5 $\mu$m
mosaics in the five CANDELS fields \citep{ashby15}, which achieve a total
integration time of at least 50 hr for most of the area used in this
study.  Because IRAC images have significantly lower spatial
resolution (PSF FWHM $\sim$ 2\arcsec) than {\it HST}/WFC3 ($\sim$
0.2\arcsec), multiple sources in the {\it HST}/WFC3 imaging may appear
blended together in the IRAC imaging.
To mitigate the effect of source confusion in IRAC on flux estimation,
we performed PSF-matched IRAC photometry using information in a
high-resolution image as a prior. Specifically, we exploited the {\tt
  T-PHOT} software \citep{merlin15,merlin16}, using our $H_{\rm 160}$
detection image as a prior and the IRAC PSF as
transfer kernel between the {\it HST} $H_{160}$-band and IRAC
data, following the procedure outlined in \citet{song16a}.

The IRAC PSF was constructed by identifying and stacking point sources
in the IRAC mosaics. Similar to \S 2.4, we first identified point
sources in each band in a half-light radius versus magnitude diagram
in IRAC imaging. Then, we visually inspected them in the  {\it
  HST}/$H_{160}$ and IRAC data, to make sure that they appear to be
point sources in the {\it HST} imaging and not to be severely blended
with nearby sources in either IRAC band.
``Clean'' point-sources for each field identified from the above step
were then over-sampled to the pixel scale of the {\it HST}/$H_{160}$
image, registered, normalized, and median-combined to create the final
IRAC PSFs in each band and in each field.
 
Before feeding the {\it HST} segmentation map into {\tt T-PHOT}, we
enlarged the segmentation map using the program {\tt dilate}
\citep{desantis07}.  This is to account for the previous finding that
Source Extractor tends to underestimate the isophotal area of fainter
objects more \citep{galametz13,guo13}. To prevent their IRAC fluxes
measured based on isophotes from being underestimated, we thus
enlarged the segmentation map using the factor devised by the CANDELS
team \citep{galametz13,guo13}.
 
The IRAC fluxes for each source detected in the {\it HST} imaging were then measured with {\tt T-PHOT}. 
In {\tt T-PHOT}, models for the IRAC data for each source were generated by convolving the enlarged segmentation map with a transfer kernel, leaving the IRAC flux as free parameter.
The best-fit IRAC fluxes were then obtained by solving for the solution that best-matches the observed IRAC data.
We refer the reader to \citet{song16a} for full details.

\section{Sample Selection}

One concern when selecting such distant galaxies is sample
contamination.  The number of $z >$ 8 galaxies in these {\it HST} data are likely few, and contaminating objects at lower redshifts with similar colors 
likely outnumber our galaxies of interest.  With this in mind,
previous work in this epoch typically employs a highly conservative sample
selection, including strict color-cuts designed to rule out these interlopers. 
While the upcoming {\it James Webb Space Telescope}, with its larger
collecting area and longer wavelength baseline will significantly
improve our knowledge of galaxy evolution in this epoch, we can make
gains now by using all available imaging when selecting galaxies.
Specifically, {\it Spitzer}/IRAC photometry is not often used as a
primary observable when selecting galaxies, due to its shallower
depth, and larger point-response function size.  However, the CANDELS fields all now have deep 50 hr
imaging at both 3.6 and 4.5 $\mu$m from the S-CANDELS survey (\S 2.5.3), extending to $m_{3\sigma} \sim$ 26.5, reaching the
potential rest-optical brightnesses expected for our objects of interest.

\begin{figure*}[!t]
\epsscale{1.15}
\plotone{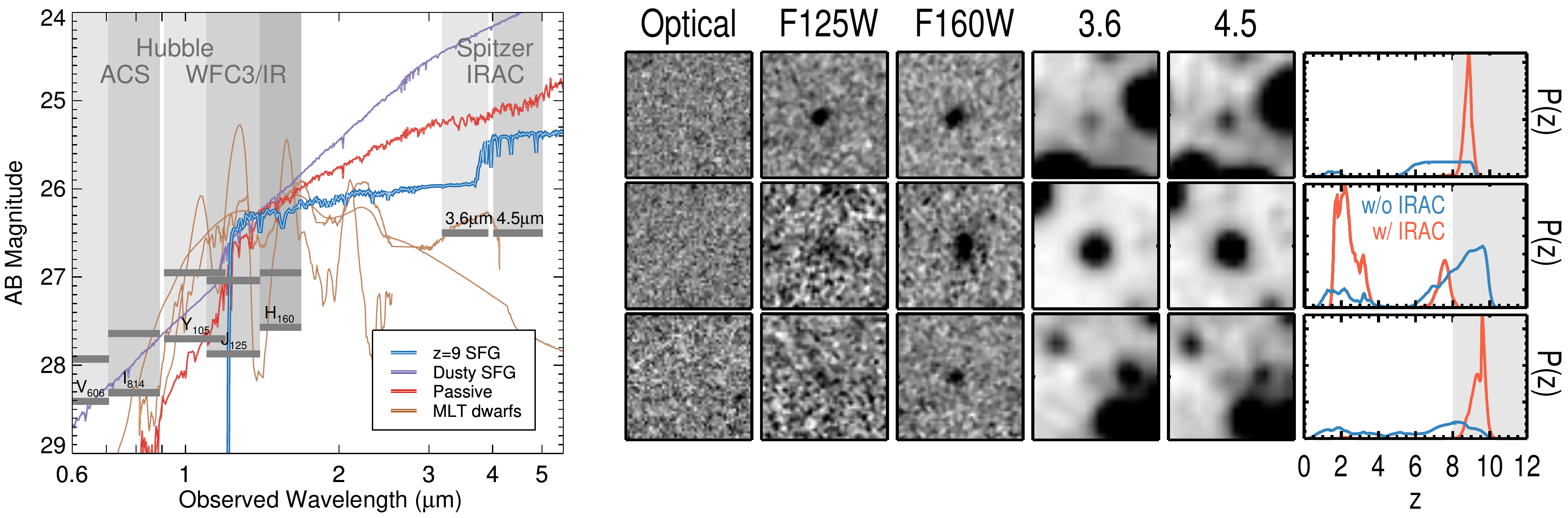}
\vspace{-2mm}
\caption{\emph{Left}) Model spectra of a star-forming galaxy at $z =$
  9 (log[M$_{\ast}$/M\sol] = 9.9; E(B-V)=0.2; SFR=90 M\sol\
  yr$^{-1}$), compared to a number of potentially contaminating
  objects, including a passive and dusty star-forming galaxy, both at
  $z =$ 2.2 (log[M$_{\ast}$/M\sol] $=$ 9.3, to roughly match the flux
  of the high-redshift model in the WFC3 bands), and example M, L and T
brown dwarfs \citep{burgasser14}.  The background gray shading denotes the filters used
common in all fields (the shallower F140W is not shown), and the gray
bars show the 5$\sigma$ limiting magnitudes for the CANDELS Wide (top)
and deep (bottom) surveys.  All objects look roughly similar in the
WFC3/IR bands, but diverge rapidly by {\it Spitzer}/IRAC wavelengths.
This highlights the utility of using these longer-wavelength data when
selecting high-redshift galaxies.
\emph{Right})  Stamp images (3\arcs\ in {\it HST}, 10\arcs\ in IRAC) of
three example galaxies in the {\it HST} and IRAC filters (the optical is a stack of
F606W$+$F814W).  In {\it HST} only, all objects show no optical flux,
and a red $J-H$ color, consistent with a $z >$ 8 solution, indicated
by the blue P(z) curve in the right panel.  The inclusion of IRAC data
either solidifies the high-redshift solution (top and bottom rows), or
shows the low-redshift solution to be more likely (middle row).  The
inclusion of IRAC imaging, even with a faint or no detection, can thus
significantly increase the fidelity of photometric redshift measurements.}
\label{fig:iracfig}
\end{figure*}

In the left panel of Figure~\ref{fig:iracfig}, we highlight this issue further,
showing a model spectrum of a $z =$ 9 galaxy, compared to likely
contaminants, including a passive galaxy at $z =$ 2.2, a
dusty-star-forming galaxy \citep{casey14} at $z =$ 2.2, and example M,
L and T brown dwarfs.  This figure highlights that these objects all
have very similar observed colors at $\sim$1.2--1.8 $\mu$m, the
wavelengths where we detect light from true $z >$ 8 galaxies with {\it
  HST}.  While these contaminants all have flux at shorter wavelengths, it is weak, and
often not detectable at typical {\it HST} depths.  However, moving to
redder wavelengths, these objects have divergent colors.  The spectral
energy distribution (SED) of a true
high-redshift galaxy is mostly flat, with a potential small jump due to
the Balmer break (depending on the age of the galaxy, e.g., \citealt{roberts-borsani20,laporte21}) or nebular
emission lines (not shown for clarity).  The $z =$ 2.2 galaxies are
both much brighter due to their intrinsically red SEDs, while the potential stellar contaminants become
blue at these wavelengths, as we are probing molecular absorption
bands near the peak of their thermal emission.

In the right panel of Figure~\ref{fig:iracfig}, we illustrate the utility of this IRAC
imaging using three examples.  All three rows show objects which have a
significant probability of being at $z >$ 8 when measuring photometric
redshifts with only {\it HST} imaging.  When adding IRAC imaging to
the first object, the photometric-redshift solution peaks up tightly
at $z \sim$ 9, as the IRAC flux is comparable to the {\it HST} flux
(e.g., a roughly flat SED), and the IRAC [3.6]$-$[4.5] color is red,
indicative of either a Balmer break, or, more likely, strong
[\ion{O}{3}] emission in the redder band.  The third row has a similar
result, though in this case the IRAC detection is very low
signal-to-noise, illustrating that even the lack of an IRAC detection is
highly useful.  The middle row shows an object which is now more
likely at $z \sim$ 2, due to the very bright IRAC fluxes.
Importantly, this source did have a $z \sim$ 2 potential solution even
with {\it HST}-only data, but the inclusion of IRAC makes this the
dominant redshift solution.

In this work, we are thus motivated to use photometric redshifts as
our primary sample selection tool (as opposed to first applying
color-cuts, and then a photometric redshift analysis) to make full use
of the available {\it Spitzer}/IRAC data in these fields.  We note
that these fields also contain an abundance of ground-based imaging.  We
first select our galaxy sample using space-based data alone, but in \S
4 we vet this sample using the available ground-based imaging.

\subsection{Photometric Redshift Measurements}

We use photometric redshifts to select our galaxy sample,
following techniques similar to our previous work
\citep[e.g.][]{finkelstein15}.  While photometric redshift
calculations at these high-redshifts primarily key off of the Lyman
break, and are in principle similar to color-color selection
\citep[e.g.,][]{steidel93,giavalisco04,bouwens15,bridge19}, this method has the advantage that it
simultaneously uses all available photometric information.  This
simplifies the selection process and results in a
more inclusive sample, including objects lying just outside
color-selection windows \citep[e.g.,][]{mclure09,finkelstein10,finkelstein15,bowler12,bowler14,bowler15,atek15,livermore17,bouwens19}.  Additionally, these codes typically provide
a redshift probability distribution function (PDF), which we will
denote $\powerset(z)$, which provides more information on the redshift
than the binary in/out of color-selection.

\begin{figure*}[!t]
\epsscale{1.15}
\plotone{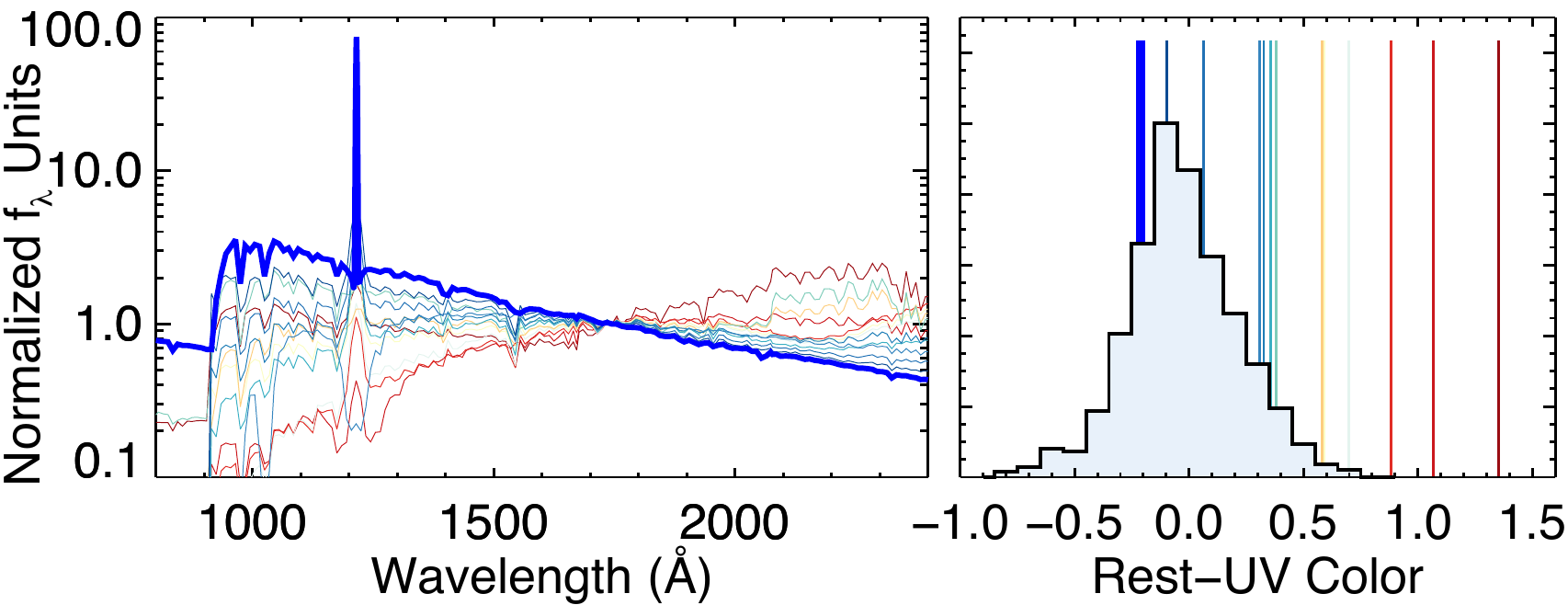}
\vspace{-2mm}
\caption{\emph{Left}) The rest-frame ultraviolet region of the EAZY
  template set used in our analysis to measure photometric redshifts.  The
thin lines show the latest standard template set
(tweak\_fsps\_QSF\_v12\_v3), while the thick line shows an additional
template we added from the low-mass, star-forming galaxy BX418
\citep{erb10}, which is very blue, and has high nebular-line EWs.  All
templates are normalized to their flux density at 1750 \AA.
{\it Right}) The histogram shows the distribution of $J-H$ colors for
$z =$ 6--8 galaxies from the catalog of \citet{finkelstein15}.  The
solid lines show the rest-UV color of the templates in the left panel,
using the ratio of the 2000/1500 \AA\ flux to calculate the color,
which is approximately the rest-frame wavelength probed by $J-H$ at $z
\sim$ 7.  The bluest EAZY template only reaches a rest-UV color of
$-$0.1, while roughly half of the comparison high-redshift sample has
bluer colors (and we expect $z >$ 8 galaxies to be at least as blue).  The BX418 model extends our template set another $\sim$0.1
mag bluer in color.  This full template set can reproduce the colors
of $\sim$85\% of the \citet{finkelstein15} $z =$ 6--8 high-redshift sample.}
\label{fig:templates}
\end{figure*}   

We use the photometric-redshift (photo-$z$) code $\tt{EAZY}$
\citep{brammer08} to perform our redshift estimation, assuming a
flat luminosity prior as the epoch we are probing has not been well
characterized.  Use uses all available photometry and compares it to a series of templates,
allowing non-linear combinations of any number of provided templates.
We make use of the latest EAZY template set, known as
``tweak\_fsps\_QSF\_v12\_v3'', which is based on the Flexible Stellar
Population Synthesis code \citep[FSPS, ][]{conroy10}, including a more
representative set of templates (inclusive of emission lines) than the original EAZY v1.0
templates.  This template set has further been corrected (or
``tweaked'') for systematic offsets observed between data and the
models \footnote{Further information on EAZY and these templates can
  be found at https://github.com/gbrammer/eazy-photoz/}.  

Figure~\ref{fig:templates} shows the full slate of EAZY templates from this
set.  In the right-hand panel, we compare the rest-UV color of these
templates (measured from the ratio of the 1500-to-2000 \AA\ flux
density in f$_{\nu}$ units) to the distribution of similar
rest-wavelength colors from $z >$ 6 galaxies in the CANDELS fields
(these are the $J-H$ colors for all galaxies with $z_{phot} >$ 6 from
\citealt{finkelstein15}).  Most of the standard templates are much
redder than typical high-redshift galaxies, and the bluest
template is only as blue as the median high-redshift galaxy.  For this
reason, similar to our previous work we add as an additional template
the $z =$ 2.3 galaxy BX418,
which is young, low-mass and blue \citep{erb10}.
This galaxy's color is 0.12 mag bluer than the bluest standard template, and
has a color bluer than 85\% of the known high-redshift galaxies.  We
add two versions of this template; one with the observed Ly$\alpha$
emission as shown in Figure~\ref{fig:templates}, and one where we
remove the Ly$\alpha$ emission, to account for blue galaxies whose
Ly$\alpha$ has been absorbed from a potentially neutral IGM
\citep[e.g.,][]{miralda-escude98,malhotra06,dijkstra14}.

For each field, we perform two runs of EAZY; one with all photometric
bands, and a second excluding the IRAC bands.  This second run is
included to allow true high-redshift galaxies which have unreliable
IRAC photometry due to poor deblending to still potentially be
selected based on their {\it HST} photometry only (see also \S 3.3.4).

\subsection{Sample Selection Criteria}
We use the results from EAZY in tandem with our photometric catalogs
to select our sample of $z \sim$ 9--11 galaxy candidates.  Following
our previous work \citep{finkelstein15}, we elect to use information present in the
$\powerset(z)$ to select our sample, rather than using the
best-fitting photometric redshift alone.
Our sample selection criteria are (where $\land$ and $\lor$ represent the logical
{\tt and} and {\tt or} operators, respectively):
\begin{enumerate}[label={[\arabic*]}]
\item S/N$_{H}$ $>$ 7 $\lor$ S/N$_{J}$ $>$ 7
\begin{itemize}[leftmargin=*]
\item[] This requires a significant detection in either the $J_{125}$ or $H_{160}$
  bands in a 0.4\arcs\ diameter photometric aperture.  We allow
  objects satisfying this significance level in either bad to make our sample, as
  blue galaxies at the lower edge of our desired redshift range ($z
  \sim$ 8.5) may have a higher significance in the $J_{125}$-band.  We also
  note that our requirement of a 7$\sigma$ detection is more
  conservative than the typically used 5$\sigma$.  However, with the
  limited photometric information available, we wanted to ensure our
  sample was free from spurious sources.  Finally, we use a 0.4\arcs\
  diameter aperture for all sources to consistently measure the
  significance of the peak of an object's emission, as the occasionally larger
  Kron aperture would result in a lower S/N.
\end{itemize}

\item RMS value in $V_{606}$, $I_{814}$, $J_{125}$, $H_{160}$ $<$ 10$^3$ [counts/sec]
\begin{itemize}[leftmargin=*]
\item[] This requires photometric coverage in all key {\it HST}
  filters, as in our reduced mosaics pixels with no exposure time have
  a flag value of  $RMS=$10$^4$.  
\item[] In the GOODS fields, we add the $B$-band to this
  requirement (and use $i_{775}$ rather than $I_{814}$, as it covers
  more area).
\end{itemize}

\item (S/N$_{V}$ $<$ 2 $\lor$ S/N$_{I}$ $<$ 2) $\land$ min(S/N$_{V}$,
  S/N$_{I}$) $<$ 1.4)
\begin{itemize}[leftmargin=*]
\item[] This allows a maximum 2$\sigma$ significance detection in
  either of the $V_{606}$ and $I_{814}$ optical bands, and also requires
  $<$1.4$\sigma$ significance in the weaker of the two bands.
\item[] In the GOODS fields, we also require less than 2$\sigma$
  significance detections in $i_{775}$, $z_{850}$ and $Y_{105}$, and
  we add $B_{435}$, $i_{775}$ and $z_{850}$ to the $<$1.4$\sigma$
  significance criterion.
\item[] In principle these requirements are superfluous given the
  photometric redshift requirements below, but we employ them to keep
  our sample selection conservative.
\end{itemize}

\item $\int$ $\powerset(z > 8)$ $>$ 0.6 $\land$ $\Delta \chi^2 = \chi^2_{low-z} - \chi^2_{high-z} >$ 3.5
\begin{itemize}[leftmargin=*]
\item[] The first criterion requires $>$60\% of probability to reside
  at $z >$ 8 (integrated out to the maximum redshift considered by EAZY of 15),
  allowing $<$40\% to be present in a low-redshift solution.  The
  second criterion requires the goodness-of-fit of any secondary
  low-redshift solution to be significantly worse than that of the
  high-redshift solution; we chose a threshold just less than a 95\%
  confidence requirement (which would be $\Delta \chi^2 >$ 4) to
  ensure we explore objects which fall just below this threshold.  We
  note that only one object has 3.5 $< \Delta \chi^2 < 4$
  (EGS\_z910\_26890, with $\Delta \chi^2 =$ 3.51), and that this
  increases to 12.9 with the inclusion of additional {\it HST} imaging
  (\S 4.3).
\end{itemize}

\item $\chi^2_{EAZY} <$ 60
\begin{itemize}[leftmargin=*]
\item[] This requires EAZY to have found a reasonably good fit,
  rejecting objects where even the best-fitting solution is not a good
  match to the observed photometry.
\end{itemize}

\item $\int\powerset(z\sim$ 9 $\lor$ 10 $\lor$ 11) $>$ $\int\powerset(z\sim$
  5 $\lor$ 6 $\lor$ 7 $\lor$ 8)
\begin{itemize}[leftmargin=*]
\item[] This requires the integral of $\powerset(z)$ in $\Delta z=$1
  bins centered at $z\!\!=$9, 10 or 11 to be greater than those centered
  at $z\!\!=$5, 6, 7 or 8.
\end{itemize}

\item r$_{h}$ $>$ 1.1 pixels $\lor$ Stellarity $<$ 0.9
\begin{itemize}[leftmargin=*]
\item[] These criteria are designed to remove potentially lingering
  image artifacts such as hot pixels or cosmic rays.  As objects in
  our redshift range of interest may be single $H_{160}$-band detections,
  these artifacts can satisfy the sample selection.  We thus require a
  half-light radius of greater than 1.1 pixels (following the analysis
  done in \citealt{rojasruiz20}), or a Source Extractor stellarity value of $<$ 0.9.
\end{itemize}

\item 22 $<$ $H_{160}$-band magnitude $<$ 26.6
\begin{itemize}[leftmargin=*]
\item[] The brighter cut corresponds to $M_{UV} \sim -$25 at $z =$ 9,
  and thus is much brighter than any source we expect to find in our
  sample, but imposing this cut removes interlopers such as bright
  stars which can contaminate the sample.  The fainter cut is
  approximately the 7$\sigma$ limit for our shallowest field (Table
  2), reflecting our goal of studying the brighter galaxy population
  in this epoch.  True candidates certainly exist in these data
  fainter than this limit, but as our goal here is to study bright
  galaxies, imposing this limit reduces the number of faint
  contaminants needing to be analyzed.
\end{itemize}

\end{enumerate}

Together, these criteria select sources which are photometrically
robust, modestly bright, have high photometric redshift likelihoods of residing at
$z >$ 8.5, and rely minimally on human intervention, limiting potential biases in our final sample.

\subsection{Contaminant Rejection}

Our initial sample as described in the previous subsection consists of
140 sources: 19 in EGS, 20 in COSMOS,
67 in UDS, 28 in GOODS-N and 6 in GOODS-S.  This is significantly
larger than any reasonable expectation of the number density of such
distant galaxies, thus we expect this initial sample to be dominated by
contaminant objects.  In this subsection, we detail the several types
of contaminants we screen for and remove.  During this screening we
rely on automated quantitative cuts whenever possible.  \edit1{We
  implement these steps in this order to remove the most obvious
  contaminants (e.g., persistence, diffraction spikes) first, prior to
more subjective removal.} We provide
images in the appendix for any objects removed based on a subjective visual inspection.

\subsubsection{Persistence}

Persistence from previously observed
bright targets could masquerade as high-redshift galaxies.   This
problem is exacerbated as CANDELS observed the $H_{160}$-band
before the $J_{125}$-band, and, as persistence fades with time,
the resulting photometric color could mimic
a $J_{125}$-band dropout.  
For each of our initial candidates, we searched the
MAST archive for observations done in the 24 hours previous to each of
the individual images (FLTs) which went into the CANDELS mosaics,
recording the maximum value in a 10 $\times$ 10 pixel box around the position of our object.  
We did this for both the CANDELS $J_{125}$ and $H_{160}$ images, as well
as the 3D-HST $JH_{140}$ imaging.  The latter were taken at a
different time, and so give an independent measure of whether a source
which is detected in multiple bands is a real source, or persistence.

If all of the images which went into the
$H_{160}$ mosaic had a fluence $<$100,000 e$^-$ at the same position as a
given candidate measured in an image from the preceding 24 hours, we consider the object to be not significantly
affected by persistence.  This threshold was devised empirically, to ensure all potentially spurious sources were
removed.  For objects with a $H_{160}$-band fluence $>$ 100,000 e$^-$, if they have
significant detections in the $J_{125}$ and/or $JH_{140}$ bands \emph{and} those
bands have a measured fluence $<$ 100,000 e$^-$, then we consider them
to be real objects and keep them in our sample.  However, in the more
common case, objects are undetected in $JH_{140}$, and have a fluence $>$
100,000 e$^-$ in both $J_{125}$ and $H_{160}$; these objects are removed from the sample.

This process removed 53 galaxies from our sample, 42 of which were
from the heavily affected UDS field as the previous
observations in that resulted in a large number of persistence
artifacts of varying brightness levels.
Our sample at this stage thus consisted of 87 galaxies.

\subsubsection{{\it HST} Visual Inspection}

We performed a detailed visual inspection of every object in this
initial sample to remove image
artifacts, examining each source in each band to determine whether a source is a
diffraction spike, an oversplit region of a nearby (usually) bright galaxy, near an
image edge, or any other obvious image artifact.  Through
this process, we identified 21 objects as being diffraction spikes, 11
as being oversplit regions of nearby galaxies, and 13 objects as other
artifacts. As this process is unavoidably subjective, we include
Figure~\ref{fig:spurious} in the Appendix showing $H_{160}$-band image
cutouts of objects removed in this process.  Our sample
size following this step was 42 galaxies.

\begin{figure*}[!t]
\epsscale{0.54}
\plotone{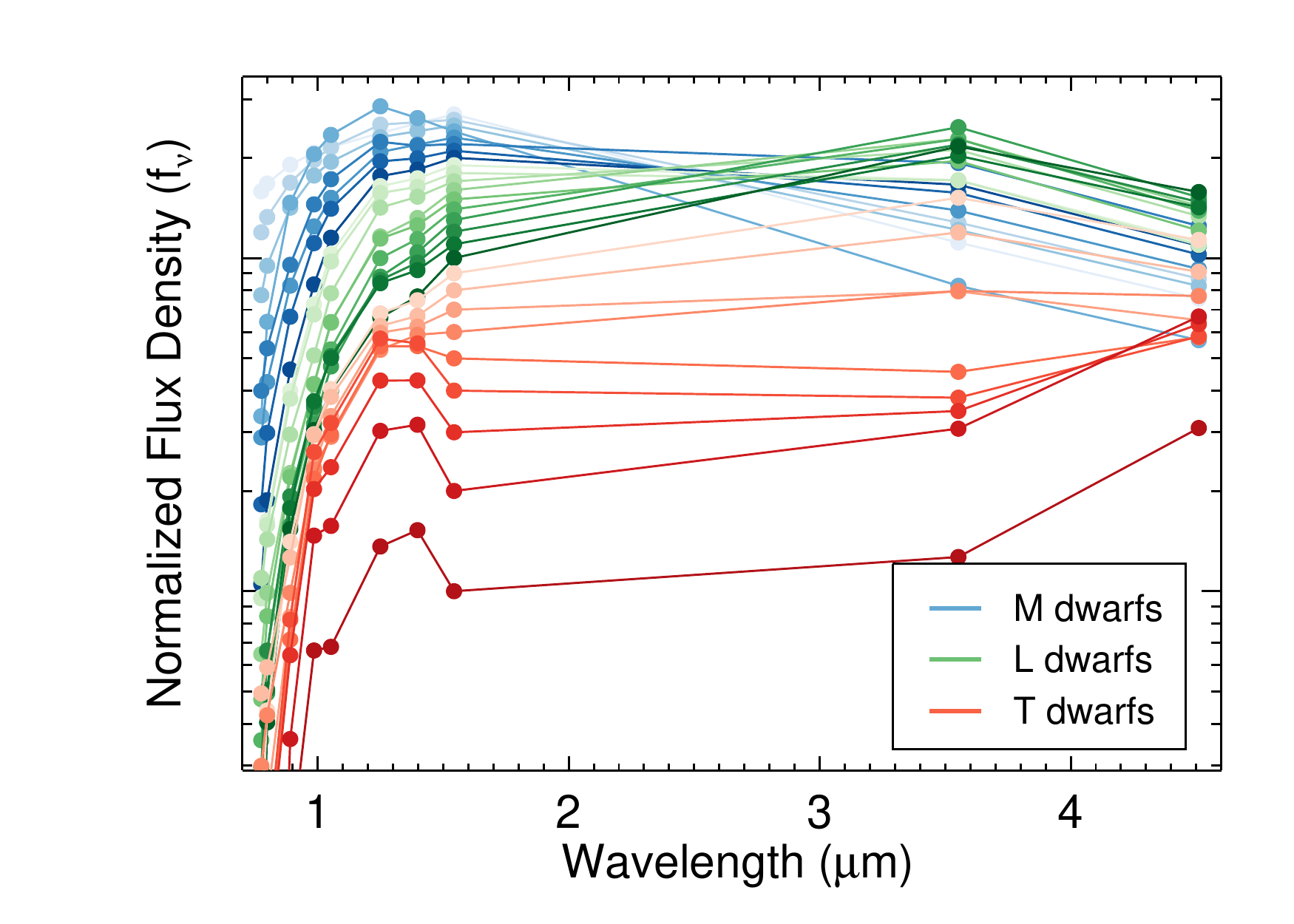}
\hspace{5mm}
\epsscale{0.52}
\plotone{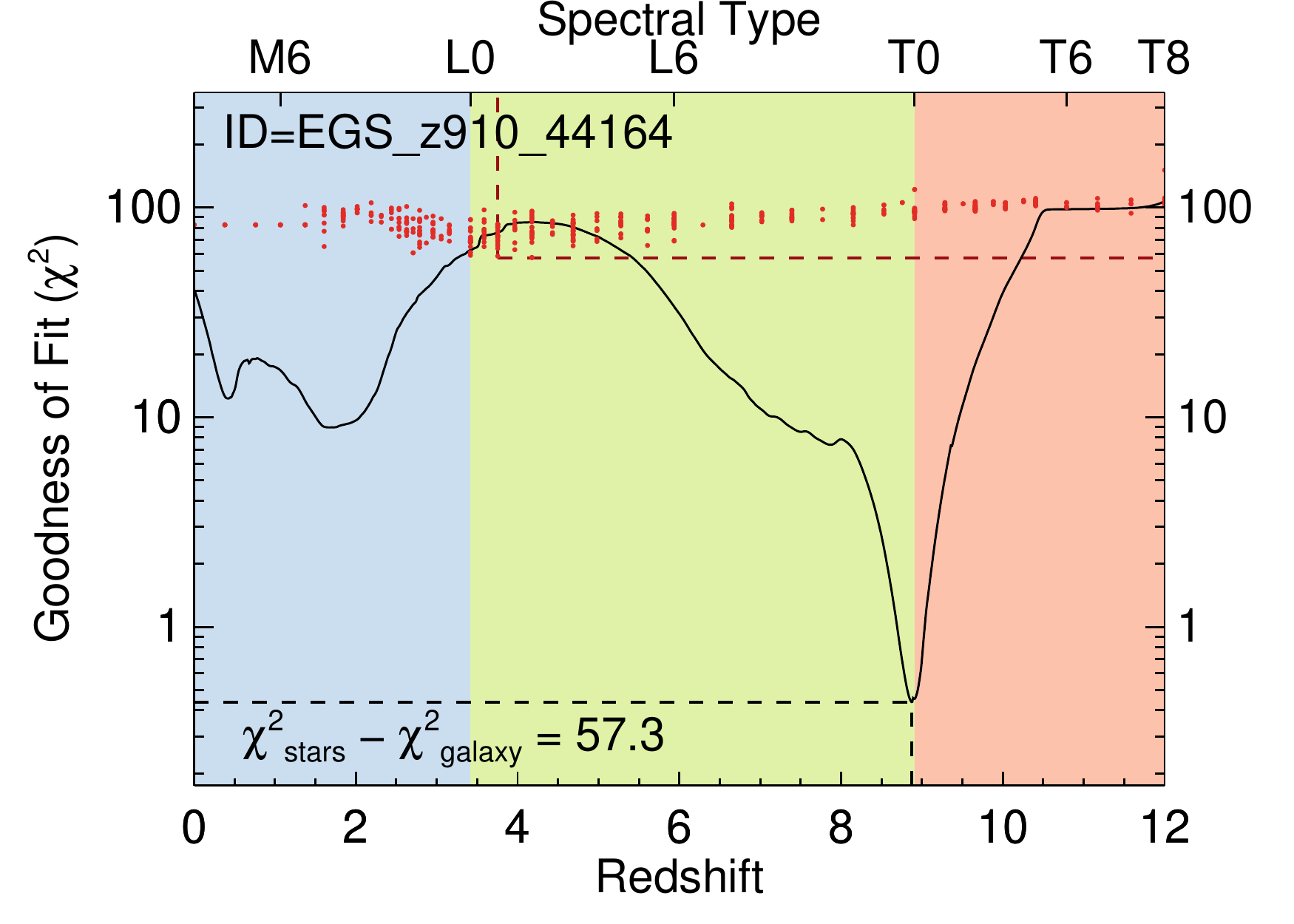}
\vspace{-2mm}
\caption{\emph{Left}) The stellar SEDs derived for MLT dwarf stars.
  The near-IR photometry was obtained by integrating the SpeX prism
  library spectra through the HST bandpasses, while the IRAC
  photometry was obtained by matching SpeX stars to IRAC-observed
  stars from \citet{patten06} at the same spectral type, normalizing
  by 2MASS $K$-band magnitude.  The full set of stellar SEDs we
  derived numbered 570; we show here a subset of 27 spanning the full
  spectral range used, of M3.5 -- T8.  {\it Right}) The results of the
  stellar template fitting for one candidate galaxy in the EGS field.
  Each red dot denotes the $\chi^2$ value for a given stellar template
  when compared to our observed photometry for this galaxy, while the
  black line shows the $\chi^2$ for the galaxy templates from EAZY.
  The blue, green and red shaded regions correspond to the top-axis M,
  L and T spectral types, respectively.
  The dashed lines show the best-fitting values, corresponding to $z
  =$ 8.9 for the best-fitting galaxy template, and a spectral type of
  L1 for the best-fitting stellar template.  As
  the galaxy solution is a significantly better fit ($\Delta \chi^2
  \gg$ 4), we conclude this source is not a stellar contaminant.  We
  find similar results ($\Delta \chi^2
  >$ 4) for all 14 candidates in the initial galaxy sample.
 }
\label{fig:bds}
\end{figure*}

\subsubsection{Local Noise Measurement}

While our noise calculation takes into account the difference in rms
map value at the position of our source compared to the median of the
image, it is still possible that the region surrounding an object of
interest has higher noise properties, rendering an object which
formally passes our S/N cuts in truth less significant.  To explore if
this affected any of our objects, we calculated the local noise in the
$H_{160}$-band in a
region around each of our sources.  We did this by randomly placing 0.4\arcs\ diameter apertures over a
region 450 pixels wide around a given source of interest.  The
random positions were constrained to be non-overlapping and also to
avoid real objects (using the Source Extractor segmentation map).  The
local noise was then taken as a sigma-clipped standard deviation of
the measured values.  We used this noise in tandem with the measured
0.4\arcs\ diameter flux values of the source to create a measure of the local signal
to noise.  We removed any object in our catalog which had a local S/N
$<$ 5.0 in the $H_{160}$-band.  This resulted in the removal of two objects total, both in
the GOODS-S field.

\subsubsection{Poor IRAC Deblending}

Inspection of the remaining 40 galaxies showed that several sources
were crowded in the IRAC bands.  While our use of the TPHOT photometry
deblending software was done to mitigate inaccurate photometry due to
crowding, these methods can fail when bright sources are sufficiently
close.  For this reason, we did another round of visual inspection on
the remaining sources, exploring whether the TPHOT residuals in the
vicinity of our objects of interest were sufficiently small. For any
source where the residuals were large, we then considered the results
of our EAZY run without the IRAC photometry.  If a given source
satisfied our sample selection criteria when the IRAC photometry was
not used with EAZY, then it remained in our sample, else it was
removed.  This process removed 26 objects from our sample, all of
which are shown in Figure~\ref{fig:baddeblend} in the Appendix.
We note that ID 44740 in GOODS-S formally did satisfy the selection
criteria without IRAC.  However, upon inspection of the IRAC image, it
was clear that significant source flux is coming from this object,
leading to a very red SED.  We thus concluded that it is not likely to
be a high-redshift galaxy, removed it at this stage, and also show
it in Figure~\ref{fig:baddeblend}.  Following this step, our sample consisted of 14 galaxies: 7, 3, 3,
1 and 0 in the EGS, COSMOS, UDS, GOODS-N and GOODS-S fields, respectively.

\subsubsection{Stellar Contamination Screening}

As shown in Figure~\ref{fig:iracfig}, low-mass stars and brown dwarfs
can have colors similar to very high redshift galaxies, and models of
the Milky Way stellar distribution predict that our observed fields
may have a few such objects at similar magnitudes as our target
galaxies, though the majority will have $J_{125} <$ 25 as they are primarily
in the Galactic disk \citep{ryan16}.  To explore
whether any of our candidates are likely to be a stellar contaminant,
we calculate the goodness-of-fit ($\chi^2$) of each object compared to
synthetic photometry created from a suite of stellar spectra,
considering an object for removal if it had $\chi^2_{Stellar} <
\chi^2_{EAZY}$, where the latter is the value from the best-fitting
EAZY galaxy template.

We use the IRTF SpeX stellar library \citep{burgasser14}, which has
empirical spectra for 165 M-dwarfs, 272 L-dwarfs, and 133 T-dwarfs,
spanning 0.6--2.5 $\mu$m in wavelength, and the full temperature range
for each spectral type.
We integrated each of these 570 spectra through the {\it HST} filter
set used in our study to derive bandpass-averaged fluxes.  However, as
these spectra only extend to 2.5$\mu$m, we cannot use them directly to
derive the photometry in the IRAC bands.  To do this, we make use of
the tabulated 2MASS and IRAC photometry for 84 MLT stellar dwarfs
(spanning M3.5 -- T8) from \citet{patten06}.  For each of the SpeX
stars, we find all stars in the \citet{patten06} sample with a similar
spectral type ($\Delta$type $<$ 1).  In two cases, there were not at
least three stars within a unit spectral type; in these cases we
extended the search to $\Delta$type $<$ 3, which yielded three stars
in both cases.

We estimate the IRAC photometry for the SpeX stars from these stars of
matched spectral type as
\begin{align} 
[3.6]_{SpeX}&=K_{SpeX} - \left<K_{Patten}-[3.6]_{Patten}\right>\\
[4.5]_{SpeX}&=[3.6]_{SpeX} - \left<[3.6]_{Patten} - [4.5]_{Patten}\right >
\end{align}
where the $K$-band magnitudes come from 2MASS, and all magnitudes are
converted from Vega to AB ($m_{AB} = m_{Vega}+$0.91, 1.39, 1.85, 2.79
and 3.26 for $J$, $H$, $K$, [3.6] and [4.5], respectively).  Combining
these IRAC fluxes with our calculated {\it HST}-filter fluxes provides
us with photometry for these 570 MLT dwarfs in the same filter set as
our observed galaxy candidates.

\begin{figure*}[!t]
\epsscale{1.15}
\plotone{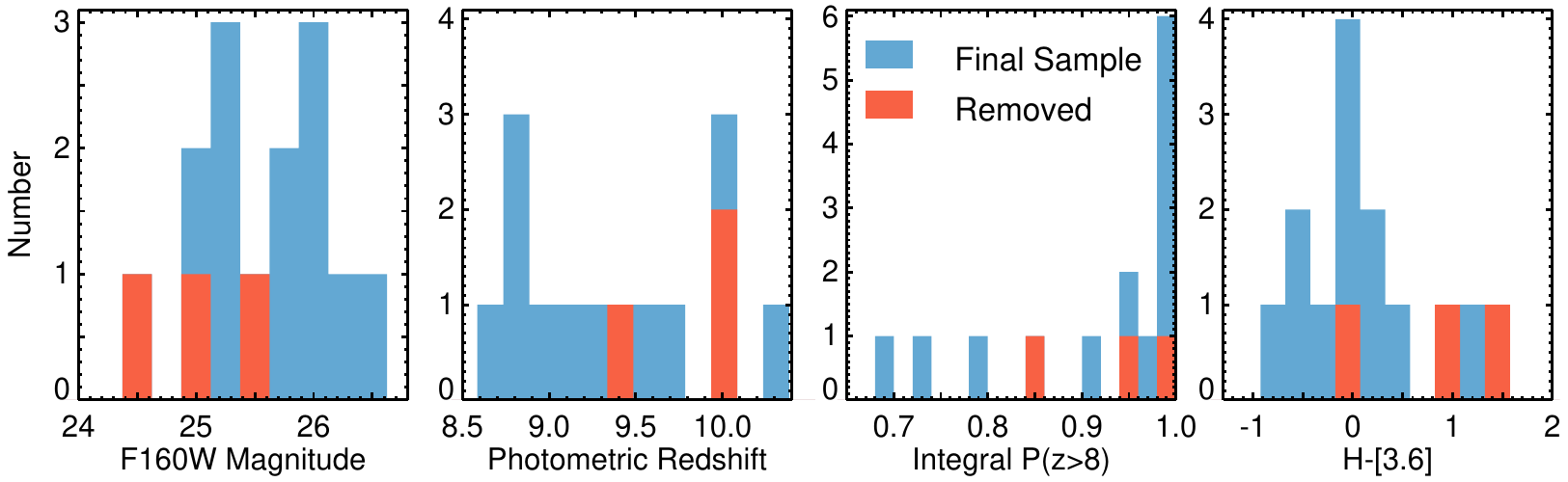}
\vspace{-2mm}
\caption{The distribution of properties from our {\it HST}$+${\it
    Spitzer}-selected sample.  Objects which are removed via further
  analysis in \S 4 are shown in red, while those which
  remain in our final sample are shown in blue.  From left-to-right,
  the plots are: the total $H_{160}$-band apparent magnitude, the best-fitting
  photometric redshift, $\int$$\powerset(z>8)$, and the $H_{160}$-[3.6] color.}
\label{fig:quad}
\end{figure*}   

\begin{deluxetable*}{ccccccc}
\vspace{2mm}
\tabletypesize{\small}
\tablecaption{Initial Candidate Bright ($H <$ 26.6) Galaxies at $z >$ 8.5}
\tablewidth{\textwidth}
\tablehead{
\colhead{ID} & \colhead{RA} & \colhead{Dec} & \colhead{F160W Mag} &
\colhead{$\int$ $\powerset(z>8)$} & \colhead{z$_{best}$} &
\colhead{Half-light radius}\\
\colhead{$ $}  & \colhead{J2000} & \colhead{J2000} & \colhead{} & \colhead{$ $} & \colhead{$ $} & \colhead{(arcsec)}}
\startdata
EGS\_z910\_6811&215.035385&52.890666&25.16 $\pm$ 0.05&1.00&8.84$^{+0.12}_{-0.25}$&0.138\\
EGS\_z910\_44164&215.218737&53.069859&25.41 $\pm$ 0.07&0.96&8.87$^{+0.17}_{-0.32}$&0.174\\
EGS\_z910\_68560&214.809021&52.838405&25.76 $\pm$ 0.08&1.00&9.16$^{+0.17}_{-0.36}$&0.111\\
EGS\_z910\_20381&215.188415&53.033644&26.05 $\pm$ 0.10&0.74&8.83$^{+0.15}_{-1.20}$&0.125\\
EGS\_z910\_26890&214.967536&52.932966&26.09 $\pm$ 0.07&0.80&8.99$^{+0.19}_{-6.84}$&0.099\\
EGS\_z910\_26816&215.097775&53.025095&26.11 $\pm$ 0.10&0.92&9.40$^{+0.24}_{-0.55}$&0.102\\
EGS\_z910\_40898&214.882993&52.840414&26.50 $\pm$ 0.11&0.69&8.76$^{+0.23}_{-1.33}$&0.103\\
COSMOS\_z910\_14822$^{\dagger}$&150.145769&2.233625&24.51 $\pm$ 0.04&1.00&9.44$^{+0.04}_{-0.34}$&0.140\\
COSMOS\_z910\_20646&150.081846&2.262751&25.42 $\pm$ 0.08&1.00&9.82$^{+0.19}_{-0.59}$&0.108\\
COSMOS\_z910\_47074&150.126386&2.383777&26.32 $\pm$ 0.10&0.99&9.59$^{+0.12}_{-0.60}$&0.103\\
UDS\_z910\_731$^{\dagger}$&34.317089&-5.275935&25.02 $\pm$ 0.12&0.85&10.10$^{+1.72}_{-0.65}$&0.252\\
UDS\_z910\_18697&34.255636&-5.166606&25.32 $\pm$ 0.09&1.00&10.04$^{+0.38}_{-0.22}$&0.174\\
UDS\_z910\_7815$^{\dagger}$&34.392823&-5.259911&25.82 $\pm$ 0.15&0.95&10.03$^{+0.98}_{-0.55}$&0.174\\
GOODSN\_z910\_35589&189.106061&62.242040&25.82 $\pm$ 0.05&1.00&10.41$^{+0.30}_{-0.07}$&0.121
\enddata
\tablecomments{The full list of candidate $z >$ 8.5 galaxies selected
  with {\it HST}$+${\it Spitzer} data.  $^{\dagger}$ These three objects
  were later removed from the sample based on ground-based and
  additional {\it HST} photometry,
  which implied they likely reside at $z <$ 8.  The positions have
  been corrected from the original CANDELS astrometry to PAN-STARRS
  DR1, which was tied to {\it Gaia} DR1. The reported half-light
  radii are measured by Source Extractor on the F160W image.}
\label{tab:tab3}
\end{deluxetable*}

\begin{figure*}[!th]
\epsscale{1.0}
\plotone{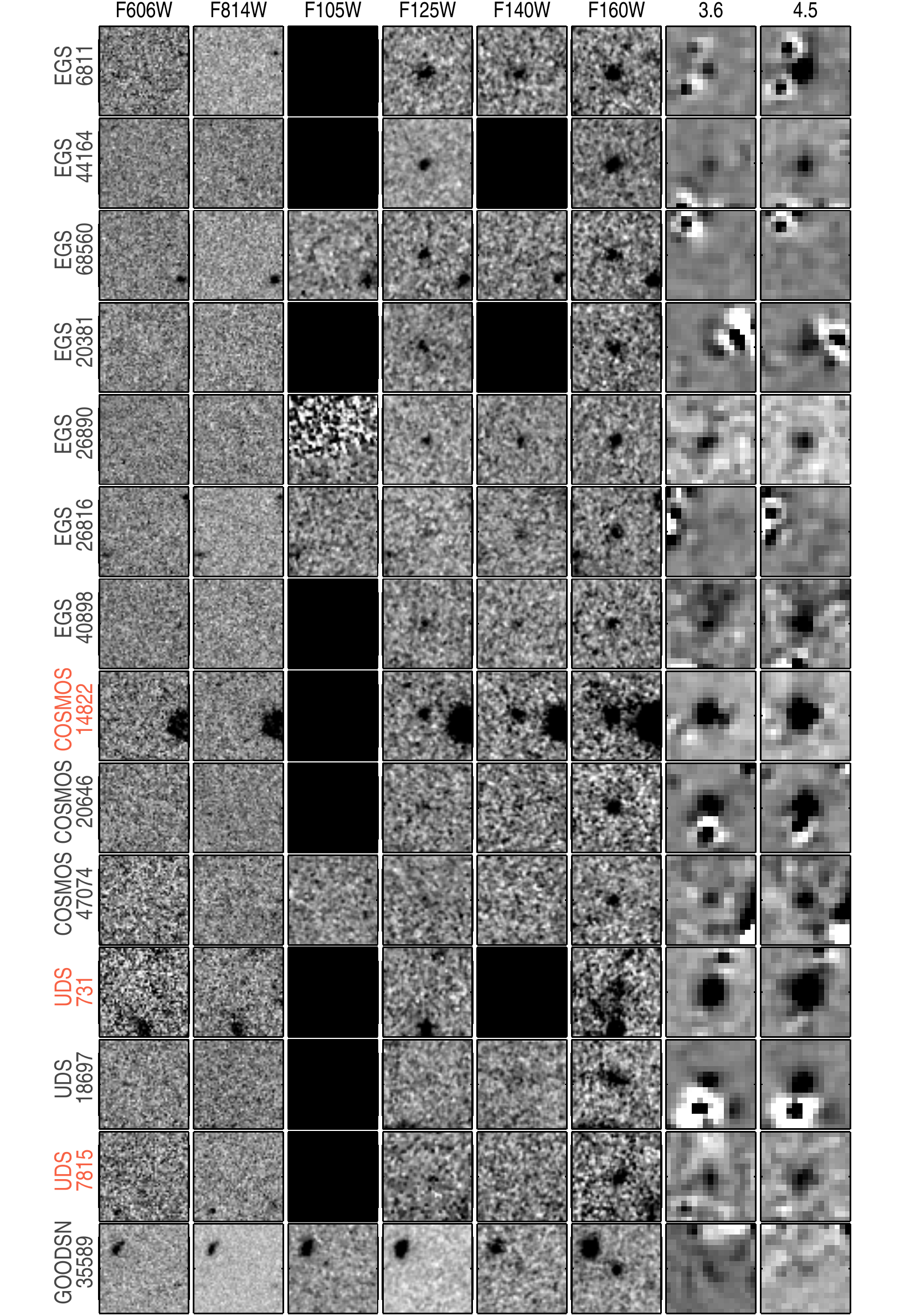}
\vspace{-2mm}
\caption{Cutout images of our {\it HST}$+${\it
    Spitzer}-selected sample, centered on the candidates.  The IRAC cutouts have the nearby neighbors
  subtracted with TPHOT.  The {\it HST} cutouts are 3\arcs\
  $\times$ 3\arcs, while the IRAC cutouts are 10\arcs\ $\times$
  10\arcs.  The three sources removed during our vetting process have
  their labels in red.}
\label{fig:stampfig}
\end{figure*}

The left panel of Figure~\ref{fig:bds} shows the range of stellar SEDs
we derived from this process, highlighting that these stellar objects
have a range of SED shapes and features.  To explore whether these
objects could be contaminating our galaxy sample, we calculated the
goodness-of-fit $\chi^2$ statistic between our candidate galaxies and
each of these 570 stellar templates, where the stars were normalized
to the galaxy fluxes using a weighted mean of the flux ratios in all
bands.  We give an example of this process in the right panel of
Figure~\ref{fig:bds}, showing the EAZY $\chi^2$(z) alongside the
stellar $\chi^2$ as a function of spectral type.  For this object,
which has a best-fitting $z_{phot}\!\!=$8.9 with $\chi^2_{EAZY}\!\!=$0.4,
the best-fitting stellar model has a spectral type of L1, with
$\chi^2_{stars}=$57.7, with $\Delta \chi^2= \chi^2_{stars} -
\chi^2_{EAZY} =$ 57.3.  We consider an object to be a potential
stellar contaminant if $\Delta \chi^2 <$ 4, such that the stellar SED
cannot be ruled out at $>$95\% confidence.  We find that none of our
14 candidate galaxies have $\Delta \chi^2 <$ 4 (only EGS\_z910\_26890 has
$\Delta \chi^2 <$ 15, with $\Delta \chi^2 <$ 5.3).
We thus conclude that none of the 14 objects in this initial sample are likely to
be stellar in nature.

\subsubsection{Active Galactic Nuclei}
Although unlikely to be found in our relatively small area at such
extreme redshifts, we still consider whether our sources could be
dominated by emission from an accreting supermassive black hole.  The
easiest way to diagnose the presence of a dominant active galactic
nucleus (AGN) is via X-ray emission.  We searched the published X-ray
catalogs in these fields \citep{kocevski18,xue16,nandra15,elvis09} to see if there was a detection coincident
with our candidates, finding no such matches (the closest X-ray source
in the catalogs was at 6--30\arcs\ from our candidates).  This
unsurprisingly rules out bright unobscured AGNs as the power source
for our objects.  This does not however rule out fainter AGN
contribution to our source luminosities,
which should be easily discernable via emission-line ratios {\it JWST} spectroscopy.

\subsection{Summary of Initial {\it HST}$+${\it Spitzer}-Selected Sample}

Through the processes described above, we have selected a sample of 14
candidate galaxies at $z \sim$ 8.5--11 across all CANDELS fields at $H <$
26.6: 7, 3, 3, 1 and 0 in the EGS, COSMOS, UDS,
GOODS-N and GOODS-S fields, respectively.  We list properties of this
sample in Table 3, and summarize these properties in
Figure~\ref{fig:quad}.  By construction, this sample is well
constrained to have $z >$ 8, with 11/14 objects having $\int
\powerset(z > 8) \geq$ 0.85, and best-fitting photometric redshifts
from 8.8 $< z <$ 10.4.  Also by construction these sources are bright,
with one exceptionally bright source at $H =$ 24.5, which we consider
further below.   Image cutouts of these candidates are shown
in Figure~\ref{fig:stampfig}, highlighting the optical non-detection,
and robust F160W detection.

\subsection{Comparison to Color-Color Selection Methods}
\begin{figure}[!t]
\epsscale{1.1}
\plotone{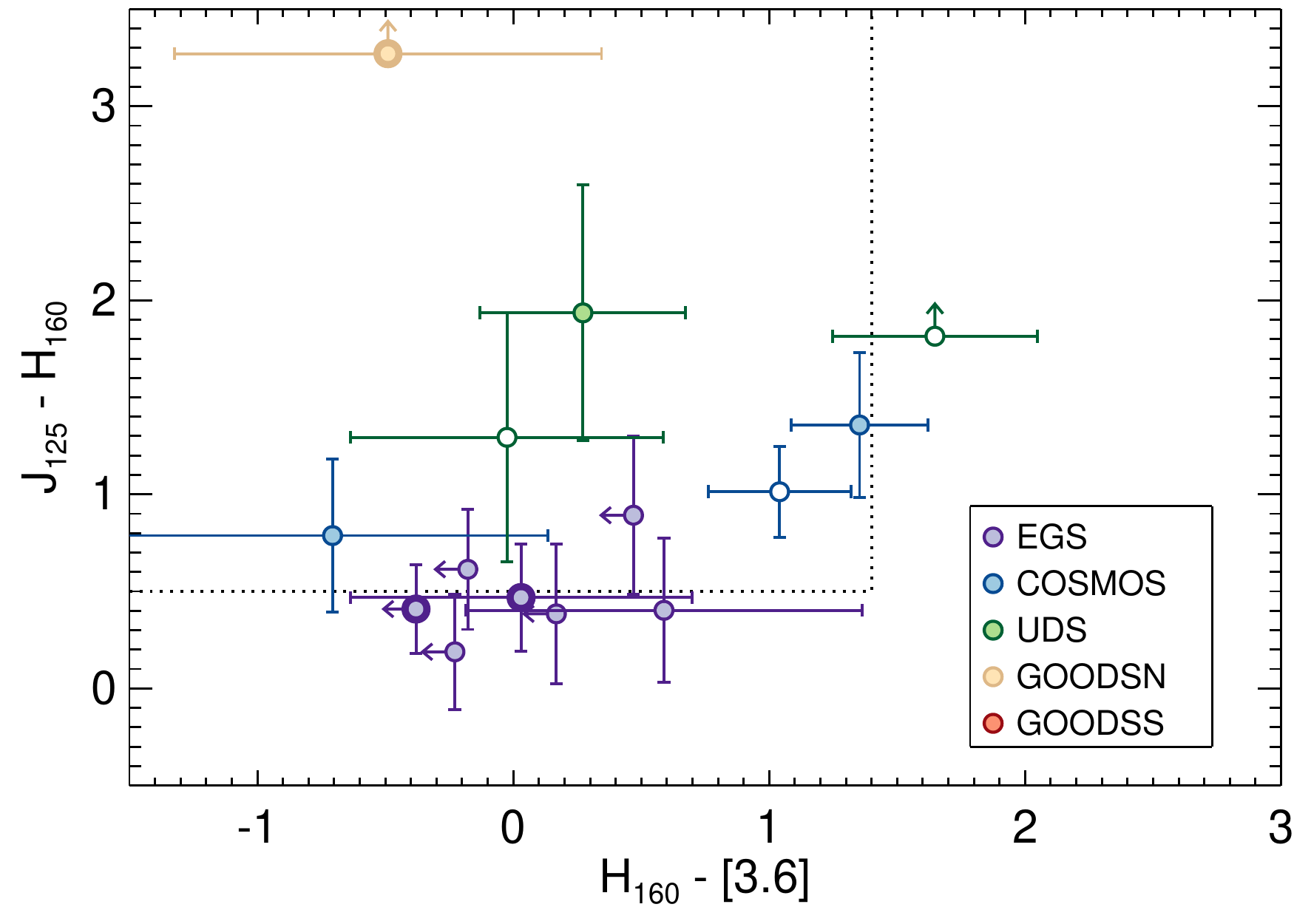}
\vspace{-2mm}
\caption{A color-color diagram showing the colors of our {\it HST}$+${\it
    Spitzer}-selected sample.   Objects which are removed via further
  analysis in the next section are shown as white-filled circles.
  Objects where a given band is detected at $<$1.5$\sigma$
  significance are replaced by 1.5$\sigma$ limits.  The dotted lines
  show the color-selection criteria employed by \citet{bouwens15c}.
  This color selection would have excluded six galaxies from this
  initial sample, five which have slightly bluer $J-H$ colors,
  indicating $z \sim$ 8.5--9.  The two sources with known
  spectroscopic confirmations are denoted by thicker circles.}
\label{fig:colcol}
\end{figure}

 We show the colors spanned by these 14 candidates in
Figure~\ref{fig:colcol}, color-coded by their field.  We compare 
the color selection originally employed by \citet{bouwens15c}, of
$J-H>$0.5 and $H-$[3.6]$<$1.4.  We find that our candidates all lie
either in this selection box, or within 1$\sigma$ of a given color.
We note that had we employed this color selection to our sample in
addition to our photometric redshift selection, we would have
excluded six galaxies, a significant fraction of our total sample
size.  The majority of those candidates outside the box have slightly
bluer $J-H$ colors than the \citet{bouwens15c} threshold, which likely
indicates that they reside in the low side of our redshift bin.
Indeed, of the five candidates, all in the EGS field, with $J-H <$
0.5, four have $z_{best} <$ 9.0, indicating a potential overdensity of
sources at $z \sim$ 8.7 in this field, which we discuss further in \S 7.

\subsection{Spectroscopic Confirmation}
Three of our 14 candidates are already
spectroscopically confirmed to lie at $z >$ 8.5.  \citet{zitrin15}
measured Ly$\alpha$ emission at $z =$ 8.68 from the second brightest
source in our initial sample, EGS\_z910\_6811 at $H =$ 25.2 (known as EGSY8p7 in \citealt{zitrin15}), consistent
with our measured photometric redshift of 8.84$^{+0.07}_{-0.18}$.
\citet{mainali18} also present a 4.6$\sigma$ detection of \ion{N}{5}
$\lambda$1243 for this source, indicating it may be powered in part by
an active galactic nucleus (AGN), perhaps explaining its high rest-UV
luminosity.  Our source EGS\_z910\_44164, the next brightest source in
the EGS field with $H =$ 25.4, has a 7$\sigma$ detection of
Ly$\alpha$ at $z =$ 8.665 $\pm$ 0.001, also from MOSFIRE (Larson et
al.\ 2021).  Our source GOODSN\_z910\_35589 with $H_{160}$=25.8 has a measured redshift of
$z =$ 11.09$^{+0.08}_{-0.12}$ from low-resolution {\it HST} grism spectroscopy
\citep[][also known as GNz11]{oesch16}, and recently \ion{C}{3}] was
detected from this source at $z =$ 10.957 \citep{jiang21}, with the
\ion{C}{3}] strength also implying a potential AGN contribution.  This
redshift is $\sim$2$\sigma$ higher than our photometric
redshift of 10.41$^{+0.30}_{-0.07}$, though our measurement is
consistent with the photmetric redshift of $z =$ 10.2 from \citet{oesch16}.
Larson et al. (2021) has spectroscopically observed many of the
remaining 11 candidate galaxies with no significant line detections,
thus {\it JWST} (or perhaps ALMA) will be required to
spectroscopically confirm these candidates.

\section{Vetting the Sample with Additional Imaging}

\subsection{The Brightest Source: COSMOS\_z910\_14822}
\begin{figure}[!t]
\epsscale{1.15}
\plotone{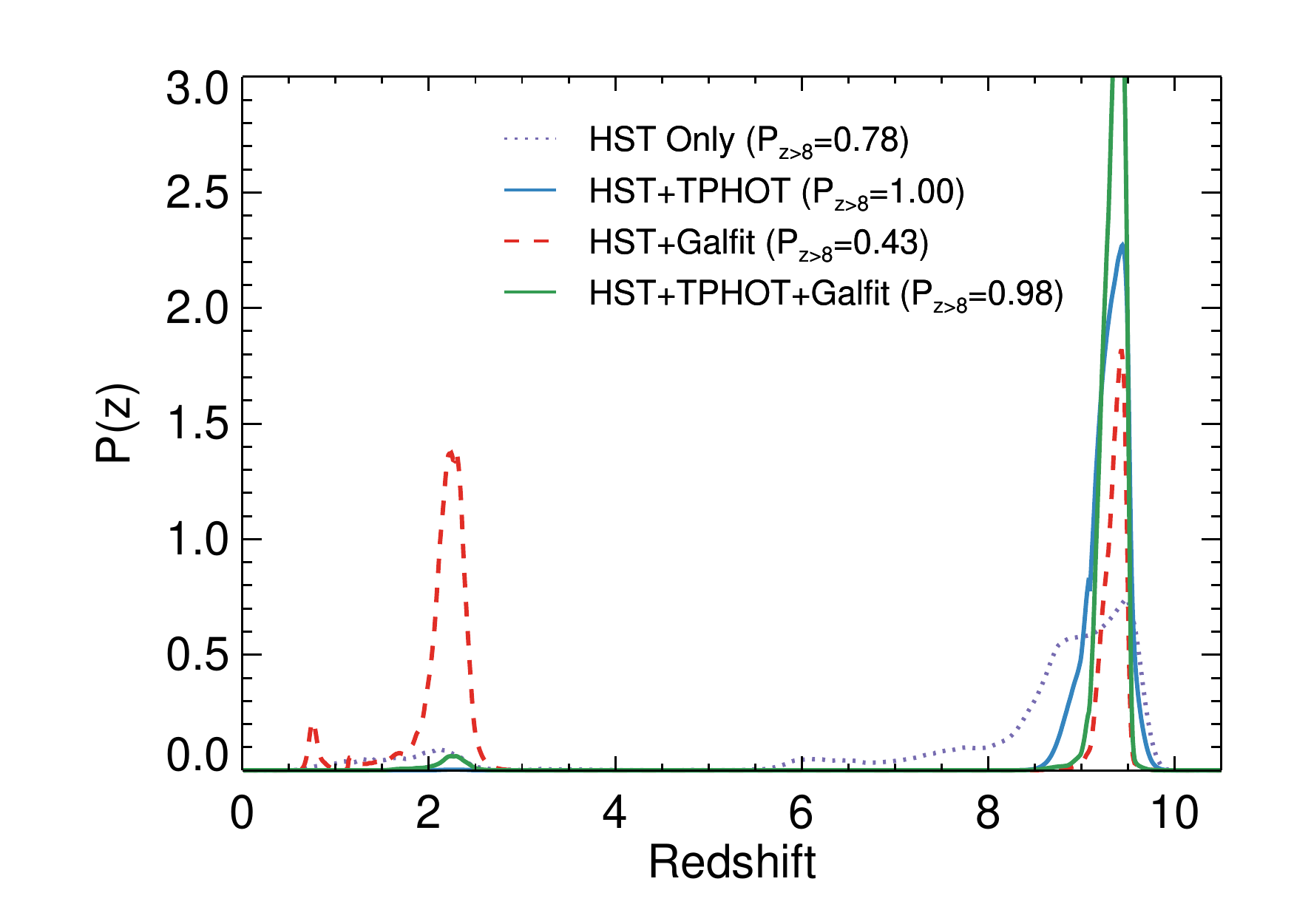}
\vspace{-2mm}
\caption{The photometric redshift probability distribution function
  (PDF) for the brightest source in our {\it HST}$+$IRAC sample,
  COSMOS\_z910\_14822.  The blue solid line shows our fiducial photometric
  redshift, combining {\it HST} photometry with TPHOT-deblended $\it
  Spitzer$/IRAC photometry, which has 100\% of the PDF at $z >$ 8.
  However, due to a nearby bright neighbor, the IRAC photometry for
  this source is the most affected by deblending uncertainties.
  Fitting the \textit{HST} photometry only (purple dotted) still prefers a
  high-redshift solution, although with less (78\%) of its PDF at $z
  >$ 8.  Photometric redshift results with an independent deblending
  technique using {\it {\sc Galfit}} (red dashed) change the IRAC color
  enough to result in the low-redshift solution being nearly equally preferred.
  Combining the {\it HST} photometry with both IRAC measures (green solid)
  still prefers the high-redshift solution.  Due to the uncertainty in
  the IRAC fluxes the nature of this source is thus uncertain with space-based
  data alone.}
\label{fig:noground}
\end{figure}

Of our 14 candidates, only COSMOS\_z910\_14822 has $H\!\!<$
25.0, with $H\!\!=$24.51 $\pm$ 0.04 and $z_{best} =$
9.44$^{+0.04}_{-0.34}$.  In Figure~\ref{fig:noground}, we show our initial
photometric redshift PDF for this source.  Our photometric
redshift with our default photometry, either with (in blue) our
without (in purple) the
deblended TPHOT IRAC photometry, overwhelmingly prefers a $z >$ 9
solution, although a small probability exists of a $z \sim$ 2 solution.

However, as shown in Figure~\ref{fig:stampfig}, this
source is the only candidate in our sample which has a bright neighbor
visible within the 3\arcs\ {\it HST} image cutout.  While the IRAC residual
image shows that this neighbor seems to be well subtracted, here we
consider how reliant the photometric redshift is on the IRAC
photometry deblending method.  With our original TPHOT photometry, we
measure $m_{3.6} =$ 23.47 $\pm$ 0.08, and $m_{4.5} =$ 22.98 $\pm$
0.05, giving [3.6]$-$[4.5] $=$ 0.49.  This red IRAC color is
consistent with a high-redshift solution, where the red color is
caused by either strong rest-optical emission lines, or (perhaps less
likely) a strong 4000 \AA\ break.

To explore the dependence of the preferred redshift on the IRAC
photometry, we calculate an independent set of deblended photometry
using the methodology of \citet{kfinkelstein15}.  This uses
{\sc Galfit} \citep{peng02} to model and subtract all neighboring sources,
and to measure the photometry of our source of interest.  With this
method we measure
$m_{3.6} =$ 23.22 $\pm$ 0.11 and $m_{4.5} =$ 23.28 $\pm$ 0.14, giving
[3.6]$-$[4.5] $=$ $-$0.06.  A flat color is more consistent with
spanning the peak of the stellar SED at $\lambda_{rest} \sim$
1.6$\mu$m, thus when combining our {\it HST} photometry with the
{\sc Galfit}-based IRAC photometry, the $z \sim$ 2 solution is preferred.
 We also perform a run of EAZY including both the
 TPHOT and {\sc Galfit} IRAC constraints.  As shown in Figure~\ref{fig:noground}, this
result tacks closer to the TPHOT-only result, though now a small $z
\sim$ 2 solution is present. 
However, given the uncertainty about the IRAC fluxes, the redshift of this source
remains uncertain from space-based data alone.

To investigate this further, we explore whether any additional
constraints can be gained from the abundant, although often shallower,
ground-based photometry in this field.  
We make use of the optical CFHT/MegaPrime $u^{\ast}$, $g^{\ast}$,
$r^{\ast}$, $i^{\ast}$, $z^{\ast}$ and Subaru/SuprimeCam
$B$, $g^{+}$, $V$, $r^{+}$, $z^{+}$ photometry as published in the catalog of
\citet{nayyeri17}.  As this catalog also includes {\it HST} photometry, we scale
all measured fluxes by the ratio of our measured {\it HST} $H_{160}$-band
to that measured by \citet{nayyeri17}.  For this source this ratio is nearly
unity (0.992), lending credence to our aperture correction methods.
We find no $>$1.5$\sigma$ significance detections in
these optical bands.  There are 1.49$\sigma$, 1.2$\sigma$ and
1.28$\sigma$ measurements in the CFHT $z^{\ast}$, Subaru $V$ and
Subaru $z^{+}$ bands, respectively.  Cutout images of this source in
these bands are shown in Figure~\ref{fig:all14822}.  The Subaru $V$ band shows
a positive low-significance peak, but it is offset from the position
of this source.  While the CFHT $z^{\ast}$-band image shows nothing above
the noise, the Subaru $z^{+}$ image shows some positive flux close to
the position of this source, though it is weak, and visually appears
consistent with our $\sim$1.3$\sigma$ significance measurement.  This
is thus consistent with just being random noise, though it could also
represent weak flux from a source at $z <$ 7.

\begin{deluxetable}{cccc}
\vspace{2mm}
\tabletypesize{\small}
\tablecaption{COSMOS\_z910\_14822 Photometry}
\tablewidth{\textwidth}
\tablehead{
\colhead{ID} & \colhead{Flux (nJy)} & \colhead{Error (nJy)} & \colhead{S/N}}
\startdata
CANDELS F606W&-5.2&13.8&-0.38\\
CANDELS F814W&-0.8&20.6&-0.04\\
CANDELS F125W&223.4&23.5&9.52\\
CANDELS F140W&498.2&37.6&13.25\\
CANDELS F160W&568.0&22.0&25.84\\
TPHOT 3.6&1481.2&104.0&14.24\\
TPHOT 4.5&2326.8&104.9&22.18\\
Galfit 3.6&1870.7&182.8&10.23\\
Galfit 4.5&1770.1&230.7&7.67\\
\hline
Cycle 26 F098M&57.7&26.7&2.16\\
\hline
CFHT $u^{\ast}$&9.7&12.0&0.81\\
CFHT $g^{\ast}$&-3.3&8.0&-0.42\\
CFHT $r^{\ast}$&4.8&10.0&0.48\\
CFHT $i^{\ast}$&3.8&15.2&0.25\\
CFHT $z^{\ast}$&51.4&34.6&1.49\\
Subaru $B$&0.9&4.6&0.19\\
Subaru $g^{+}$&13.0&15.9&0.82\\
Subaru $V$&14.6&12.2&1.20\\
Subaru $r^{+}$&2.9&10.9&0.26\\
Subaru $z^{+}$&57.5&45.0&1.28\\
\hline
UVISTA $Y$&141.1&46.5&3.04\\
UVISTA $J$&253.7&46.4&5.46\\
UVISTA $H$&519.0&53.2&9.75\\
UVISTA $K$&855.9&72.0&11.89\\
zFourGE $J1$&32.8&48.9&0.67\\
zFourGE $J2$&152.2&57.1&2.66\\
zFourGE $J3$&293.0&58.0&5.05\\
zFourGE $H_s$&516.5&107.9&4.79\\
zFourGE $H_l$&618.4&112.4&5.50\\
zFourGE $K_s$&888.0&75.7&11.73
\enddata
\tablecomments{Measured photometry for the brightest source in our
  sample, COSMOS\_z910\_14822.  The optical ground-based photometry is
  taken from \citet{nayyeri17}, while we measure photometry in all
  other filters (see \S 2.6  and \S 3.5.1 for details).}
\label{tab:tab4}
\end{deluxetable}

As this catalog does not
include the latest UltraVISTA \citep{mccracken12} data release, nor the photometry from
the ZFOURGE medium-band survey \citep{spitler12,straatman14}, we perform our own
near-infrared ground-based photometry.  We obtained the public
UltraVISTA DR4 imaging in the $Y$, $J$, $H$ and $K$ bands (these
images have a prefix of
``UVISTA\_X\_19\_11\_18\_allpaw\_skysub\_015\_dr4\_rc\_v2'', where ``X'' denotes
the filter name).  For ZFOURGE, we downloaded the latest public version
of the data, which was v0.9.3, in the $J1$, $J2$, $J3$, $H_{s}$,
$H_l$, and $K_s$ bands.  As these images are seeing limited, we do
not expect our sources to be resolved.  We thus devised a custom
point-source photometry code, which executes the following steps.
\begin{enumerate}
\item Taking the position of the source of interest, cuts out a region
  of the image around this source 5$^{\prime}$ on a side, with the
  source of interest at the center.  If this source is $<$5$^{\prime}$
  from the edge, it offsets the center of the cutout to keep it 5$^{\prime}$ in size.

\item Runs Source Extractor with this image as both the detection and
  measurement image to map the background, as well as create an initial
  catalog sufficient to
  identify stars in the image.  As our source may not be detected in
  all bands, this does not provide our needed photometry.  We visually
  inspect the Source Extractor results, to ensure the detection
  parameters are identifying all robust sources.  We subtract this
  Source Extractor background map from the science image to create a
  background-subtracted image we use to perform our object
  photometry.

\begin{figure*}[!t]
\epsscale{1.15}
\plotone{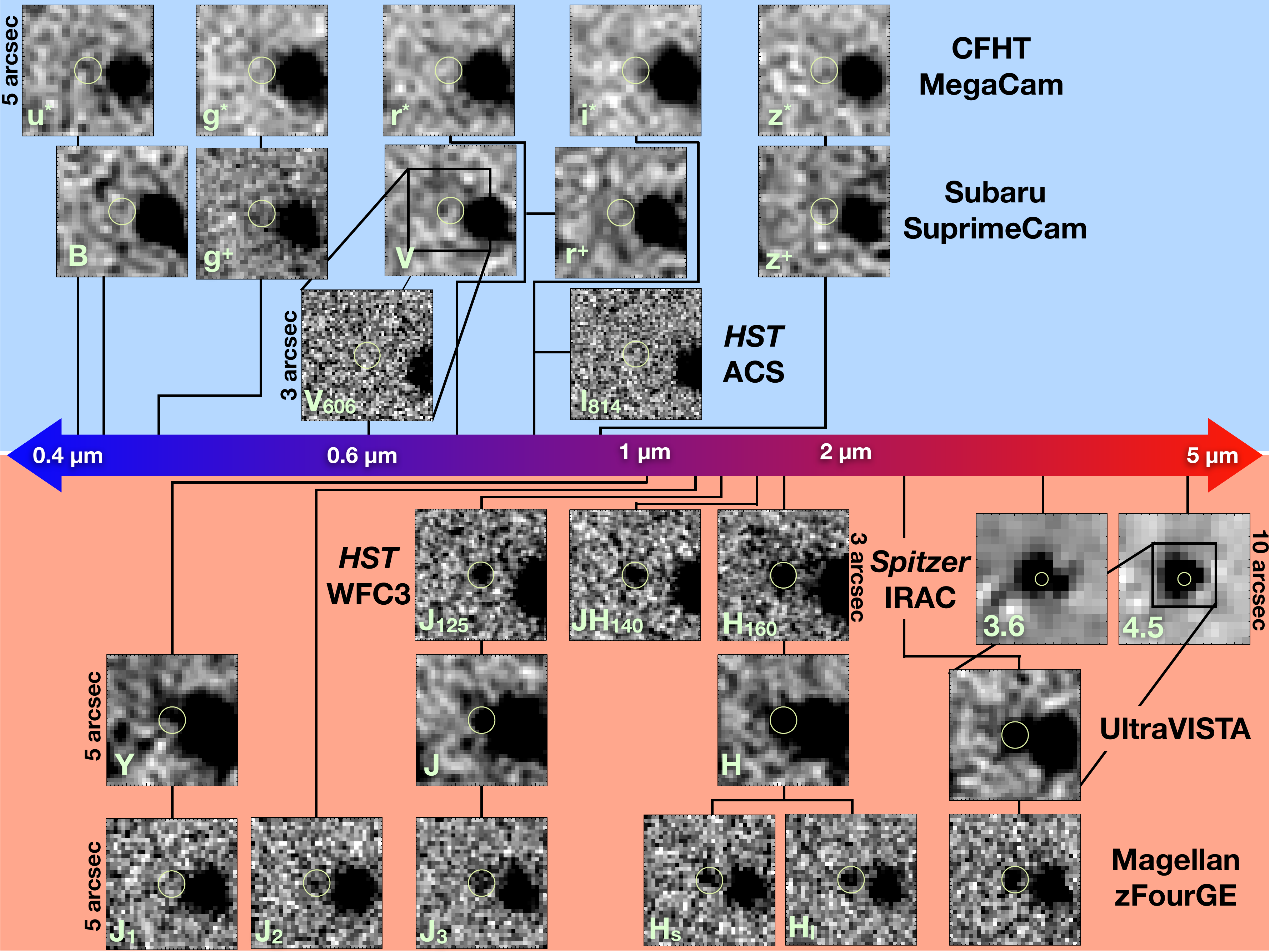}
\vspace{-2mm}
\caption{Cutout images of COSMOS\_z910\_14822 in all space and
  ground-based bands used in this
study.  Images are 3\arcs, 5\arcs and 10\arcs\ in size for {\it HST},
ground-based, and {\it Spitzer} images, respectively.  Most bands are
as expected for a $z >$ 9 galaxy, except the UltraVISTA Y band.
However, the significant emission in this band appears offset
spatially to the North, thus the nature of the source remained unclear
even after the inclusion of all ground-based data.}
\label{fig:all14822}
\end{figure*}

\item We use the initial Source Extractor catalog to identify stars by
  identifying their position in a plot of source magnitude versus
  spatial full-width at half-maximum (FWHM).  We measure a histogram of the FWHM of sources brighter than $m <$ 20 (with
some variation allowed for the depth of the image), and take the
default stellar (e.g. unresolved) FWHM to be the peak of this
histogram. We then identify stars as those objects with 17 $< m <$ 24 and
a FWHM of $\pm$ 0.5 pixels of this stellar FWHM.
  Finally, we select
  a fiducial sample of stars as the 6th--26th brightest objects in
  this list (skipping the first five to ensure we do not use any
  saturated objects, which should be omitted in any case due to the $m
  >$ 17 limit).  We then measure the curves-of-growth of these stars,
  to identify the median radius which encloses 70\% of the total
  flux.  We use this radius below to calculate aperture photometry,
  and then apply an aperture correction of 1/0.7 to correct both the
  flux and noise to total
  (again assuming a point source).

\item We correct for any astrometric offset between a given image and
  the {\it HST} CANDELS images by matching sources between this Source
  Extractor catalog and our {\it HST} catalog for a given field, and
  calculating the median offset in right ascension and declination.
  We then apply these offsets to the position of our source of
  interest to calculate their corrected position in this ground-based
  image.

\item We measure the noise in our 70\% flux-enclosed aperture using
  the same local noise routine discussed in \S 2.5.2 above (correcting
  to total).

\item We measure aperture photometry in this 70\% flux-enclosed
  aperture at the corrected position of our source on the
  background-subtracted image using the IDL
  routine $aper$ (correcting to total).
\end{enumerate}

We used this routine to measure photometry at the position of
COSMOS\_z910\_14822 on all ZFOURGE and UltraVISTA images.  The values
of our measured fluxes and
flux uncertainties in all bands are given in Table~\ref{tab:tab4}, and cutout images
of this source in all bands are shown in Figure~\ref{fig:all14822}.
If this source was truly at $z >$ 8.5, we would expect to see no flux
blueward of 1.155 $\mu$m, thus we should see no flux in the VISTA
$Y$-band and the ZFOURGE $J1$ filters, which have red edges of 1.066
and 1.104 $\mu$m, respectively (ZFOURGE $J2$ has a red edge of 1.212
$\mu$m, thus can contain flux for sources up to $z=$ 8.97).  While the
ZFOURGE $J1$ band has no significant flux, there is a 3.0$\sigma$
significance detection in the UltraVISTA $Y$-band.  Investigating this
image in Figure~\ref{fig:all14822}, the apparent significant flux is
offset from the position of this source in the UltraVISTA $JHK$ bands.
It is thus unclear if the flux
causing the $Y$-band ``detection'' is truly from this source, is
noise, or is flux from the near neighbor, which is heavily blended in
this seeing-limited data.  Thus, even with the addition of
ground-based data, the nature of this source is still uncertain.

\begin{figure*}[!t]
\epsscale{1.15}
\plotone{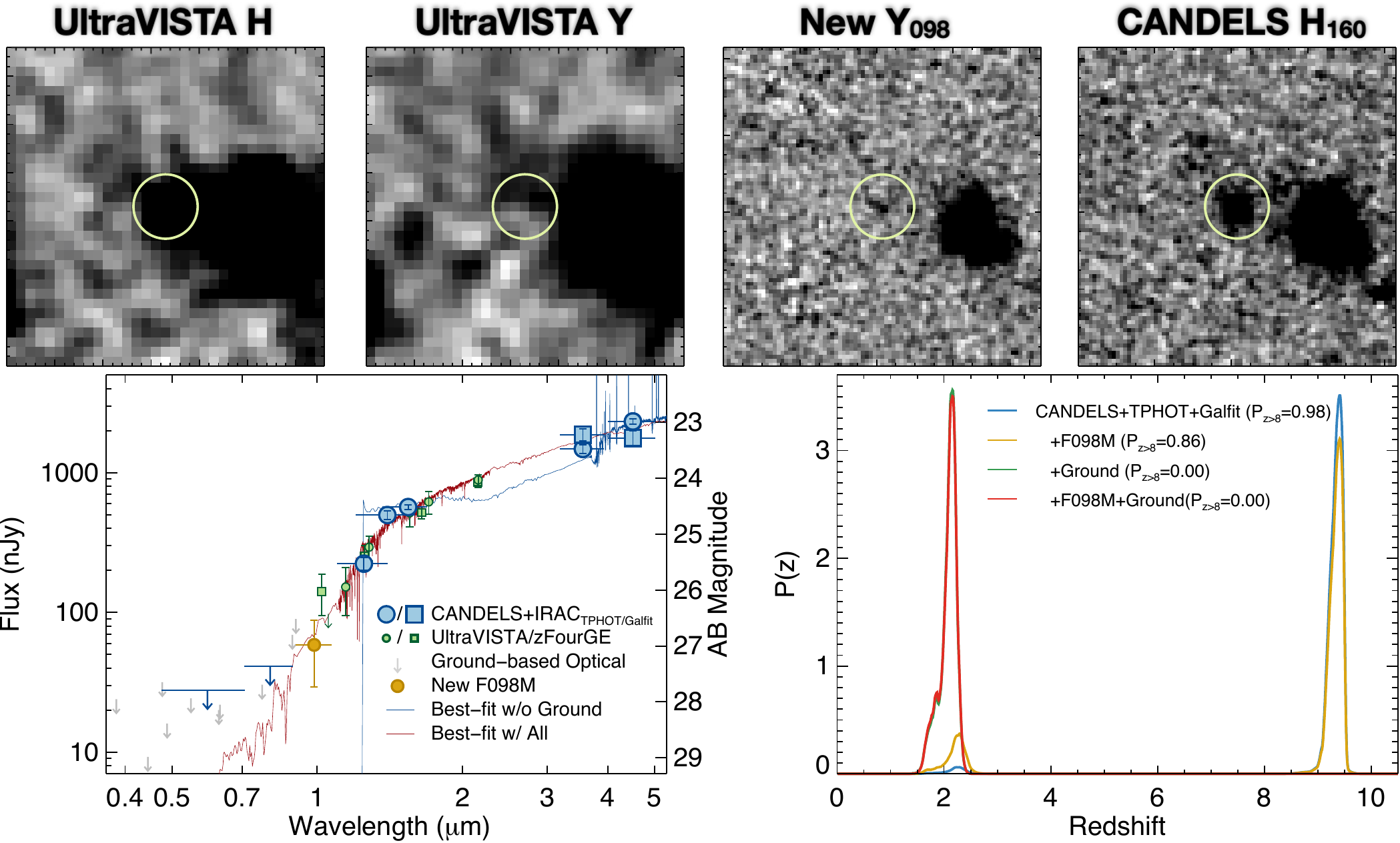}
\vspace{-2mm}
\caption{Top: 5\arcs\ $\times$ 5\arcs\ images of COSMOS\_z910\_14822 in
  the UltraVISTA H and Y bands, and the {\it HST} $Y_{098}$ and
  $H_{160}$ bands.  While the significant flux measured in the
  UltraVISTA Y-band image is not coincident with the $H$ band source,
  in our new F098M imaging we do see a 2.0$\sigma$ detection
  coincident with the position of the $H_{160}$ source.  
  Bottom: The SED of this source is shown on the left, and EAZY
  $\powerset(z)$ results on the right.  The addition of the $Y_{098}$
  photometry to our fiducial {\it HST}$+$IRAC fit does not change the
  preference for a high redshift solution (blue and yellow lines), due to the marginal nature
  of the $Y_{098}$ measurement.  However, adding in the ground-based
  photometry (green and red lines) results in a strong preference for
  a low-redshift solution, predominantly due to the high-significance
  $K$-band measurements showing a red $H-K$ and $K-$[3.6] color.  We
  thus remove this object from our sample for the remainder of our analysis.}
\label{fig:14822_098}
\end{figure*}   

We thus proposed for and were awarded two orbits of Cycle 26 mid-cycle {\it
  HST}/WFC3 F098M imaging (PID 15697; PI Finkelstein).  This probes
similar wavelengths as the UltraVISTA $Y$-band, but at a deeper depth
and much higher resolution.  This filter
probes 0.90 -- 1.07 $\mu$m, and a significant detection would restrict
the Lyman break to be at $z <$ 7.8.  These data were obtained in May 2019, using a 4-point dither pattern
within each orbit ("WFC3-IR-DITHER-BOX-MIN", adopting the default
parameters for half-pixel sub-sampling), with a larger 2-point dither
pattern across the two orbits ("WFC3-IR-DITHER-BLOB", adopting the
default parameters for moving across detector defects), and specifying
the SPARS25 read-out pattern with NSAMP=13, designed to provide robust
up-the-ramp cosmic ray rejection. These exposures were processed
beyond default calibration, following the methods outlined in
\citet{koekemoer11}, including improved background sky correction and
removal of bad pixels along with other detector-level defects, as well
as absolute astrometric alignment to the existing 60mas/pixel CANDELS
F160W mosaics in this field.

We re-ran Source Extractor following the methodology used in \S 2 for
our fiducial catalog, adding F098M as a measurement image (including
PSF-matching this image to the $H_{160}$-band, correcting for Galactic
extinction, and applying an aperture correction to estimate the total flux).
In our fiducial corrected Kron aperture, we measure a F098M flux of 58.6
$\pm$ 29.3 nJy, for a detection significance of 2.0$\sigma$ (there is a
more significant 2.7$\sigma$ detection [28.0 $\pm$ 10.3 nJy] in a 0.4\arcs\ -diameter aperture).  We show a
cutout of this low-significance detection in
Figure~\ref{fig:14822_098}.  A faint, but likely real, source is
visible in the F098M image coincident with the position of the
$H_{160}$ source.  The morphology is odd in that it is streak-like,
but that may be expected at such low signal-to-noise.  Interestingly,
as shown in this figure this flux does not line up spatially with the
faint flux visible in the UltraVISTA $Y$-band.

\begin{figure*}[!t]
\epsscale{1.1}
\plotone{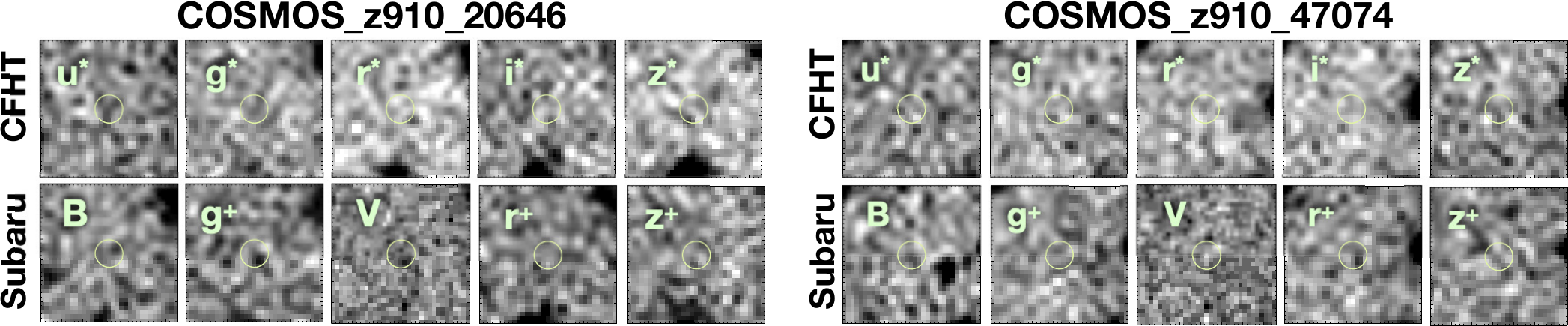}
\vspace{-2mm}
\caption{5\arcs\ $\times$ 5\arcs\ images of the two remaining COSMOS
  sources (20646 on the left, and 47074 on the right) in the ground-based optical bands.  Very weak detections are
visible in a few bands, but offset from the position of the source.
When including these fluxes in the photometric redshift modeling, the
preference is still for a high-redshift solution, although a low
redshift probability is present for both sources at a low ($\sim$10\%) level.}
\label{fig:cosmos_gb_stamps}
\end{figure*}
 
To explore what constraints on the redshift these new data play, we re-ran EAZY adding this new
datapoint to our fiducial set of space-based data (including both
the TPHOT and {\sc Galfit} IRAC measurements in the fit).  These results are
shown in the lower-panel of Figure~\ref{fig:14822_098}.  With these
space-based data alone, a high-redshift fit is still preferred, with
$\int \powerset(z>8) =$ 0.86.  We then ran EAZY including the ground-based
photometry discussed above.  Regardless of whether the new F098M data
is included, the fits with the ground-based imaging strongly prefer a
low-redshift solution.  Since this doesn't change when the F098M
imaging is removed, this implies that the ground-based photometry is
driving the change in redshift.  Since, as described above, the positive flux driving this
shift is spatially offset from the source of interest, it is unclear
whether these photometry points should be allowed to drive the fit.
However, inspecting the SED in the bottom-left of
Figure~\ref{fig:14822_098}, one can see that both the UltraVISTA and
ZFOURGE $K_s$-band photometry are more consistent with the red
continuum slope expected for the lower-redshift solution.  As these
data have much higher S/N values than the optical points, it is likely
that they drive the fits (and this is confirmed when running EAZY
excluding the UltraVISTA $Y$-band, which does not change the result).

This analysis, including the new F098M imaging slight detection,
places enough doubt on the high redshift nature of this object
that we remove it from our sample for the analysis below (similar
conclusios were reached in \citealt{bouwens19} for this object [COS910-8]).  Future
observations of the COSMOS field with the {\it James Webb Space
  Telescope} (expected from COSMOS-Webb [PID 1727, PIs Kartaltepe and
Casey] and PRIMER [PID 1837, PI Dunlop]) should better elucidate the nature of this enigmatic source.

\subsection{Ground-based Photometry for All Sources} \label{sect:ground}
While we elected to use space-based imaging to select our sources due
to the high sensitivity and small {\it HST} PSF,
the preceding subsection highlighted how the abundance of ground-based imaging in these fields can
further constrain our photometric redshifts.  Here we explore how,
if at all, our photometric redshift results change when we include
available ground-based photometry for the remainder of our sample.
The bulk of our ground-based photometry comes from the published
CANDELS-team catalogs in these fields, always scaled by $H_{160}$
magnitude to match our catalog as discussed above.  Details on the
catalog measurements, and citations for specific datasets, can be
found in the catalog papers cited below.  Where appropriate,
we measure our own photometry in more recent datasets.
\vspace{1mm}

\noindent \underline{\emph{EGS}}:  In the EGS field, we make use of the ground-based photometry available
in the CANDELS catalog, published in \citet{stefanon17}.  This
includes optical imaging ($u^{\prime}, g^{\prime}, r^{\prime},
i^{\prime}$ and $z^{\prime}$) from CFHT/MegaCam as part of the CFHT Legacy
Survey, as well as two $K$-band photometric
measurements, from KPNO/NEWFIRM and CFHT/WIRCAM.  All optical
measurements have S/N $<$ 1.5, except for the $z$-band for object
EGS\_z910\_26890, which has S/N$_{z} =$ 1.79.  Five of the seven
objects have 1--2$\sigma$ significance detections in the CFHT $K$-band
imaging.  We re-ran EAZY including all ground-based photometry and
found no significant change in the photometric redshift results for these seven
objects.  The marginal $K$-band detections are consistent with the
fiducial SEDs, and the single $z$-band measurement is low enough
significance that EAZY still strongly prefers a high-redshift solution
(due to the very strong $I_{814} - J_{125}$ break, and red
[3.6]$-$[4.5] color).
\vspace{2mm}

\begin{figure*}[!t]
\epsscale{1.15}
\plotone{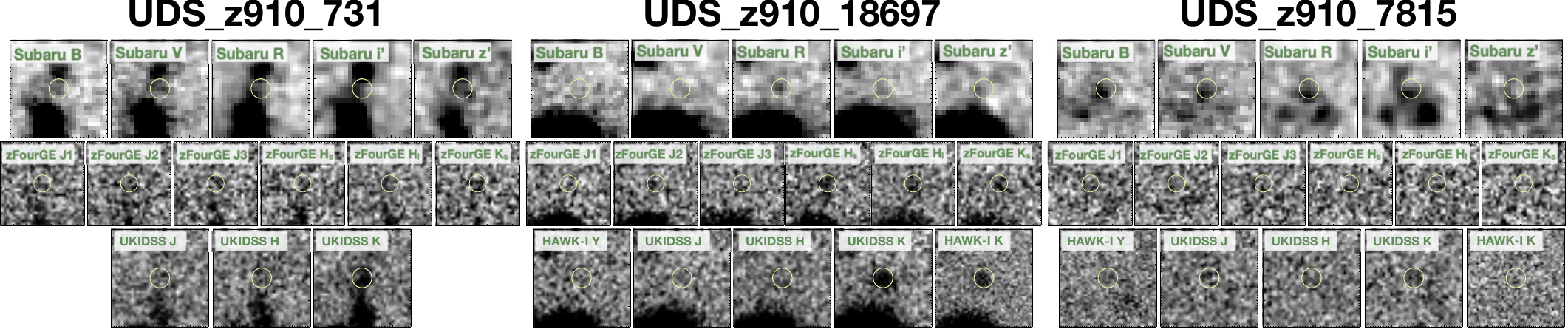}
\vspace{-2mm}
\caption{5\arcs\ $\times$ 5\arcs\ images of the three candidates
  galaxies in the UDS fields in all available ground-based images.  
Weak optical flux is seen in many of the optical images; as discussed
in the text, we attribute this flux to nearby neighbors in
UDS\_z910\_18697 and UDS\_z910\_7815, removing those data from the
updated photometric redshift fit.  With the inclusion of the ground-based data, the
updated results give $\int \powerset(z>8)$ = 0.0, 1.00 and 0.92, for sources
731, 18697, and 7815, respectively.  We thus remove object
UDS\_z910\_731 from our final sample at this stage.}
\label{fig:UDS_gb_stamps}
\end{figure*}

\noindent \underline{\emph{GOODS-N}}: The single object in our GOODS-N sample,
GOODSN\_z910\_35589, has a previously published emission-line redshift
of $z =$ 10.957 from \citet{jiang21}.  Nonetheless, we explore the existing ground-based
photometry, to see how it affects the photometric redshift, though as
this object benefits from the 11 bands of {\it HST} imaging available
in the GOODS-N field, the photometric redshift is already fairly
secure.  We make use of the CANDELS catalog of
\citet{barro19}, where this object has measured non-detections in the
KPNO/MOSAIC and LBT/LBC $u$-bands, and a detection in Subaru/MOIRCS
$K$-band imaging.  We reran EAZY, and found that these additional
photometric data points did not change the previous results, which
prefer $z \sim$ 10.4 (the spectroscopic redshift of 10.957 is contained within
the 99.5\%/3$\sigma$ confidence range of the $\powerset[z]$).
\vspace{2mm}

\noindent \underline{\emph{COSMOS}}:  In the COSMOS field we make use of the same ground-based data in the
preceding subsection, including measuring fluxes in the ZFOURGE and
UltraVISTA DR4 data.  Neither of the two remaining objects in this
field have detections in the UltraVISTA $Y$ or ZFOURGE $J1$ or $J2$ bands.
The object COSMOS\_z910\_20646 does have a measured $>$2$\sigma$
significance detection in two optical bands: CFHT $i^{\ast}$ (33.3 $\pm$ 11.0
nJy; 3.0$\sigma$), and Subaru $g^+$ (34.8 $\pm$ 14.3 nJy;
2.4$\sigma$).  inspecting these images, there is barely visible positive flux, but it
is shifted relative to the expected position of the source
($\sim$0.5\arcs\ to the SE in the CFHT $i^{\ast}$ image, and
$\sim$0.5\arcs\ S in the Subaru $g^+$ image).
Correspondingly, adding these ground-based data to EAZY does not significantly change
the best-fit redshift, though it does introduce a small secondary peak
at $z \sim$ 2.5.  This is consistent with the fact that there is no significant detection at the position of this
object in the {\it HST} $V$ and $I$ bands (with 1$\sigma$ upper limits at the
position of this object of 8.6 and 10.5 nJy, respectively), nor in the
other ground-based optical bands.  Optical measurements for the object
COSMOS\_z910\_47074 exceed 2$\sigma$ only in the CFHT $r^{\ast}$ band,
at 2.4$\sigma$.  Visually inspecting this image shows positive flux at
this position, albeit visually consistent with the noise level.  The
photometric redshift continues to prefer the high-redshift solution
even when including these data, dominated by the stringent
non-detections in the {\it HST} $V$, $I$ and $Y$ bands (and flat
$H-$[3.6] color and non-detection in the ground-based $K$-bands),
though the weak optical $r^{\ast}$ flux does again lead to a small
low-redshift solution.  Including this ground-based photometry and
re-running EAZY yields photometric redshifts which still
satisfy our sample selection, with $\int$ $\powerset(z > 8) =$ 0.96
and 0.87, and $\Delta \chi^2 =$ 5.2 and 3.9 for COSMOS\_z910\_20646
and COSMOS\_z910\_47074, respectively.
The ground-based optical cutout images for these two
sources are shown in Figure~\ref{fig:cosmos_gb_stamps}.
\vspace{3mm}

\noindent \underline{\emph{UDS}}:  In the UDS field, we initially make use of the ground-based photometry available
in the CANDELS catalog, published in \citet{galametz13}. These include
$u$-band imaging from CFHT/MegaCam, $B$, $V$, $R_c$, $i^{\prime}$ and
$z^{\prime}$ imaging from Subaru/Suprime-Cam, $Y$ and $K$-band imaging
from VLT/HAWK-I from the HUGS survey \citep{fontana14}, and $J$, $H$, and $K$-band
imaging from UKIRT/WFCAM from the UKIDSS Ultra Deep
Survey \citep{lawrence07}.  We find matches to all three of our UDS candidates
within a matching radius of 0.4\arcs.  

Exploring the optical
photometry, we find that the CANDELS catalog has significant
detections for all three objects in multiple Subaru filters ($>$2$\sigma$; up
to 10$\sigma$ in one case).  This is surprising as these images in
most cases have similar sensitivities to the CANDELS {\it HST} optical imaging used in
the selection of these objects, where there was no significant flux
measured.  We thus investigated the position of these candidates
directly in the Subaru imaging, and found that all three have nearby
neighbors which may affect their measured ground-based photometry.  We
thus performed our own photometry at the position of these sources,
following the methodology laid out in \S 3.5.1.  In addition to the
Subaru bands, we also measured photometry in the UKIDSS DR11
images, the HUGS (natural seeing) images, and the ZFOURGE near-infrared images in
this field. 

Starting with object UDS\_z910\_731, and with the filters in common
between Subaru and {\it HST}/ACS, we find a Subaru $V$-band flux density of 48.0 $\pm$
10.6 nJy and $R$-band flux density of 68.9$\pm$ 11.5 nJy, compared to 4.6
$\pm$ 37 nJy in the {\it HST}/ACS $V_{606}$ band.  In the
$I$-band, we find 58.6 $\pm$ 13.7 nJy with Subaru, and 37.7 $\pm$ 42.0
nJy with ACS.  These values are consistent with what is seen in the
images, shown in Figures~\ref{fig:stampfig} and \ref{fig:UDS_gb_stamps}, where faint flux is visible in the Subaru bands, with no
significant flux visible in the {\it HST} bands.  However, this candidate is in a somewhat
noisier than average region of the ACS imaging, thus the measured
Subaru flux would only be detected at the $\sim$1--2$\sigma$ level in
the ACS imaging.  Therefore the absense of flux in the ACS imaging is
not completely inconsistent with the weak measured flux in the ground-based
bands.  A limiting factor here is the nearby
galaxy $\sim$1\arcs\ away which does not affect the {\it HST}
photometry, but may be contributing the measured flux visible in the
Subaru imaging.  Nonetheless, as we cannot be sure this flux belongs
to the neighbor, we conservatively include all manually measured
photometry in our updated photometric redshift fitting for this source
below.

\begin{figure*}[t]
\epsscale{1.0}
\plotone{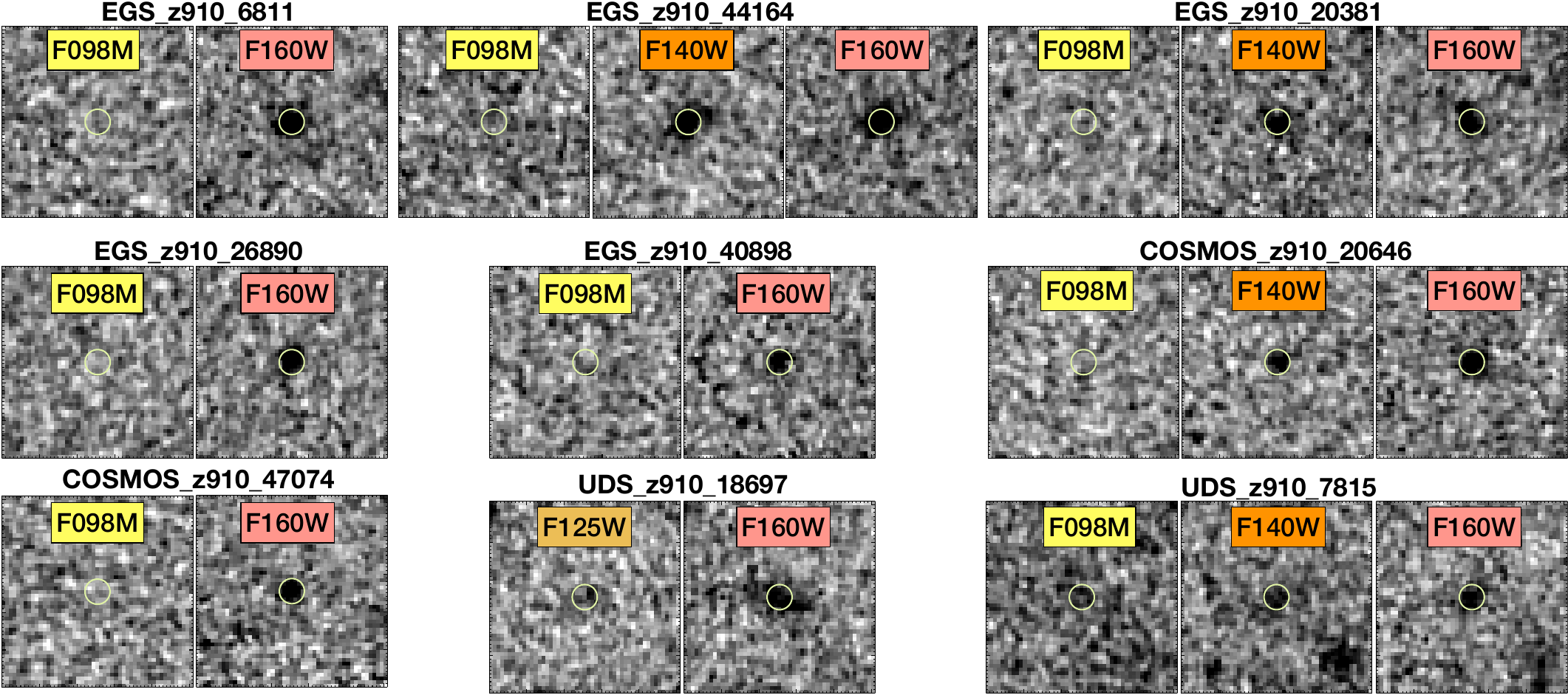}
\vspace{-2mm}
\caption{3\arcs\ $\times$ 3\arcs\ images of the candidates which
  were observed with Cycle 27 {\it HST} imaging (F098M, F125W, and/or
  F140W).  There is no significant ($<$1.5$\sigma$) detection in F098M
  for all sources, except UDS\_z910\_7815 (2.0$\sigma$).
  UDS\_z910\_18697 shows a weak (3.8$\sigma$) detection in the new
  deeper F125W image, which due to its low significance is not
  inconsistent with $z >$ 10.  With the exception of UDS\_z910\_7815,
  which is now equally likely to be a low-redshift interloper as a
  high-redshift galaxy, these new data increase the likelihood of the
  remaining candidates being true $z =$ 9--10 galaxies.}
\label{fig:cyc27stamps}
\end{figure*}

Object UDS\_z910\_18697 has measured flux densities in the Subaru $V$, $R$
and $i^{\prime}$ bands of 41.9 $\pm$ 8.1 nJy, 64.5 $\pm$ 10.8 nJy and 102.3
$\pm$ 13.2 nJy,
respectively.  In the {\it HST} imaging, this source has measured flux
densities of -9.5 $\pm$ 14.5 nJy and 1.3 $\pm$ 15.4 nJy in the $V_{606}$
and $I_{814}$ bands, respectively.  The {\it HST} measurements for
this object are more sensitive than 731 discussed above, thus the {\it
  HST} non-detections are inconsistent with the Subaru measurements,
at the 5$\sigma$ level in the $I$-band.  Inspecting
Figures~\ref{fig:stampfig} and \ref{fig:UDS_gb_stamps}
around this position, the reason for this discrepancy is apparent --
there is a very bright large spiral galaxy nearby, again $\sim$1\arcs\ away.  In
the {\it HST} imaging, even the faintest isophote from this object is
significantly far from the position of the high-redshift candidate,
while in the Subaru imaging, the candidate's position is on the wings
of this source.  We therefore conclude that the measured flux in the
Subaru bands belongs to the bright neighbor, and we exclude the five Subaru
bands from our updated photometric redshift analysis.  In the
ground-based near-infrared imaging, significant flux is seen in some bands at the
position of this source, but as the seeing in these images is much
better than the Subaru images, this flux is well-separated from the
neighboring galaxy, thus we conclude it belongs to our candidate.

Object UDS\_z910\_7815 has measured flux densities in the Subaru $V$, $R$
and $i^{\prime}$ bands of 34.5 $\pm$ 10.6, 43.8 $\pm$ 11.5 and 41.6
$\pm$ 13.7 nJy,
respectively.  In the {\it HST} imaging, this source has measured flux
densities of 18.5 $\pm$ 14.1 and 4.6 $\pm$ 15.8 nJy in the $V_{606}$
and $I_{814}$ bands, respectively.  Similar to the previous object,
the {\it HST} and Subaru photometry are seemingly inconsistent,
although less so, at the 2$\sigma$ level in the $I$-band.  Inspecting
the images in Figures~\ref{fig:stampfig} and \ref{fig:UDS_gb_stamps}, this
discrepancy is explained as the source of the flux is
offset by $\sim$0.5\arcs\ to the SE from the candidate galaxy's
position.  There is low-level positive flux at this offset position in
the {\it HST} images, thus we conclude that this is a nearby neighbor,
and we exclude these Subaru bands from the updated photometric
redshift fitting.  Weak flux, at the 1.3--1.8$\sigma$ level is 
recorded in the ZFOURGE $J$1--3 bands, and is also visible at this offset
position, thus we also do not include the ZFOURGE data in the updated
fit for this object.

We re-ran EAZY for these objects to obtain an updated photometric
redshift fit using these ground based data (omitting data where
discussed above).  Unsurprisingly, candidate UDS\_z910\_731 is
now much better fit by a lower redshift ($z \sim$ 4) solution, due to
its likely detections in the ground-based imaging.  This object is
removed from our sample.   Candidate UDS\_z910\_18697 still firmly satisfies our sample
selection criteria with $\int \powerset(z > 8) =$ 1.00 and $\Delta
\chi^2 =$ 84.4.  UDS\_z910\_7815 does have $\int$ $\powerset(z > 8) =$
0.92, however it now has $\Delta \chi^2 =$ 3.3, just below our
threshold of 3.5 for our initial sample selection.  We revisit this
object in \S 4.3.

In summary, adding the ground-based photometry removed two sources
from our sample: COSMOS\_z910\_14822 (with further evidence provided
by the weak F098M detection) and UDS\_z910\_731, leaving us with 12
candidate bright $z =$ 9--10 galaxies.  This highlights the
utility of examining \emph{all} available data when building a sample
of such difficult to identify galaxies.

\begin{figure*}[!t]
\epsscale{1.1}
\plotone{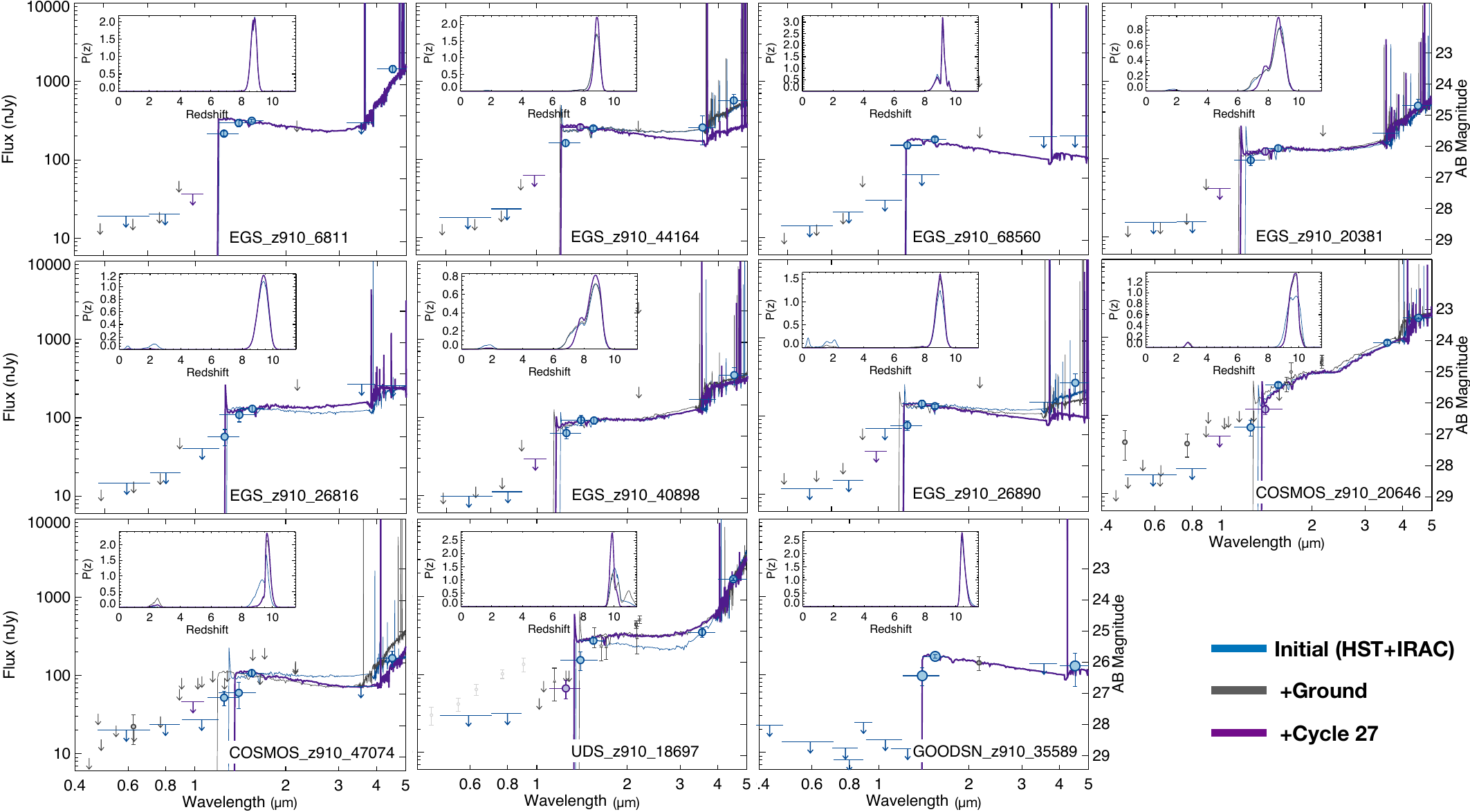}
\vspace{-2mm}
\caption{The observed spectral energy distributions for the 11 $z =$
  9--10 galaxy candidates in our final vetted sample.  Blue symbols
  show the initial {\it HST} and {\it Spitzer}/IRAC photometry, while
  gray symbols show ground-based photometry.  Purple symbols denote
  additional photometry obtained from Cycle 27 followup programs.  The
best-fitting EAZY model is shown by the blue, black and purple lines
for the initial, initial$+$ground, and initial$+$ground$+$Cycle 27
($=$final) cases.  The final photometric redshift PDF shows nearly no
low-redshift probability in all sources.}
\label{fig:final_seds}
\end{figure*}

 \begin{deluxetable*}{ccccccc}
\vspace{2mm}
\tabletypesize{\small}
\tablecaption{Photometric Redshift Constraints with Additional Data}
\tablewidth{\textwidth}
\tablehead{
\multicolumn{1}{c}{ID} & \multicolumn{3}{|c|}{$\int$ $\powerset(z>8)$}
& \multicolumn{3}{|c|}{$\Delta \chi^2_{lowz - highz}$}\\
 \multicolumn{1}{c}{$ $} & \multicolumn{1}{|c}{Initial} & \multicolumn{1}{c}{$+$Ground} & \multicolumn{1}{c|}{$+$Cycle 27} & \multicolumn{1}{|c}{Initial} & \multicolumn{1}{c}{$+$Ground} & \multicolumn{1}{c|}{$+$Cycle 27}}
\startdata
EGS\_z910\_6811&0.99&0.99&1.00&26.5&40.9&52.5\\
EGS\_z910\_44164&0.95&0.97&0.99&8.5&18.2&23.5\\
EGS\_z910\_68560&1.00&1.00&---&24.5&37.8&---\\
EGS\_z910\_20381&0.74&0.74&0.82&7.1&8.3&9.3\\
EGS\_z910\_26890&0.80&0.94&0.99&3.5&6.5&12.9\\
EGS\_z910\_26816&0.92&0.99&---&5.0&9.0&---\\
EGS\_z910\_40898&0.69&0.71&0.80&5.4&8.0&8.3\\
COSMOS\_z910\_20646&1.00&0.96&0.97&14.3&5.2&5.4\\
COSMOS\_z910\_47074&0.99&0.87&0.96&9.2&3.9&6.7\\
UDS\_z910\_18697&1.00&1.00&1.00&103.1&84.4&84.1\\
GOODSN\_z910\_35589&1.00&1.00&---&84.1&84.5&---\\
\hline
\hline
COSMOS\_z910\_14822&1.00&0.06&0.01&13.3&0.0&0.0\\
UDS\_z910\_731&0.85&0.00&---&4.5&0.0&---\\
UDS\_z910\_7815&0.95&0.92&0.79&5.1&3.3&0.9
\enddata
\tablecomments{This table shows how the two key photometric redshift
  constraints (the integral of the PDF at $z >$ 8, and the difference
  in $\chi^2$ between the low and high-redshift solutions) change as
  additional data is added.  The first column is based on the initial
  {\it HST}$+$IRAC selection.  The second column adds ground-based
  data, while the third column adds additional {\it HST} imaging from
  Cycles 26 and 27.  The three objects removed from our sample
  following this process are listed below the horizontal lines.}
\label{tab:tab5}
\end{deluxetable*}

\subsection{Cycle 27 {\it HST} Imaging}

Although the sample of 12 candidate $z =$ 9--10 galaxies still
satisfies our sample selection criteria even after the inclusion of
all available ground-based imaging, there are still often low-redshift
solutions present in their $\powerset(z)$'s.  We can increase the
fidelity of this sample in two ways.  The first is with deeper
$Y$-band imaging, which falls just below the Lyman-$\alpha$ break at
$z >$ 8.  Because 11/12 candidates are in the CANDELS wide fields
which lack this imaging (with the exception of a few objects which
were included in previous followup programs), additional $Y$-band
imaging can provide higher confidence in their high-redshift nature.
Secondly, while most of our objects contain some F140W imaging, it
is extremely shallow.  Deeper F140W imaging can both provide a second
detection band at these redshifts for $z >$ 10 galaxies, and can also
better separate $z \sim$ 9 from $z \sim$ 10 galaxies.

We thus obtained additional {\it HST} imaging for the nine of our
candidates which lacked any $Y$-band imaging from
a 14-orbit Cycle 27 proposal (PID 15862; PI Finkelstein).  Similar to
the Cycle 26 mid-cycle observations of COSMOS\_z910\_14822, we elected
to use F098M rather than F105W because photometric redshift simulations
found that F098M is more constraining. Although F098M is slightly
shallower in a fixed exposure time, a faint detection in F098M would
rule out a $z >$ 8 solution, while the Lyman-$\alpha$ break is within the FWHM
of the F105W filter at $z \sim$ 8.7.
Sources targeted in
F098M were: EGS\_z910\_6811, 44164, 20381, 26890 and 40898, both
sources in the COSMOS field, and source UDS\_z910\_7815.  For source UDS\_z910\_18697 the lack of significant flux in the
CANDELS F125W image points to a potential $z >$ 10.5 solution, so
rather than F098M we obtained additional data in F125W.  Our two
additional orbits increased the F125W exposure time by a factor of
four over CANDELS alone.
 Finally, we obtained F140W imaging for sources which either had no
 previous F140W coverage, or had a {\it HST} detection in only F160W.
 This was the case for objects EGS\_z910\_44164,
 EGS\_z910\_20381, COSMOS\_z910\_20646 and UDS\_z910\_7815.

These data were obtained between Oct 2019 and Oct 2020.  The images were obtained with a similar exposure pattern
and reduced in the same way as our Cycle 26 data (described in \S
4.1); any existing images in these filters were included in our newly
reduced data.   We show 3\arcs\ cutout images of our candidates in these new images
(alongside the existing F160W images) in
Figure~\ref{fig:cyc27stamps}.  For the EGS and COSMOS objects, we
see no significant flux in F098M, providing even stronger evidence for
a $z >$ 8 solution for these objects.  We also see strong detections
in F140W for the three galaxies in those two fields, implying these
galaxies lie closer to $z \sim$ 9 than 10.  In object UDS\_z910\_18697
we elected to observe in F125W rather than F098M, and we see a very
faint signal coincident with this object, with a significant detection
in F140W.  This near non-detection in
F125W implies $z >$ 9.5, and is consistent with $z \sim$ 10.5
as the F125W filter has $\sim$4\% throughput at 1.398 $\mu$m (the
observed wavelength of a $z =$ 10.5 Lyman-$\alpha$ break).  Finally, for object
UDS\_z910\_7815, there does appear to be faint flux in the F098M image
at the position of this object.

\begin{figure*}[t]
\epsscale{1.1}
\plotone{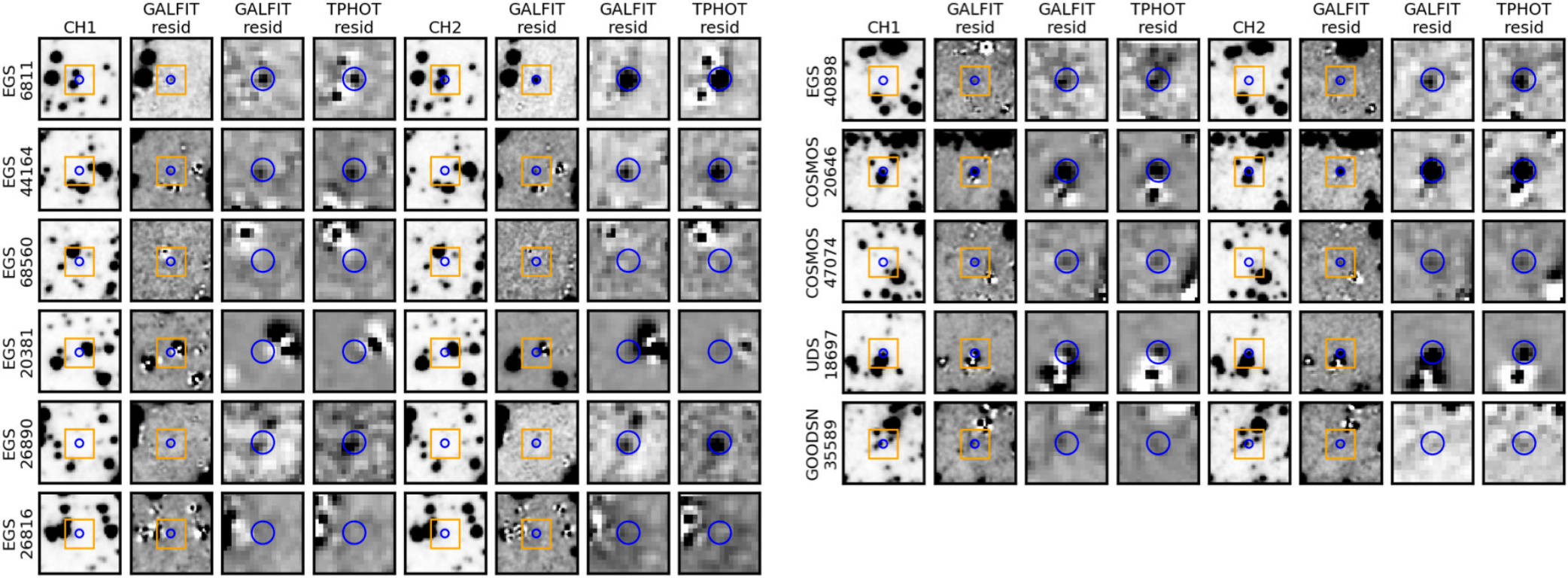}
\vspace{-2mm}
\caption{A comparison of IRAC deblending results between TPHOT and
  {\sc Galfit}.  Each row contains one object.  The first four columns are
  for the 3.6$\mu$m band.  The first two columns show the IRAC image
  and the {\sc Galfit} residual image (the source of interest is left behind
  in all residual images; only neighbors have been subtracted) and are
 30.6\arcs\ on a side.  The third and fourth columns show 10.2\arcs\
  zoom-ins on the {\sc Galfit} and TPHOT residual images.  Columns 5-8 are
  the same for the 4.5$\mu$m band.  Isolated sources yield similar
  results between {\sc Galfit} and TPHOT.  Unsurprisingly, heavily
  crowded sources have differences in the measured photometry which
  can affect the photometric redshift.  Two of our sample of 11
  sources have best-fit solutions at $z <$ 8 when combining these two
  photometric measurements.  Due to the deblending uncertainty we ultimately keep these sources in our
  fiducial sample.}
\label{fig:galfit}
\end{figure*}
 
\begin{deluxetable*}{ccccccccc}
\vspace{2mm}
\tabletypesize{\small}
\tablecaption{IRAC Photometry Comparison}
\tablewidth{1.1\textwidth}
\tablehead{
\multicolumn{1}{|c|}{$ $} & \multicolumn{4}{|c|}{Galfit} &
\multicolumn{4}{|c|}{TPHOT}\\  \cline{2-9}
\multicolumn{1}{|c|}{$ $} & \multicolumn{2}{|c|}{3.6$\mu$m} &
\multicolumn{2}{|c|}{4.5$\mu$m} & \multicolumn{2}{|c|}{3.6$\mu$m} &
\multicolumn{2}{|c|}{4.5$\mu$m}\\
\hline
\multicolumn{1}{|c|}{ID} & \multicolumn{1}{|c|}{Flux} & \multicolumn{1}{|c|}{Error} &
\multicolumn{1}{|c|}{Flux} & \multicolumn{1}{|c|}{Error} & \multicolumn{1}{|c|}{Flux} & \multicolumn{1}{|c|}{Error} & \multicolumn{1}{|c|}{Flux} & \multicolumn{1}{|c|}{Error}}
\startdata
EGS\_z910\_6811&260.8&39.1&1153.5&3.5&201.1&147.7&1433.0&161.7\\
EGS\_z910\_44164&125.2&61.6&238.9&72.3&256.3&104.0&568.0&105.5\\
EGS\_z910\_68560&259.0&84.8&287.9&13.9&-97.1&97.4&-83.8&99.0\\
EGS\_z910\_20381&2032.4&371.7&848.1&151.3&157.6&106.6&482.2&107.5\\
EGS\_z910\_26890&239.3&94.0&36.0&38.0&107.8&75.5&265.7&77.5\\
EGS\_z910\_26816&121.8&146.4&121.3&84.2&158.3&133.9&231.2&129.6\\
EGS\_z910\_40898&121.4&34.4&196.5&37.2&156.7&84.8&347.0&85.4\\
COSMOS\_z910\_20646&1177.2&1.7&1557.7&45.6&855.4&51.7&1782.8&50.7\\
COSMOS\_z910\_47074&74.9&3.5&180.7&38.4&56.2&36.0&167.4&37.4\\
UDS\_z910\_18697&1132.6&102.6&3026.1&93.8&349.7&49.5&1680.9&56.8\\
GOODSN\_z910\_35589&116.7&2208.6&13.9&489.6&109.0&69.7&129.3&58.4
\enddata
\tablecomments{All fluxes are in nJy.}
\label{tab:tab6}
\end{deluxetable*}

To obtain robust photometry from these new images, we first PSF-matched these images to the
F160W image using the same procedures as our full catalog.  As these
single-WFC3-field images lack sufficient numbers of stars to create a
robust PSF, we use the PSF kernels derived in the GOODS-S field (which
has a large area covered by all three filters we obtained new data
in).  We then performed photometry with Source Extractor in an
identical way as our initial cataloguing, using the F160W image as the
detection image, measuring and applying the empirical noise function,
and applying a F160W-based aperture correction.  Our photometric
measurements validate what we observed in
Figure~\ref{fig:cyc27stamps}.  We measure $<$1.5$\sigma$ significance
detections in F098M for all objects except UDS\_z910\_7815, where we
measure a 2.0$\sigma$ significance detection (2.9$\sigma$
significance when measured in a smaller 0.4\arcs\ -diameter
aperture).  The three objects observed in F140W have measured flux at
5--10$\sigma$ significance.  UDS\_z910\_18697 has a 3.8$\sigma$
detection in the F125W filter.

To explore how these additional photometric constraints affect the
photometric redshift fitting, we perform another run of EAZY, adding
these new measurements to those used in the previous run (initial {\it
  HST}, IRAC and ground-based).  We summarize the key results in
Table~\ref{tab:tab5}, which broadly follow the interpretation gleaned
from Figure~\ref{fig:cyc27stamps}, in that when there is no
significant detection in F098M, the value of both $\int
\powerset(z>8)$ and $\Delta \chi^2$ increases.  For the objects
with F140W imaging, the measured flux did not significantly alter the
best-fitting redshift.
Interestingly, in the case of UDS\_z910\_7815, which had the 2$\sigma$
F098M detection, the EAZY analysis still prefers a high-redshift
solution with $\int
\powerset(z>8) =$ 0.79; lower than the fit without these new data, but
still satisfying this sample selection criterion ($>$0.6).  However,
the best-fitting low-redshift solution is almost as good a match to
the data as the high-redshift solution, giving $\Delta \chi^2 =$ 0.9,
below our threshold of 3.5.  We thus consider UDS\_z910\_7815 as at least
a plausible low-redshift interloper, and remove it from our sample for
the remainder of the analysis.  Finally, for UDS\_z910\_18697 the
addition of the deep F125W imaging, with its weak detection, did not
give any further evidence against this object being a true $z \sim$ 10
galaxy, as both $\int \powerset(z>8)$ and $\Delta \chi^2$ had
essentially no change with the addition of these data.  Approved {\it
  JWST} Cycle 1 NIRSpec observations (PID 1758, PI Finkelstein) will
soon provide a precise spectroscopic redshift for this source.  We show the final SEDs and
photometric-redshift PDFs for our final sample of 11 candidate $z =$ 9--10
galaxies in Figure~\ref{fig:final_seds}.

\subsection{Custom IRAC Deblending with {\sc Galfit}}

In Section 4.1 we showed that different methods of IRAC deblending
can result in modest changes to the photometry, which in the case of
COSMOS\_z910\_14822, resulted in significant changes to the
photometric redshift.  As one final in-depth investigation into our
sample of galaxies, we ran {\sc Galfit} in a similar manner on the 11
remaining sources (excluding also UDS\_z910\_731 and UDS\_z910\_7815 from the previous
subsection).  In Figure~\ref{fig:galfit} we show the results of
this additional deblending effort, showing both the original image,
and the neighbor-subtracted images for both methods in both bands, and
we compare photometry in Table~\ref{tab:tab6}.
We recalculated the photometric redshifts using a
combination of the TPHOT and {\sc Galfit} flux measurements, where we used
the mean of the two measurements as the flux estimate, and the maximum
of the two uncertainty measurements as the flux uncertainty.  We found
that for nine of the 11 sources, the photometric redshifts did not
significantly change, with all nine continuing to satisfy our stringent
selection criteria.  

For two sources, the TPHOT+{\sc Galfit} photometric redshift preferred $z
<$ 8 solutions.  For one source, EGS\_z910\_20381, this shift was
mild, with EAZY now preferring $z =$ 7.4 rather than 8.7, with $\int
\powerset(z>7) =$ 0.88, albeit with $\int \powerset(z>8) =$ 0.32.
Inspecting the deblending residuals in Figure~\ref{fig:galfit} for
this source, it does appear that TPHOT does a better job of removing
the bright neighboring source, therefore we do not believe the {\sc Galfit}
results nullify this object as a candidate $z =$ 9--10 galaxy.  

However, the {\sc Galfit} IRAC measurements of COSMOS\_z910\_20646 strongly prefer a
$z \sim$ 2.5 solution, with no significant high-redshift peak.  This
object is also heavily crowded, with both TPHOT and {\sc Galfit} struggling
to subtract its neighbor, thus the differences in measured fluxes represent
differences in the failure to fully remove the nearby bright galaxy,
rather than true differences in the SEDs of the objects.
Looking further at the photometry in Table~\ref{tab:tab6}, COSMOS\_z910\_20646
had a significant red [3.6]-[4.5] color measured with TPHOT,
while the color is less red with {\sc Galfit}, explaining the shift in $\powerset(z)$ to lower redshift.
These differences depending on which photometric method is used
highlight a systematic uncertainty when making use of deblended
photometry from low-resolution imaging, something which will be
alleviated soon with {\it JWST}.  Given the uncertainty regarding
which method is more accurate, we keep this source in our sample,
though we caution the reader that this nature of this galaxy is still
somewhat uncertain.

\subsection{Gravitational Lensing Magnification}
Due to the bright nature of our sample of candidate $z =$ 9--10
galaxies, we must consider the possibility that their apparent
brightness has been magnified by sources near to the line-of-sight.
While sources along all lines of sight are subject to weak lensing by
the total mass distribution along the line of sight, the correction
for this is statistical in nature, and is likely small
for blank fields such as the CANDELS fields our candidate galaxies
reside in.  Here we thus consider strong lensing, which we can
estimate and correct for on a source-by-source basis.  We follow the
methodology of \citet{mason15}, who derive a relationship between the
observed brightness of $z <$ 3 bright (m $<$ 23) galaxies and their
velocity dispersions.  Using our observed fluxes and derived
photometric redshifts, we use the \citet{mason15} relations to estimate the
velocity dispersions for galaxies within 10$^{\prime\prime}$ of our
high-redshift candidates, restricting potential lenses to those with
best-fit photometric redshifts of $z <$ 3.

This process provides an estimate of the lensing
potential of any source near to an object if interest, which can be used with Equation
4 from \citet{mason15} to calculate the size of the Einstein radius.
Using the measured separation between our object of interest and the
nearby potentially lensing galaxy, this radius can be used to
calculate the magnification.
This magnification is dependent on the photometric redshift of both
the source and the lens object, and the observed flux of the lens
object, all of which are measured at varying levels of precision.
Additionally, the \citet{mason15} velocity dispersion -- apparent
magnitude relations also have sizable uncertainties.  We account for
these sources of error by sampling all uncertainties via Monte Carlo
simulations, deriving the magnification for each nearby source 1000
times, sampling the photometric
redshift PDF for both the source and lens, perturbing the \hb\-band
flux of the lens within its uncertainty, and perturbing the velocity dispersion -- magnitude relations
within their fit uncertainties.  The fiducial magnification and its uncertainty is the median and
standard deviation of these 1000 values.

Considering all sources within 10\arcs\ and taking the results at face
value implied significant magnifications for all 11 objects, with a
median magnification for each candidate of $\mu$ = 1.8, obtained by
combining all neighbors per candidate. This high magnification is
surprising as even intermediate lenses should be somewhat rare. To
explore whether this procedure results in a bias towards higher
magnification factors, we ran the following test. For each of our 11
candidates, we simulated placing the candidate at 1000 different
positions in its respective CANDELS field by randomly selecting
similarly-bright sources (i.e., $m \sim$ 25--26.5) and replacing the $\powerset(z)$ of
the random source with that of our candidate. After repeating our
magnification calculation as described above, we found that the median
magnification for all 1000 positions was measured to be $\mu$ $\approx$ 1.4. As the
physical expectation is unity, clearly considering all nearby sources
results in the potential for an excess positive bias in the
magnification. Additionally, this methodology was designed to
determine magnifications for intermediate ($\mu >$ 1.4) and strong ($\mu >$ 2)
lensing, whereas all of the magnifications for our candidates (when
calculated individually for each close neighbor) are consistent with
weak ($\mu <$ 1.4) lensing. Based on these results, we consider only
potential lenses which result in median magnifications of $\mu >$ 1.4.

We found that none of our 11 candidates had a nearby source within 10\arcs\
with a median magnification of $\mu >$ 1.4, thus we do not apply any
magnification correction to our sample.  We note that two sources
do have close companions which have estimated 1.3 $< \mu <$ 1.4.
EGS\_z910\_20381 has a $V=$19.3 galaxy at $z_{phot,lens} =$ 0.3 at a separation of
3.6\arcs, with a median
magnification of $\mu =$ 1.32, while UDS\_z910\_18697 has a $H_{160}$=20.8 galaxy at
$z_{lens} =$ 1.1 with a separation of 3.0\arcs, with a median
magnification of $\mu =$ 1.35.  Both of these magnifications have
sizable uncertainties, thus it is possible that both sources are being
significantly magnified.  While within our present uncertainties we
cannot conclusively determine this magnification, should future work
obtain spectroscopic redshifts for both source and lens, and perform
more robust mass modeling of the lens, this magnification could be
better determined.

\begin{deluxetable*}{cccccccc}
\vspace{2mm}
\tabletypesize{\small}
\tablecaption{Final Sample Summary}
\tablewidth{1.1\textwidth}
\tablehead{
\colhead{ID} & \colhead{$\int$ $\powerset(z>8)$} &
\colhead{z$_{best}$} & \colhead{z$_{spec}$} & \colhead{$\int$
  $\powerset(z \approx 9)$} & \colhead{$\int$ $\powerset(z \approx
  10)$} & \colhead{Sample} & \colhead{M$_{UV}$}}
\startdata
EGS\_z910\_6811&1.00&8.84$^{+0.12}_{-0.25}$&8.683$^{+0.001}_{-0.004}$$^{\dagger}$&0.92&0.00&9&-22.13$^{+0.05}_{-0.03}$\\
EGS\_z910\_44164&0.99&8.87$^{+0.19}_{-0.17}$&8.665$\pm0.001$$^{\ddagger}$&0.97&0.00&9&-21.87$^{+0.05}_{-0.05}$\\
EGS\_z910\_68560&1.00&9.15$^{+0.17}_{-0.34}$&&0.92&0.05&9&-21.49$^{+0.11}_{-0.09}$\\
EGS\_z910\_20381&0.82&8.67$^{+0.32}_{-0.74}$&&0.56&0.01&9&-21.20$^{+0.14}_{-0.12}$\\
EGS\_z910\_26890&0.99&8.99$^{+0.22}_{-0.29}$&&0.94&0.01&9&-21.26$^{+0.11}_{-0.09}$\\
EGS\_z910\_26816&0.99&9.38$^{+0.28}_{-0.39}$&&0.66&0.32&9&-21.28$^{+0.14}_{-0.12}$\\
EGS\_z910\_40898&0.80&8.77$^{+0.25}_{-0.90}$&&0.55&0.01&9&-20.80$^{+0.18}_{-0.13}$\\
COSMOS\_z910\_20646&0.97&9.80$^{+0.10}_{-0.46}$&&0.29&0.68&10&-22.26$^{+0.29}_{-0.04}$\\
COSMOS\_z910\_47074&0.96&9.64$^{+0.22}_{-0.13}$&&0.11&0.85&10&-21.11$^{+0.10}_{-0.11}$\\
UDS\_z910\_18697&1.00&9.89$^{+0.16}_{-0.15}$&&0.00&0.99&10&-22.25$^{+0.12}_{-0.10}$\\
GOODSN\_z910\_35589&1.00&10.41$^{+0.24}_{-0.08}$&10.957$\pm$0.001$^{\dagger\dagger}$&0.00&0.58&10&-21.91$^{+0.03}_{-0.11}$
\enddata
\tablecomments{A summary of the final sample of candidate $z >$ 8.5
  galaxies.  The photometric redshift quantities shown include
  photometry from ground-based data and Cycle 27 {\it HST} imaging, as discussed in \S 4.  The fifth and sixth
  columns show the integral of the redshift $\powerset(z)$ in $\Delta
  z =$ 1 bins centered at $z =$ 9 and 10, respectively.  The higher of
  these two quantities sets whether the galaxy is placed in the $z =$
  9 or $z =$ 10 sample.  The
  UV absolute magnitude is derived via SED fitting, with the
  uncertainty inclusive of both photometric and photometric redshift uncertainties.
  $^{\dagger}$Spectroscopic redshift measurement from \citet{zitrin15}
based on Ly$\alpha$ emission.  $^{\ddagger}$Spectroscopic redshift
measurment from Larson et al.\ (2021) based on Ly$\alpha$
emission.  $^{\dagger\dagger}$Spectroscopic redshift
measurement from \citet{jiang21} based on rest-UV metal lines,
consistent with the \citet{oesch16} grism redshift from the
Ly$\alpha$ break.}
\label{tab:tab7}
\end{deluxetable*}

\begin{deluxetable*}{cccccccccc}
\vspace{2mm}
\tabletypesize{\footnotesize}
\tablecaption{Final Sample Photometry}
\tablewidth{\textwidth}
\tablehead{
\colhead{ID} & \colhead{F606W} & \colhead{F814W} & \colhead{F098M} & \colhead{F105W} & \colhead{F125W} & 
\colhead{F140W} & \colhead{F160W} & \colhead{3.6$\mu$m} & \colhead{4.5$\mu$m}}
\startdata
EGS\_z910\_6811&\phantom{$-$}2.3$\pm$9.6&4.3$\pm$10.3&$-$11.0$\pm$18.4&---&215.8$\pm$15.1&294.9$\pm$30.9&314.5$\pm$13.3&201.1$\pm$147.7&1433.0$\pm$161.7\\
EGS\_z910\_44164&$-$3.6$\pm$9.0&1.2$\pm$11.7&41.7$\pm$31.2&---&161.9$\pm$14.0&257.7$\pm$19.2&249.2$\pm$15.5&256.3$\pm$104.0&568.0$\pm$105.5\\
EGS\_z910\_68560&$-$4.0$\pm$7.0&$-$3.7$\pm$10.6&---&$-$30.0$\pm$15.0&151.5$\pm$15.3&8.1$\pm$31.9&180.2$\pm$12.7&$-$97.1$\pm$97.4&$-$83.8$\pm$99.0\\
EGS\_z910\_20381&$-$9.4$\pm$7.6&$-$2.7$\pm$7.8&8.1$\pm$20.9&---&96.2$\pm$14.3&124.7$\pm$13.3&137.0$\pm$13.1&157.6$\pm$106.6&482.2$\pm$107.5\\
EGS\_z910\_26890&$-$3.3$\pm$5.9&5.8$\pm$7.4&$-$18.5$\pm$17.8&62.6$\pm$34.3&75.8$\pm$10.7&143.1$\pm$17.3&133.3$\pm$8.9&107.8$\pm$75.5&265.7$\pm$77.5\\
EGS\_z910\_26816&$-$4.0$\pm$7.3&1.4$\pm$10.0&---&$-$4.5$\pm$20.4&57.3$\pm$13.6&108.8$\pm$20.2&130.3$\pm$12.1&158.3$\pm$133.9&231.2$\pm$129.6\\
EGS\_z910\_40898&$-$1.3$\pm$4.9&$-$2.2$\pm$5.6&3.8$\pm$14.5&---$-$&62.9$\pm$10.2&92.2$\pm$14.8&91.1$\pm$8.9&156.7$\pm$84.8&347.0$\pm$85.4\\
COSMOS\_z910\_20646&\phantom{$-$}3.9$\pm$8.6&2.6$\pm$10.5&21.3$\pm$26.9&---&70.5$\pm$16.1&119.8$\pm$16.9&246.0$\pm$17.6&855.4$\pm$51.7&1782.8$\pm$50.7\\
COSMOS\_z910\_47074&$-$5.2$\pm$10.0&$-$3.6$\pm$11.9&$-$22.6$\pm$23.3&4.6$\pm$13.8&52.2$\pm$11.0&60.1$\pm$22.1&107.7$\pm$10.3&56.2$\pm$36.0&167.4$\pm$37.4\\
UDS\_z910\_18697&$-$9.5$\pm$14.5&1.3$\pm$15.4&---&---$-$&66.5$\pm$17.4&152.2$\pm$41.4&270.1$\pm$22.2&349.7$\pm$49.5&1680.9$\pm$56.8\\
GOODSN\_z910\_35589&$-$1.8$\pm$6.8&1.6$\pm$4.1&---&$-$2.8$\pm$7.3&3.6$\pm$5.6&96.6$\pm$24.8&171.2$\pm$7.8&109.0$\pm$69.7&129.3$\pm$58.4\\
\enddata
\tablecomments{All fluxes are in nJy.}
\label{tab:tab8}
\end{deluxetable*}

\subsection{Calculation of UV Absolute Magnitudes}
Here we measure the rest-frame UV absolute magnitudes.  We follow convention in this field and we use 1500 \AA\ rest-frame as our reference
UV wavelength.  While at our redshifts of interest this is contained
within the $H_{160}$-band, we can more accurately estimate the flux at this
rest-frame wavelength by using all available photometry.  We do this
following the method of \citet{finkelstein15}, SED fitting the {\it HST}$+${\it Spitzer} photometry for an object, 
measuring the bandpass-averaged flux through a 100
\AA-wide top-hat filter centered at rest-frame 1500 \AA.  We then use
the photometric redshifts to convert this flux to a rest-frame UV
absolute magnitude ($M_{UV}$).  We calculate uncertainties on these
derived magnitudes via Monte Carlo simulations (see
\citealt{finkelstein15} for more details), perturbing both the
photometry and the photometric redshifts within their uncertainties.
For the purposes of this calculation we assume that all candidates are
truly at $z >$ 8, thus we remove any low-redshift solution from the
$\powerset{(z)}$ such that it does not
contribute to the $M_{UV}$ uncertainty budget.
Our inferred values of $M_{UV}$ are listed in Table~\ref{tab:tab7}.

\subsection{Final Sample Summary}
As described in the above section, our additional vetting including
all available ground-based data reduces our sample size from 14 to 11
candidate bright $z >$ 8.5 galaxies.  This final sample is summarized
by the blue-shaded histograms in Figure~\ref{fig:quad}.
Interestingly, two of the candidates which were removed were the two
brightest in our sample ($H=$24.5 and 25.0).  These two sources also both
have red $H_{160}$-[3.6] colors of $>$1 mag, implying that red
$H_{160}$-[3.6] colors may be a good discriminator against contaminants (see
also discussion in \citealt{bouwens19}).  The $H_{160}$-[3.6] color
distribution for our initial sample is shown in the right-hand panel
of Figure~\ref{fig:quad}.  There is one remaining object in our sample
with $H-$[3.6] $>$ 1, COSMOS\_z910\_20646.  As discussed above, this
object does present a potential low-redshift solution which becomes
dominant if the {\sc Galfit}-based IRAC photometry is used; with the
TPHOT-based IRAC photometry it still satisfies our sample selection
criteria with $\int \powerset(z>8) =$ 0.96.  \edit1{We also explored
  the sizes of our sources to see whether a combination of half-light
  radius and $H_{160}$-[3.6] color could help discern contaminants.
  However, we found that 13/14 initial candidates had half-light radii
  of 1--3 pixels ($<$0.2\arcsec), with the size not correlated with
  $H_{160}$-[3.6] color.  The exception is UDS\_z910\_731, which has
  the largest size ($r_h =$ 4.2 pixels) and the reddest
  $H_{160}$-[3.6] color (1.6 mag).  Thus outliers with both red
  $H_{160}$-[3.6] color and large sizes may identify potential
  contaminants in future samples.}

For the remainder of this work, we use the photometric redshift
results when including all available data (ground-based and Cycle 26
and 27 {\it HST}), with the final values
listed in Table~\ref{tab:tab7}.  In this table we split our sample into two redshift
bins, $z \sim$ 9 and $z \sim$ 10.  To decide which bin a galaxy is in,
we use the integrated redshift probability distribution functions,
where galaxies with $\int_{8.5}^{9.5} \powerset(z) >\int_{9.5}^{10.5}
\powerset(z)$ are in the $z \sim$ 9 sample, and vice versa.  As seen
in  Table~\ref{tab:tab7}, seven of our sources fall in the $z \sim$ 9
bin, and four in the $z \sim$ 10 bin.  Finally, we list the full {\it
  HST}$+$IRAC photometry for all sources in our final sample in Table~\ref{tab:tab8}.

\section{Comparison to Previous CANDELS Samples}

Previously, \citet{oesch14}, \citet{bouwens15}, \citet{bouwens16} and
\citet{bouwens19} have selected bright $z >$ 8.5 candidate galaxies in
the CANDELS fields.  We restrict our comparison primarily to
\citet{bouwens19}, as it is inclusive of the previous work listed and it has a larger number of candidates than
\citet{bouwens16} due to the adoption of photometric redshift
selection in \citet{bouwens19}, rather than strict color-color selection as
was adopted in \citet{bouwens16}.
Here we discuss each of our candidates and provide a discussion of
whether or not they were previously published. 

\subsection{Previously Published Candidates in Our Sample}

\begin{itemize} 
\item Candidate EGS\_z910\_6811 -- This candidate galaxy was originally
  published in \citet{roberts-borsani16} as $z =$ 8.6 candidate galaxy
  in particular due to its very red [3.6]$-$[4.5] color.  It was
  then spectroscopically confirmed via Ly$\alpha$ emission by
  \citet{zitrin15} to have $z_{spec} =$ 8.683, in excellent agreement
  with our photometric redshift of 8.84$^{+0.12}_{-0.25}$.  We also
  see a red IRAC color, though note that this source is heavily
  crowded in IRAC, so the true IRAC color is difficult to discern.  This object was not
  selected in the color-selection employed by \citet{bouwens16}, but
  is included (as EGS910-10) in the photometric-redshift selection of
  \citet{bouwens19}, who find a similar brightness and photometric
  redshift as we do ($m_H =$ 25.3 and $z_{phot} =$ 8.6).

\item Candidate EGS\_z910\_44164 -- This candidate galaxy was not
  published in original work in this field \citep{bouwens16}, but was included (as
  EGS910-8) in the selection of \citet{bouwens19}.  We find
  $m_H =$ 25.4 $\pm$ 0.1, which is somewhat brighter than the
  \citet{bouwens19} value of 25.7 $\pm$ 0.1.  Our photometric redshift
  of $z_{phot} =$ 8.87$^{+0.19}_{-0.17}$ is consistent with the value
  of  $z_{phot} =$ 9.1 from \citet{bouwens19}.

\item Candidate EGS\_z910\_20381 -- This candidate galaxy was also not
  published originally, but was included (as
  EGS910-9) in the recent selection of \citet{bouwens19}.  Our
  redshift estimates are consistent (8.67$^{+0.32}_{-0.74}$ for this
  work, and 9.1 for \citealt{bouwens19}), and our $H_{160}$-band magnitudes
  are identical ($H =$ 26.1).  In the appendix of
  \citet{bouwens19} this candidate is also listed as EGS910-13, with
  identical coordinates as their EGS910-9, but a brighter magnitude (25.9) and a lower
  photometric redshift (8.5).  

\item Candidate EGS\_z910\_26890 -- This candidate galaxy was not
  included in the fiducial sample of $z \sim$ 9 candidates by
  \citet{bouwens19}, but it was listed as a possible $z \sim$ 9
  candidate (as EGS910-15) as they measured $z_{phot} =$ 8.3 with
  $\int$ $\powerset(z > 8) =$ 0.38 (including our F098M data presented
  here, \citealt{bouwens21} find $\int$ $\powerset(z > 8) =$ 0.56).  Our  photometric redshift is significantly higher at 8.99$^{+0.22}_{-0.29}$, and
  correspondingly $\int$ $\powerset(z > 8)$ is shifted higher as well
  to 0.99.   Their slightly lower redshift may be driven by the weakly significant
  positive flux in the $Y_{105}$-band at the position of this galaxy
  (62.6 $\pm$ 34.3 nJy; 1.8$\sigma$).  However, this object falls on the edge
  of the F105W pointing (which was targeting another galaxy), with
  only 800 sec of integration, compared to the full depth of $>$3000
  sec, and upon visual inspection is not distinguished from several
  other nearby bright noise peaks.  Regardless, this object was
  stringently not detected in our Cycle 27 F098M imaging ($-$18.5
  $\pm$ 17.8 nJy).  We measure a strong red $J_{125}-JH_{140}$ break of
  0.7 mag, with a blue $JH_{140}-H_{160}$ color of -0.1 mag (and
  $H_{160}-$[3.6] of $-$0.2 mag).  This
  implies that the galaxy is intrinsically blue, with the break in the
  $J_{125}$-band, but mostly out of the $JH_{140}$-band.  This implies
  that the Ly$\alpha$ break must be near to the blue edge of the F140W
  filter; if it were any bluer, the $J_{125}-JH_{140}$ would not be as
  red as observed.  Likewise, if it were any redder, the
  $JH_{140}-H_{160}$ would be redder than observed.  The blue edge of
  the F140W filter is at 1.2$\mu$m, and a Ly$\alpha$ break observed at
  that wavelength would have $z \approx$ 8.9, similar to our inferred
  photometric redshift.

\item Candidate EGS\_z910\_26816 -- This candidate has been published in both
  \citet{bouwens16} and \citet{bouwens19} as EGS910-0.  Our brightness
  measures are similar (26.1--26.2), and their measurement of
  $z_{phot} =$ 9.1$^{+0.3}_{-0.4}$ is consistent with our value of
  $z_{phot} =$ 9.38$^{+0.28}_{-0.39}$.

\item Candidate COSMOS\_z910\_47074 -- This candidate has also been published in both
  \citet{bouwens16} and \citet{bouwens19} as COS910-1.  Our brightness
  measures are also similar (26.3--26.4).  Their redshift measurement
  of $z_{phot} =$ 9.0$^{+0.4}_{-0.5}$ is lower than ours of
  9.64$^{+0.22}_{-0.13}$, though the difference is not highly
  significant.

\item Candidate UDS\_z910\_18697 -- This candidate galaxy was not
  included in the fiducial sample of \citet{bouwens19}, but it was
  listed as a possible $z \sim$ 9 candidate (as UDS910-13).  Without
  ground-based imaging, they find $z_{phot} =$ 9.7, but with the
  inclusion of ground-based imaging, this drops to $z_{phot} =$ 2.3,
  based on a 4$\sigma$ detection in the ZFOURGE
  J2 band in their catalog (which would imply $z <$ 7.7).  In our
  examination of these data, we also see positive flux in J2 (Figure~\ref{fig:UDS_gb_stamps}).
  However, with our noise calculations we find this is less
  significant ($<$2$\sigma$).  Combined with the non-detections in
  ZFOURGE $J1$ and $J3$ filters, EAZY still prefers a $z >$ 9.  As
  discussed above, the ground-based photometry may be affected by a
  nearby large bright galaxy, thus the nature of this source is still
  a bit unclear.

\item Candidate GOODSN\_z910\_35589 -- This object was first published
  in \citet{oesch14} as GN-z910-1 with $z_{phot} =$ 10.2 $\pm$ 0.4, and was
  further found by \citet{oesch16} to have a detectable continuum
  break in {\it HST} grism spectroscopy consistent with $z_{grism} =$
  11.1 $\pm$ 0.1.  Recently, \citet{jiang21} have refined the redshift
  to $z =$ 10.957 $\pm$ 0.001 via weak rest-UV metal emission lines.
  Both spectroscopic measurements are consistent within $<$2$\sigma$ of our
  measurement of $z_{phot} =$ 10.41$^{+0.30}_{-0.07}$.  Our measured
  $H_{160}$-band magnitude is a little brighter at 25.82 $\pm$ 0.05, versus
  26.0 $\pm$ 0.1 (from \citealt{bouwens19}).

\end{itemize}

\subsection{Candidates in Our Sample Published for the First Time}
Three of our candidate galaxies are being published here for the first time.

\begin{itemize}
\item Candidate EGS\_z910\_68560 -- This candidate galaxy has $m_H =$
  25.8 $\pm$ 0.1, and $z_{phot} =$ 9.15$^{+0.17}_{-0.34}$, with
  $\int$ $\powerset(z > 8) =$ 1.0.  This object may have been
  excluded by other studies due to its mild $J_{125}-H_{160}$ color of
  only 0.2 mag, yet the complete non-detection in $Y_{105}$
  strongly constrains $z >$ 8.5.  Interestingly, this source does not
  appear at a significant level in the $JH_{140}$ imaging, which is
  unexpected given its brightness.  This could imply that the
  $J_{125}$ and $H_{160}$ counts are due to persistence prior to the
  CANDELS observation.  However, we examined this in \S 3.3.1, and found
  that the previous observations did not have a high fluence at this
  position, therefore this explanation is disfavored.

\item Candidate EGS\_z910\_40898 -- This candidate galaxy has $m_H =$
  26.5 $\pm$ 0.1, and $z_{phot} =$ 8.77$^{+0.25}_{-0.90}$ with  $\int$
  $\powerset(z > 8) =$ 0.80.  This galaxy has $J_{125}-H_{160} =$
  0.40, which is slightly bluer than commonly used color-selection
  thresholds ($\sim$ 0.5).  It has a red [3.6]$-$[4.5] color,
  consistent with strong [O\,{\sc iii}] emission at the derived
  photometric redshift.

\item Candidate COSMOS\_z910\_20646 -- This candidate galaxy has $m_H =$
  25.4 $\pm$ 0.1, and $z_{phot} =$ 9.80$^{+0.10}_{-0.46}$ with  $\int$
  $\powerset(z > 8) =$ 0.97.  Its red observed $J_{125}-H_{160}$ color
  of 1.4 mag is consistent with its high derived photometric
  redshift.  As discussed above, this object does have a fairly red
  SED, with $H-$[3.6] $=$ 1.35 mag, as red as the two sources removed
  as being low-redshift interlopers.  Additionally, the IRAC fluxes of
  this source are uncertain due to nearby blended neighbors.  However, even with ground-based
  imaging and a Cycle 27 F098M non-detection, this source still satisfies our selection criteria (though
  there is a small secondary redshift peak at $z \sim$ 2.5).

\end{itemize}

\subsection{Published Candidates Not in Our Sample}
Here we discuss the 12 galaxy candidates previously published by
\citet[][hereafter B19]{bouwens19} which are not in our
final sample.  Of their 12 galaxies which are not in our
sample, they measure $H >$ 26.6 for nine of them, fainter than we
consider here.  However, as noted
above, our magitudes sometimes differ, so we searched our catalog for
all 12 of these galaxies, and here we describe our measurements for
these candidates.

\begin{itemize}[leftmargin=*]

\item[] EGS910-2 (B19 $H_{160} =$ 26.70): We measure $H_{160} =$ 26.99, thus
  this object is too faint for our sample, though our measurement of
  $\int$ $\powerset(z > 8) =$ 0.53 is somewhat consistent with a $z >$ 8 solution.

\item[] EGS910-3 (B19 $H_{160} =$ 26.40): We measure $H_{160} =$
  26.58 with a $H_{160}$-band SNR=4.9, thus this object's SNR is too low for
  our sample, though our measurement of
  $\int$ $\powerset(z > 8) =$ 0.92 is very consistent with a $z >$ 8 solution.

\item[] UDS910-1 (B19 $H_{160} =$ 26.60):  We measure $H_{160} =$
  26.43, thus this object is bright enough for our sample, and its
  $\powerset(z)$ distribution clears our cutoff with $\int$
  $\powerset(z > 8) =$ 0.64.  However, the secondary low-redshift
  solution we find for this source is nearly as good as the
  high-redshift solution, with $\Delta \chi^2 =$ 1.2, thus this
  misses our $\Delta \chi^2 >$ 3.5 cutoff for the difference in
  $\chi^2$ between the low and high-redshift solutions, though it is still
  consistent with a $z >$ 8 solution.

\item[] UDS910-5 (B19 $H_{160} =$ 25.80):  We measure $H_{160} =$
  25.3, thus this object is bright enough for our sample.  However,
  our photometric redshift has a best-fit solution at $z_{phot} =$
  1.16, with $\int$ $\powerset(z > 8) =$ 0.09, failing our selection
  criteria.  We explored this further, and found that this galaxy
  lies off the eastern edge of the CANDELS ACS mosaic, thus our
  catalog lacks the necessary optical photometry to constrain the
  redshift.  B19 made use of ground-based optical imaging in this
  region to select this galaxy.

\item[] GS-z9-1 (B19 $H_{160} =$ 26.60): We measure $H_{160} =$ 26.94, thus
  this object is just too faint for our sample, though our measurement of
  $\int$ $\powerset(z > 8) =$ 0.99 is very consistent with a $z >$ 8 solution.

\item[] GS-z9-2 (B19 $H_{160} =$ 26.90): We measure $H_{160} =$ 27.21, thus
  this object is just too faint for our sample.  Our measurement of
  $\int$ $\powerset(z > 8) =$ 0.31 is not consistent with a $z >$ 8
  solution.  This object does not exhibit significant flux in our
  optical bands, but we find a very red $H-$[3.6] color, which makes
  for a very broad $\powerset(z)$.  However, this object has a near
  neighbor crowding this source in the IRAC imaging, so the deblended
  IRAC flux may not be accurate.

\item[] GS-z9-3 (B19 $H_{160} =$ 26.90): We measure $H_{160} =$ 27.08, thus
  this object is too faint for our sample, though our measurement of
  $\int$ $\powerset(z > 8) =$ 0.99 is very consistent with a $z >$ 8 solution.

\item[] GS-z9-4 (B19 $H_{160} =$ 26.8): This object is not in
our catalog as it is not in the CANDELS GOODS-S proper
region.  It is in the HUDF09-2 flanking field, which we did not
include in our analysis.  The B19 magnitude for this source was 26.8,
so it likely would not have been bright enough for inclusion in our
sample had we used these additional data.

\item[] GS-z9-5 (B19 $H_{160} =$ 26.4): We measure $H_{160} =$ 26.04,
  $\int$ $\powerset(z > 8) =$ 0.72, and $\Delta \chi^2 =$ 3.7,
  satisfying those specific selection criteria.  However, our
  photometric redshift distribution has $\int\powerset(z\sim$ 8) $>$
  $\int\powerset(z\sim$ 9).  Our results thus imply a redshift closer
  to $z \sim$ 8, though a $z \sim$ 9 solution is plausible.

\item[] GS-z10-1 (B19 $H_{160} =$ 26.90): We measure $H_{160} =$ 27.12, thus
  this object is just too faint for our sample.  Our measurement of
  $\int$ $\powerset(z > 8) =$ 0.67 clears our sample selection
  threshold of 0.6, with a best-fitting redshift is $z_{phot} =$
  9.91.  The optical photometry may be affected by a very close
  neighbor just South of this source, leading to a larger secondary
  redshift peak at $z \sim$ 2.5.

\item[] GN-z9-1 (B19 $H_{160} =$ 26.60): We measure $H_{160} =$ 26.79, thus
  this object is too faint for our sample, though our measurement of
  $\int$ $\powerset(z > 8) =$ 1.00 is very consistent with a $z >$ 8 solution.

\item[] GN-z10-2 (B19 $H_{160} =$ 26.80): We measure $H_{160} =$ 26.98, thus
  this object is too faint for our sample. Our measurement of
  $\int$ $\powerset(z > 8) =$ 0.32 is also too low for our sample,
  though the best-fitting redshift is $z_{phot} =$ 9.95.  However, the
  $\powerset(z)$ is very broad, due to the faintness of this object,
  as well as a very bright nearby galaxy affecting the IRAC photometry.

\item[] GN-z10-3 (B19 $H_{160} =$ 26.80): We measure $H_{160} =$ 26.90, thus
  this object is too faint for our sample. Our measurement of
  $\int$ $\powerset(z > 8) =$ 0.57 is just a bit too low for our sample,
  though the best-fitting redshift is $z_{phot} =$ 9.77.  This object
  has a somewhat broad $\powerset(z)$, primarily due to its faint
  nature, as nothing in the imaging is inconsistent with a high
  redshift solution (although this object does not have F140W coverage).

\end{itemize}

In summary, the majority of the B19 sources which are not in our
sample are fainter than we
considered in this study, though our measurements for most are consistent with a
high-redshift nature.  Of the two sources that are bright
enough for our sample, for only one (GS-z9-5) did we use similar data in the
selection, and this one just narrowly misses our sample.  We conclude
that our sample selection is inclusive yet accurate, as we recover
nearly all previously published sources which we should have, as well
as adding a few sources not previously published.

\section{Luminosity Function Estimation}

In this section we measure the rest-UV luminosity function of our candidate
sample of $z >$ 8.5 galaxies.  In order to do this, we need to
estimate our completeness in each field, as a function of source
brightness, which we describe in the following subsection.

\subsection{Estimating the Completeness}

To calculate our luminosity function, we use the effective volume method, where 
\begin{equation}
    V_{eff}(M)=\int \frac{dV}{dz}P(M,z)dz
\end{equation}
where $dV/dz$ is the comoving volume element, and $P(M,z)$ is the
probability that an object at a given absolute magnitude and redshift
satisfies our sample selection criteria.  We estimate $P(M,z)$ using
completeness simulations.  Our method follows our previous work
\citep[e.g.,][]{finkelstein15,rojasruiz20}.

We performed a more detailed set (compared to \S 2.2) of completeness simulations separately in each of the
five CANDELS fields.  In each field, we
ran 50 iterations of our simulation, where in every iteration we place
10,000 mock galaxies across each of the images for each field, for a
total sample of 500,000 mock galaxies per field.
We build SEDs of the mock galaxies to derive their bandpass-averaged
fluxes in each filter available in a given field.  First, we draw a random redshift
uniformly over the range 4.5 $< z <$ 11.5.  Then for each object, we draw
an $H_{160}$-band magnitude from a random distribution over 22 $< H <$ 28.5.
We use a combination of two log-normal distributions, a slowly-rising
log-normal distribution from $H=$22 to 25, and a steeper log-normal
from $H=$25 to 28.5.  This
combination results in $\sim$50\% of the simulated objects having $H
<$ 26.6, where our sample lies.  To ensure robust statistics at the
bright end, every 10th simulation iteration simulated only  22 $< H
<$ 25 galaxies with a flat distribution in magnitude.

We also draw stellar population ages (in log units), metallicities and dust
attenuation values from log-normal distributions, with typical values
of log (age/yr) $\sim$ 7.4, $Z =$ 0.2$Z$\sol, and E(B-V) $=$ 0.15.  The
combination of these values produces a UV spectral slope $\beta =
-$2.0, similar to those observed for bright $z >$ 9 galaxies
\citep[e.g.,][]{wilkins15}.  We use these properties to generate
colors from a \citet{bruzual03} model, normalizing the model to the
$H_{160}$-band magnitude for a given mock object. 

Mock galaxy images are generated via {\sc Galfit} \citep{peng02}, with
Sersic indices drawn from a log-normal distribution of the form
$n=e^{\mathcal{R}}+0.8$ (restricted to 1 $< n <$ 5), which has a peak
at $n =$ 1.2, with 60\% of the distribution at $n <$ 2.  The axis
ratios are drawn from a log-normal
distribution tilted towards high values (median $b/a=$ 0.75), and
position angles are drawn from a uniform random distribution.  We draw
galaxy half-light radii using observed relations between galaxy size
and their absolute UV magnitudes, using a relation similar to that
found by \citet{kawamata18} at $z =$ 6--7, of the form
\begin{equation}
r_{h}(M_{UV}) = 0.94 \times 10^{-0.4 (M_{UV}+21) \beta}~[kpc]
\end{equation}
where $M_{UV}$ is the absolute UV magnitude of an object, and $\beta$
is the slope of the size-luminosity relation.  We assume $\beta =$
0.25 for bright ($M < -$ 21) galaxies, and 0.5 for fainter galaxies.
We apply a scatter of 0.2 dex to these sizes to represent the
intrinsic scatter at fixed magnitude.  These {\sc Galfit} images are
normalized to the magnitude for a given object in a given filter, and
convolved with the measured point-spread function (PSF).  
As our images have all been PSF-matched to the $H_{160}$-band, we use
the $H_{160}$-band PSF derived for each field
(\S~\ref{section:psfmatching}) as the PSF for all bands in a given
field for these simulations.  Finally, the galaxy images are added to
a random position of the real image.

\begin{figure}[!t]
\epsscale{1.15}
\plotone{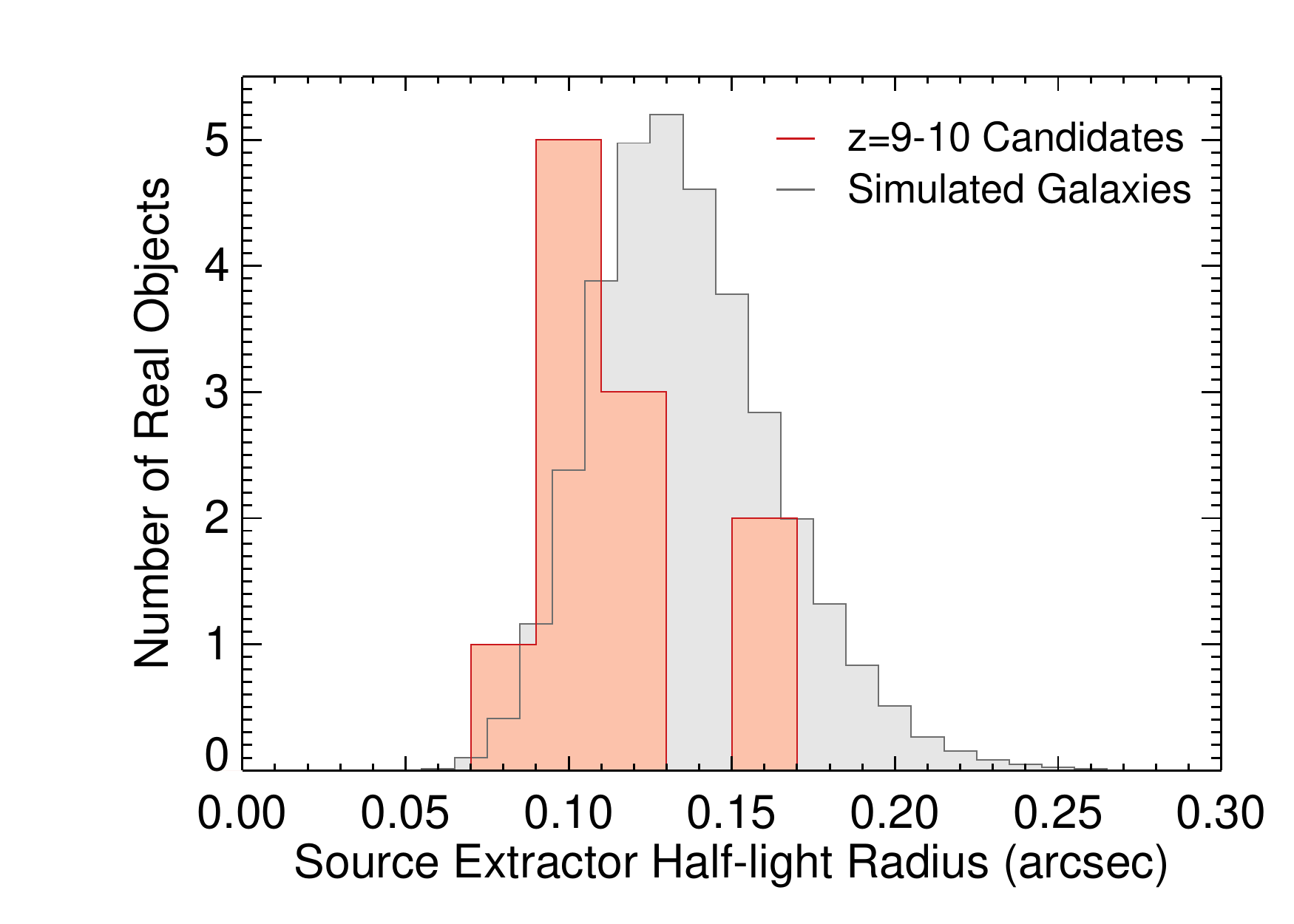}
\vspace{-2mm}
\caption{The half-light radius distribution measured by Source
  Extractor for our final list of candidates (red) compared to that
  for sources recovered from the completeness simulations (gray).  We
  tuned the later to approximately match the former to ensure that we are not
  over-correcting our observations for a population of extended
  sources which are not known to exist.  Should future surveys find
  such a population, it would increase the completeness corrections
  above what we have applied to our luminosity functions, increasing
  the derived number densities.}
\label{fig:rh}
\end{figure}

\begin{figure*}[!t]
\epsscale{0.9}
\plotone{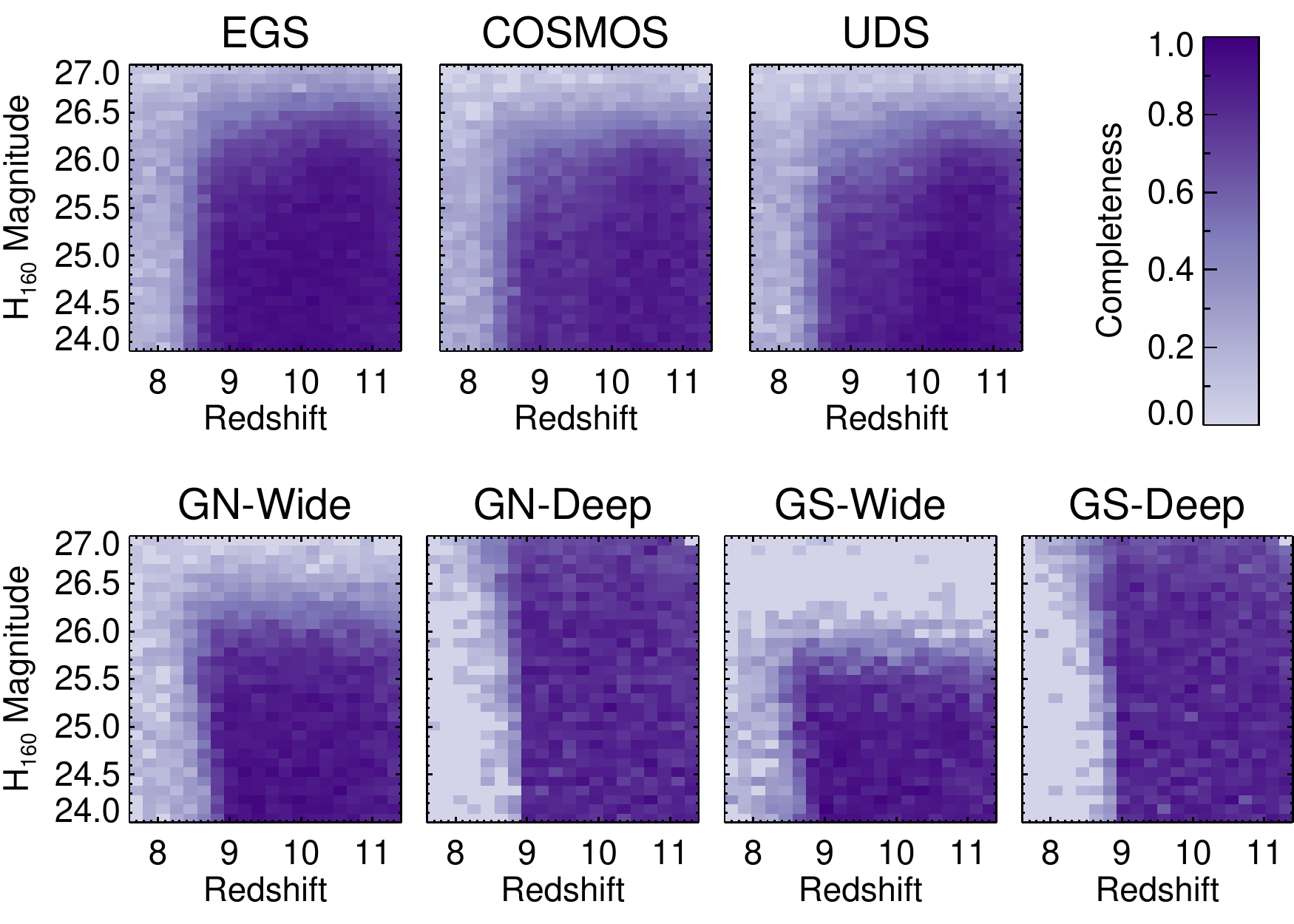}
\vspace{-2mm}
\caption{Each panel shows the results of completeness simulations from
one of our five fields.  The GOODS fields are split by their depth
(the GOODS-S ERS field is not shown, but results are in between the
Deep and Wide fields). The shading denotes the completeness as a
function of $H_{160}$-band magnitude and redshift, as shown by the color
bar.  Our selection process results in a redshift selection function
which peaks between 8.5 $< z <$ 10.5, as desired.  We see that
completeness falls off rapidly at $H >$ 26 in the CANDELS wide fields,
while the deep fields remain complete to fainter magnitudes as expected.}
\label{fig:completeness}
\end{figure*}

When we applied the half-light radius functional form
from Equation 5 directly in our simulations, our completeness values
were lower than (the near-unity) expected for very bright ($H <$ 25) simulated sources.  We inspected the
results of the simulations to determine why bright sources were
rejected, and found that it was primarily due to low S/N (failing the
S/N$_{H} >$ 7 criterion) due to somewhat large sizes.  To determine if
this was appropriate, we compared the half-light radii of our
sample of high-redshift galaxies to those recovered in our
simulations, and found that the latter were too large by a factor of
$\sim$2.  Given that our largest source in our catalog has r$_{h} =$
0.15\arcs, the volumes derived from completeness simulations where the
bulk of simulated sources are larger than that would be artificially
too small.  We thus modified Equation 5 by a single scale factor to
reduce the radii, such that the median of the recovered simulated
galaxies approximately matched those we observe.  Applying a scale
factor of 0.25 results in a recovered simulated value of r$_{h} =$
0.138 $\pm$ 0.029\arcs, compared to 0.11 $\pm$ 0.030\arcs\ for our
sample of candidate galaxies.  We show these distributions in
Figure~\ref{fig:rh}.  We thus applied this scale factor to our
simulated galaxy images to ensure that we did not artificially reduce
the effective volume.  We note that this makes the explicit assumption
that all bright $z =$ 9--10 galaxies have similar sizes to those in
our sample, which is a necessary assumption until significantly deeper
wide-field imaging is available to test this hypothesis.

We then measured photometry and photometric redshifts using Source
Extractor and EAZY, respectively, in an identical way as done on our
real science images, including measuring empirical flux uncertainties
based on the aperture sizes and positions of the sources in the
images.  For each simulation, we matched galaxies in the recovered
photometric catalog to the input catalog, counting a source as a
recovered match if it was $<$0.2$^{\prime\prime}$ from the input
position (e.g., roughly within one PSF FWHM).  Recovered objects were
then subject to the same signal-to-noise and photometric redshift
quality criteria as for our real sample.  We calculated the
completeness in bins of input magnitude and redshift as the number of
fully recovered sources (e.g., found by Source Extractor, and passing
all sample selection cuts) divided by the number of input sources per
bin.  We note that in our previous work \citep{finkelstein15} we
calculated, at significantly increased computational cost, the
completeness additionally as a function of half-light radius and UV
slope $\beta$.  As expected, we found that the completeness depends
sensitively on size, and weakly on $\beta$.  To account for these
effects in this work, we have tuned the parameters of the simulated
galaxies such that their sizes and colors roughly match the observed
sample of galaxies (see also the preceding paragraph). 

\begin{figure*}[!t]
\epsscale{0.9}
\plotone{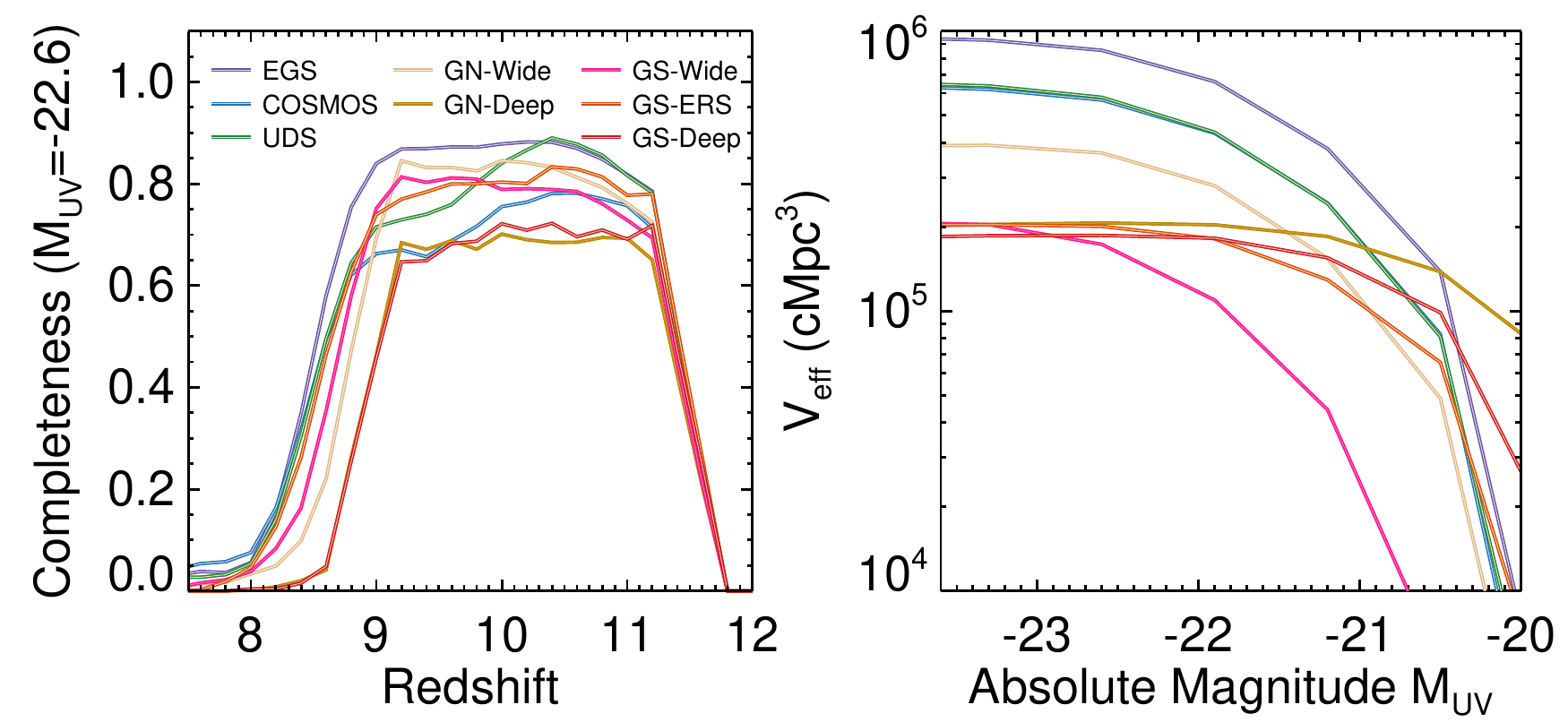}
\vspace{-2mm}
\caption{Left) The completeness as a function of redshift for
  simulated galaxies at $M_{UV} = -$22.  The redshift
  distributions are peaked at $z \sim$ 9--11, matching our selection criteria.  Right) Effective volume versus absolute
  magnitude, with the same colors as in the left
  panel.  The differences in effective volume between the fields
  depend on both the area per field (Table~\ref{tab:tab9}) and the completeness of
  a given field.}
\label{fig:volumes}
\end{figure*}

Figure~\ref{fig:completeness} summarizes these completeness
simulations.  Each panel shows the completeness in bins of redshift
and $H_{160}$-band magnitude for one of our five fields.  The completeness
peaks at $z \sim$ 8.5--10.5, which is consistent with the filter set
available.   The redshift selection functions are shown in the
left-hand panel of Figure~\ref{fig:volumes}.
In the right-hand panel of Figure~\ref{fig:volumes} we show the
effective volume as a function of absolute magnitude, calculated
following Equation 4.  We tabulate
these volumes in Table~\ref{tab:tab9}.

\begin{deluxetable*}{cccccc}
\vspace{2mm}
\tabletypesize{\small}
\tablecaption{Effective Volumes}
\tablewidth{1.1\textwidth}
\tablehead{
\multicolumn{1}{c}{Field} & \multicolumn{1}{c}{Area (arcmin$^2$)} & \multicolumn{4}{c}{Effective Volume (10$^4$ Mpc$^3$)}\\
\multicolumn{2}{c}{$ $} & \multicolumn{1}{c}{$M\!\!=\!\!-$22.6} & \multicolumn{1}{c}{$M\!\!=\!\!-$21.9} & \multicolumn{1}{c}{$M\!\!=\!\!-$21.2} & \multicolumn{1}{c}{$M\!\!=\!\!-$20.5}}
\startdata
EGS&205.9&89.43&72.86&35.76&5.42\\
COSMOS&159.8&59.09&48.28&22.07&2.59\\
UDS&153.5&60.62&48.51&21.38&2.75\\
GOODSN--Wide&101.3&39.11&31.73&13.19&1.34\\
GOODSN--Deep&67.7&20.57&20.72&19.67&15.01\\
GOODSS--Wide&53.2&19.52&11.68&1.59&0.05\\
GOODSS--ERS&49.2&20.52&19.44&14.38&4.95\\
GOODSS--Deep&60.2&18.75&18.45&17.40&10.63\\
\hline
Total&850.7&327.6&271.7&145.4&42.8
\enddata
\label{tab:tab9}
\tablecomments{The areas covered by each of our fields, and the
  effective volumes probed, calculated by our completeness simulations
  in four bins of absolute magnitude.}
\end{deluxetable*}

We compared our volumes to those calculated by \citet{bouwens19}, who
used a similar dataset, though with differences in the sample
selection and simulation methodology.  While they do not list their volumes
directly, they can be inferred from their published number densities
and the number of galaxies in their sample in each magnitude bin.  We
found that our effective volume matches the sum of their $z
\approx$ 9 and $z \approx$ 10 volumes.

\begin{figure*}[!t]
\epsscale{1.0}
\plotone{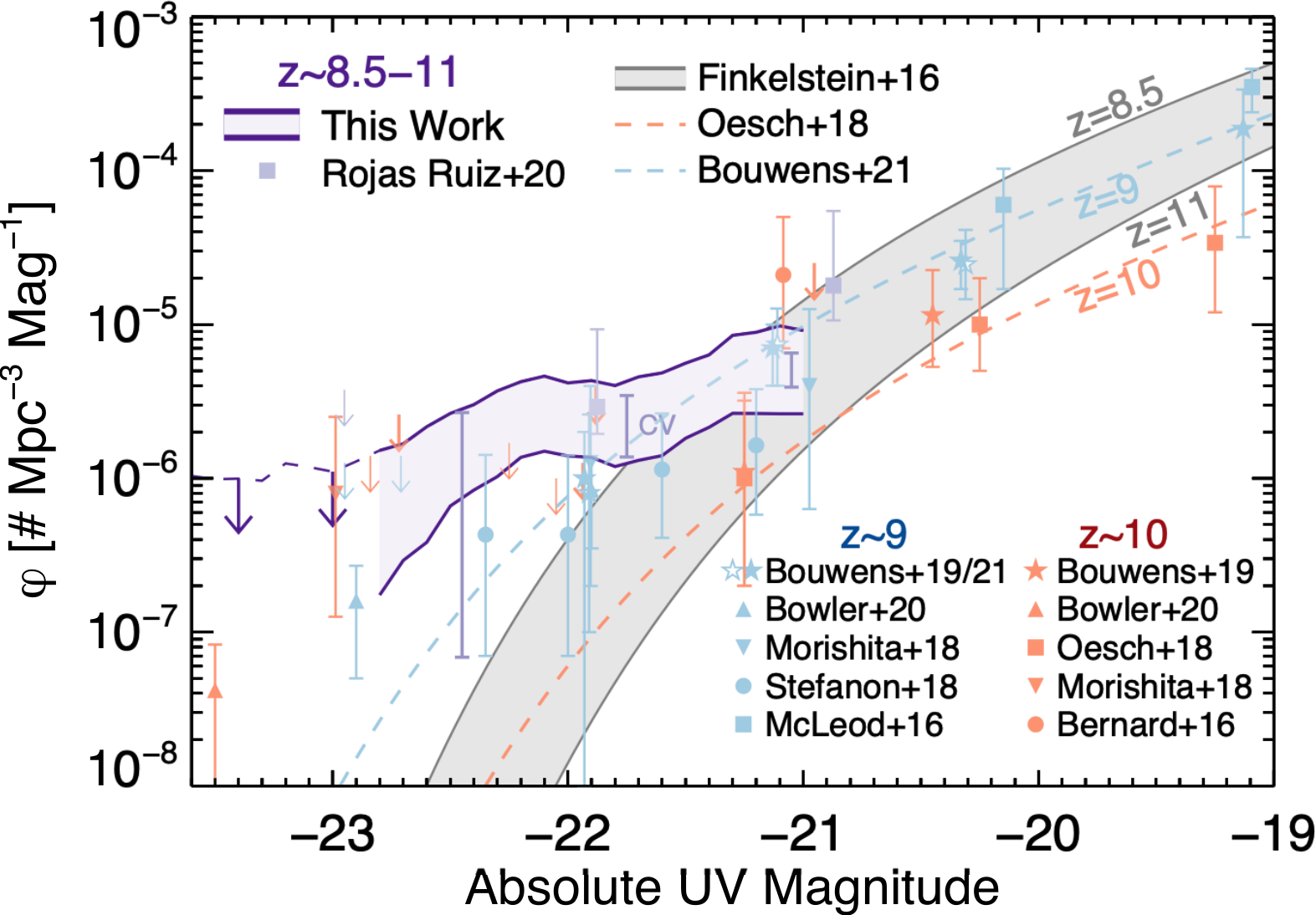}
\vspace{-2mm}
\caption{The measured number densities using our pseudo-binning
  technique on our candidate galaxy
  sample at $z \approx$ 8.5--11 (purple shaded region; the dashed
  portion shows the 84\% confidence upper limit brighter than the
  magnitude of our brightest observed source) compared to various recent literature
  results color-coded by redshift bin.  We also show the smoothly evolving
  predicted Schechter function from \citet{finkelstein16}, with the
  shaded gray region denoting its evolution from
  $z =$ 8.5--11, as well as the measured $z =$ 9 and 10 Schechter functions from
 \citet{bouwens21} and \citet{oesch18}, respectively.  The three error
 bars at either end and in the middle of our results
  represent the estimated cosmic variance uncertainty, which is
  less than the measured Poisson uncertainty except at the brightest
  end.  Our results are consistent with previously published
  luminosity functions at $M_{UV} \sim -$21, while at brighter
  luminosities our results lie
  at the upper end of some previous studies, possibly boosted by an
  apparent overdensity in the EGS field (\S 7).  Our results are
  consistent with a scenario where the
  bright end evolves less rapidly than the faint-end of the
  luminosity function, similar to what was found by \citet{bowler20}.  Due to the large uncertainties on the number
  densities, we cannot conclusively distinguish between smooth or
  rapid evolution of the UV luminosity function at $z >$ 8.}
\label{fig:lf}
\end{figure*}   

\subsection{Calculating Observed Number Densities}
Due to the small number of
galaxies in our sample, as well as the relative short time period
between the $z \sim$ 9 and 10 epochs, we elect to study our sample in
a single redshift bin, spanning $z \sim$ 8.5--11.
To calculate the observed number densities, we follow the methodology of \citet{finkelstein15}
to calculate both these values and their
associated uncertainties.
We use a Markov Chain Monte Carlo (MCMC)
method which employs a goodness-of-fit statistic ($C^2$) which models the
probability distribution as a Poissonian distribution, proper for
these small numbers, with 
\begin{equation}
\mathcal{C}^2(\phi) = -2~\mathrm{ln}~\mathcal{L}(\phi)
\end{equation}
\begin{multline}
 C^2(\phi) = -2\sum\limits_{i}\!
 \sum\limits_{j}\!~N_\mathrm{j,obs}\!~\mathrm{ln}(N_\mathrm{j,expected})
 \!~- \\
 \!~ N_\mathrm{j,expected}\!~-\!~ \mathrm{ln}(N_\mathrm{j,obs}!) 
\end{multline}
where $\mathcal{L}(\phi)$ is the likelihood that the expected number
of galaxies ($N_\mathrm{expected}$) matches that observed
($N_\mathrm{obs}$) for a given differential number density $\phi$ (in
units of number Mpc$^{-3}$ mag$^{-1}$).  The MCMC sampler used is
described in \citet{finkelstein19}.  For this work, we run a burn-in
consisting of 10$^5$ steps, which we found sufficient for the sampler
to find a minimum, and we then measured our posterior from a
subsequent 10$^4$ steps.  The median of this posterior distribution in each
redshift/magnitude bin is the fiducial number density value, while the uncertainty is taken as the 68\% central width.

The differential luminosity function is typically calculated in
magnitude bins.  This can lead to a bias where the bin centers and
width choices can significantly affect the derived number densities
with such a small sample.
While this can be avoided by assuming a specific
functional form \citep{schmidt14}, the form of
the bright end of the $z >$ 8 luminosity function is not known, thus
this choice also imparts a bias.
In an effort to avoid these biases, we implement a novel ``pseudo''-binning
procedure within our MCMC.  When we calculate the number density at a
given value of $M_{UV}$, in each step of the MCMC chain, a random
value of the binsize is chosen from 0.3 -- 1.5 mag.  We then draw an
absolute magnitude for each source by sampling the absolute magnitude posterior
distributions (see \S 4.6; this allows galaxies to effectively move bins), calculating $N_\mathrm{obs}$ for each
bin.  This procedure is repeated along each step of the MCMC chain,
allowing the derived number density to encompass uncertainties in the
absolute magnitudes and marginalizing over changes in the results due
to the bin size.  We perform this calculation in steps of 0.1 mag from
$M_{UV} = -$24 to $-$20.  We checked that these results are in
agreement with a standard binning scheme by comparing these results to
those obtained with a binsize of 0.7 mag, finding good agreement.

\begin{deluxetable}{cc}
\vspace{2mm}
\tabletypesize{\small}
\tablecaption{Measured Number Densities for $z=$9--11}
\tablewidth{2.0\textwidth}
\tablehead{
\multicolumn{1}{c}{Magnitude Bin} & \multicolumn{1}{c}{Number Density (10$^{-6}$ Mpc$^{-3}$)}}
\startdata
$-$22.7&$<$0.96\\
$-$22.6&1.050$_{-0.666}^{+1.115}$\\
$-$22.5&1.373$_{-0.712}^{+1.270}$\\
$-$22.4&1.674$_{-0.835}^{+1.338}$\\
$-$22.3&1.985$_{-0.959}^{+1.715}$\\
$-$22.2&2.389$_{-1.020}^{+1.857}$\\
$-$22.1&2.681$_{-1.180}^{+1.936}$\\
$-$22.0&2.511$_{-1.113}^{+1.657}$\\
$-$21.9&2.617$_{-1.248}^{+1.700}$\\
$-$21.8&2.414$_{-1.220}^{+1.593}$\\
$-$21.7&2.827$_{-1.517}^{+1.738}$\\
$-$21.6&2.864$_{-1.459}^{+1.981}$\\
$-$21.5&3.306$_{-1.483}^{+2.300}$\\
$-$21.4&3.789$_{-1.641}^{+2.560}$\\
$-$21.3&4.798$_{-2.151}^{+3.705}$\\
$-$21.2&4.824$_{-2.181}^{+4.075}$\\
$-$21.1&5.125$_{-2.497}^{+4.664}$\\
$-$21.0&5.218$_{-2.590}^{+3.922}$
\enddata
\tablecomments{Number densities estimated via MCMC with Poisson
  uncertainties, estimated at each magnitude over a variety of
  potential bin sizes.  We estimate the additional fractional uncertainty
  due to cosmic variance as 0.95, 0.43, 0.25 at $M_{UV} =$ $-$22.5, $-$21.75 and $-$21.0, respectively, using the estimator provided by
  \citet{bhowmick20}.} 
\label{tab:tab10}
\end{deluxetable}

We report these number density values in Table~\ref{tab:tab10}, and plot them in
Figure~\ref{fig:lf}, where we also compare to a variety of
recent results from the literature. We estimate the uncertainty due to cosmic variance via the estimator
provided by \citet{bhowmick20} based on the BlueTides simulation.  We
calculate this independently for each field using the appropriate
field geometry, combining the results in quadrature.  We find a
fractional uncertainty on the linear number density due to cosmic variance
of 0.95, 0.43, 0.25 in the $-$22.5, $-$21.75 and $-$21.0 magnitude
bins, respectively.
We plot these uncertainties as secondary error bars next to the
relevant data points in Figure~\ref{fig:lf}.
With the exception of our brightest luminosities probed, these estimated cosmic variance uncertainties are lower than the
Poisson uncertainties, implying that our results are Poisson-noise
limited.  This is not surprising given the small number of detected
galaxies, though we also note that these cosmic variance uncertainties
are dependent on this specific simulation.

\subsection{Implications}

\subsubsection{Comparison to Previous Observational Results}
Here we compare to previous results, which typically examined $z =$ 9
and $z =$ 10 separately.
Examining Figure~\ref{fig:lf}, the number density at our faintest luminosities
is consistent with previous results at both $z =$ 9 and 10 within
our uncertainties.  The middle of our probed range at M$_{UV} = -$21.9 is
comparable to what was found by \citet{rojasruiz20} over a similar
wide-redshift bin, though it is somewhat higher than found by other
studies at both $z \sim$ 9 and 10.  However, this tension is not
significant due to the large (Poisson and cosmic variance)
uncertainties.  Finally, our brightest luminosities probed have a number
density consistent with previous results by \citet{stefanon19} and
\citet{morishita18}.  This number density is also broadly consistent
with the very shallow bright-end decline suggested by
\citet{bowler20}.  However, at these bright luminosities
both the Poisson and cosmic variance uncertainties are significant.
Additionally, as we describe in \S 7, our sample does appear to
contain one overdense structure which could bias our results high,
though the galaxies in the EGS field are not the most luminous in our sample.

\begin{figure*}[!t]
\epsscale{1.18}
\plotone{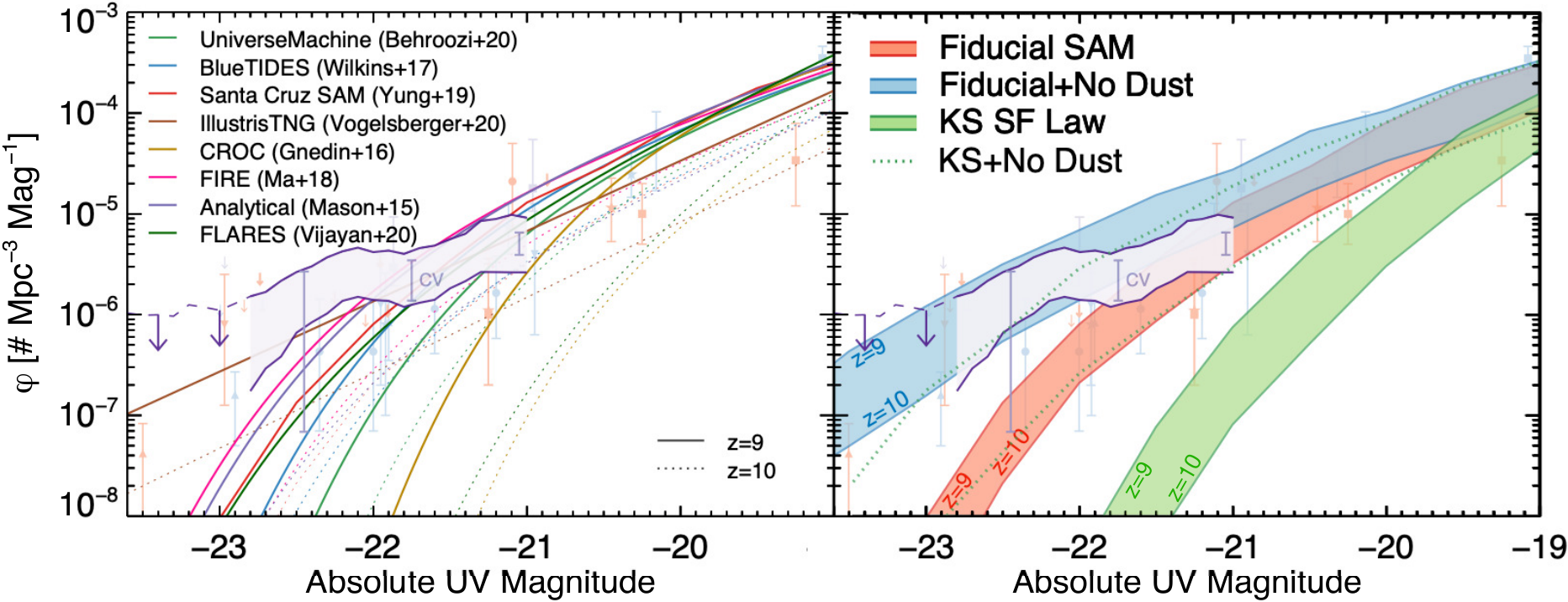}
\vspace{-2mm}
\caption{A comparison of luminosity functions from observations (with the
same symbols as in Figure~\ref{fig:lf}) to recent theoretical
predictions.  Our results are shown by the purple shaded region.  Left) Predictions from hydrodynamical
models (BlueTIDES, \citealt{wilkins17};
IllustrisTNG, \citealt{vogelsberger20}; CROC, \citealt{gnedin16};
FIRE, \citealt{ma18}; FLARES, \citealt{vijayan20}), an analytical
calculation \citep{mason15b}, an empirical model (UniverseMachine,
\citealt{behroozi20}), and a semi-analytic model (Santa Cruz
SAM, \citealt{yung19a}).  Solid (dotted) lines denote predictions at $z
=$ 9 (10).  Predictions are generally consistent with observations
within the uncertainties, except that the CROC and Universe Machine
models under-predict the number density of our brightest bin.  Right) Predictions solely from the Santa
Cruz SAM, with their fiducial model in red, a model with no dust
attenuation in blue, and a model with a shallower dependence of the SFR
on gas density in green (shaded is with dust; the dotted green lines
are with no dust).  Changing the dust attenuation and the SF
scaling dependence both significantly affect the bright end.  Our
observations are consistent with a SF with a steeper dependence on gas
density than the classic Kennicutt-Schmidt relation, though the exact
dependence is degenerate with the normalization of this relation and
the amount of dust attenuation in bright galaxies.}
\label{fig:lf_theory}
\end{figure*}

We also compare to the empirically predicted Schechter luminosity
function from \citet{finkelstein16}, shown by the gray shaded region.
This prediction is based on the assumption
that the smooth decline in the luminosity function observed at $z =$ 4
to 8 continues to higher redshift.  This is in contrast to the
accelerated decline in the luminosity function proposed by
\citet{oesch18} and \citet{bouwens19}; we show the $z =$ 9 (10) luminosity
function estimated by \citet{bouwens21} (\citealt{oesch18}) as the
blue (red) dashed line.
Our faintest results are consistent with both the smooth and
accelerated decline, while our brightest results lie above both the
smooth and accelerated decline-based luminosity functions (though
these scenarios differ the most at the faint end, which we do not probe).  This
result agrees with recent ground-based observations by
\citet{bowler20}, who propose that the bright end of the luminosity
function is evolving less rapidly than the faint-end at these redshifts.
While our brightest measurement provides weak evidence against the
accelerated decline scenario, given the uncertainties we conclude that down to our sample limit of
$H_{160}$=26.6 we cannot prove or disprove this scenario at a significant confidence level with the existing data.

\subsubsection{Physical Interpretation}
A number of physical properties can affect the bright end of the
rest-frame UV luminosity function.  It is well-known that the observed
UV light from galaxies suffers from dust attenuation, with the
specific amount of attenuation dependent on the galaxy mass or
luminosity and redshift \citep[e.g.,][]{finkelstein12a,bouwens14}.  The
abundance of UV-bright galaxies can also depend on the amount of light
coming from an accretion disk around a central supermassive black
hole, though given the paucity of known bright AGNs at $z >$ 7,
this is likely unimportant at the epoch studied here (though some of
the observed light may still be due to low-luminosity AGN within these
galaxies).  The
efficiency of the conversion of gas into stars can also affect the
bright end of the luminosity function, with a steeper relation between
the surface density of SFR and gas mass leading to a shallower
bright-end decline \citep[e.g.][]{yung19a}.  \edit1{An evolving initial mass
function (IMF) could also affect the UV luminosity function; in a scenario
where higher redshift gaalaxies have more progressively top-heavy
IMFs (presumably due to decreasing metallicity), one would expect the
shape of the UV luminiosity function to change as the
mass-to-(UV)light ratio decreases.}
Finally, the evolution of the bright end will also trace the rate at
which negative feedback effects on star formation (e.g., AGN feedback,
stellar feedback) first manifest.  If these effects are not
significant in massive galaxies in this epoch, the bright end could be out of equilibrium and
show an excess of star-forming galaxies compared to later times \citep[e.g.,][]{ypeng10}.

\begin{figure*}[!t]
\epsscale{0.5}
\plotone{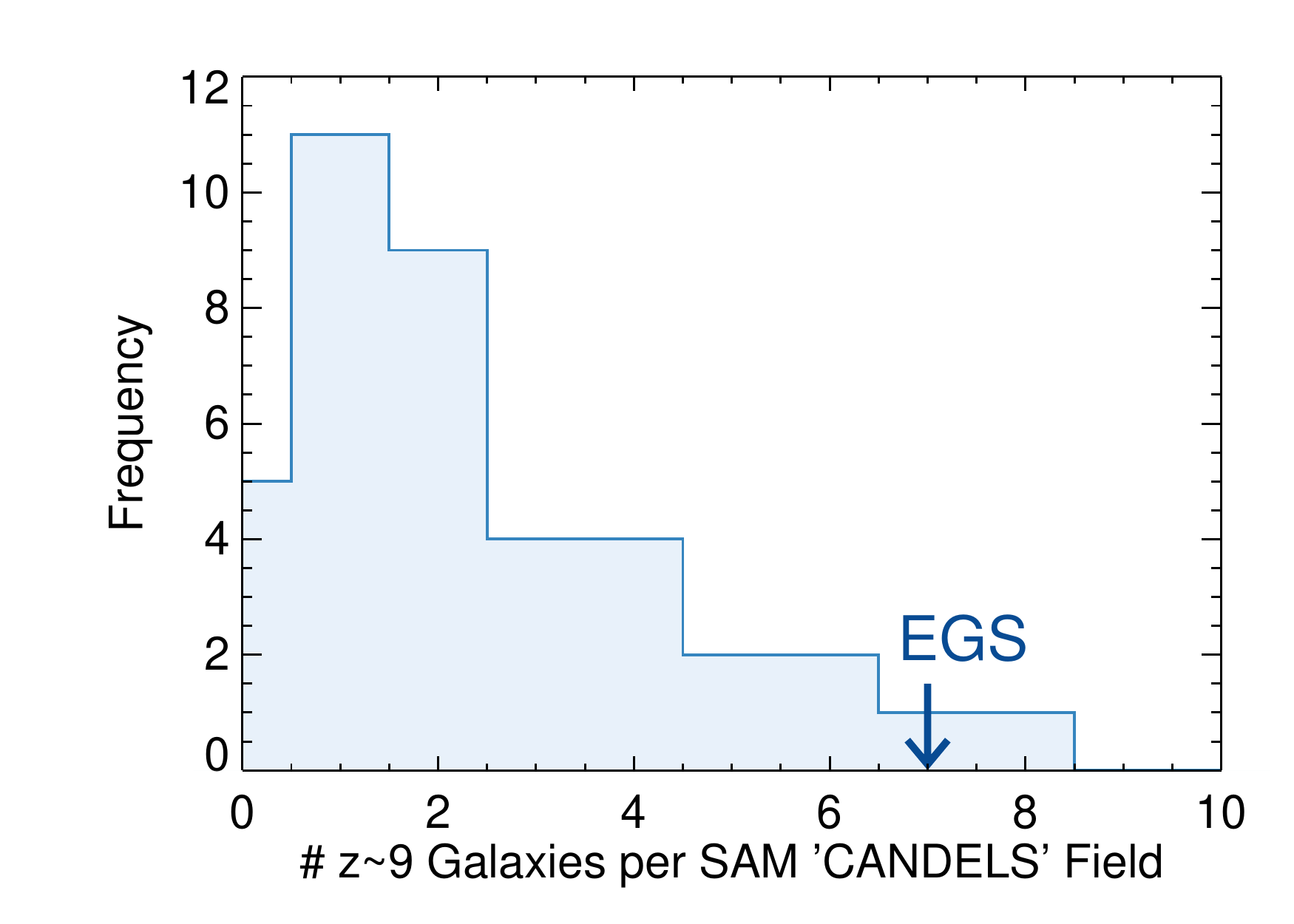}
\plotone{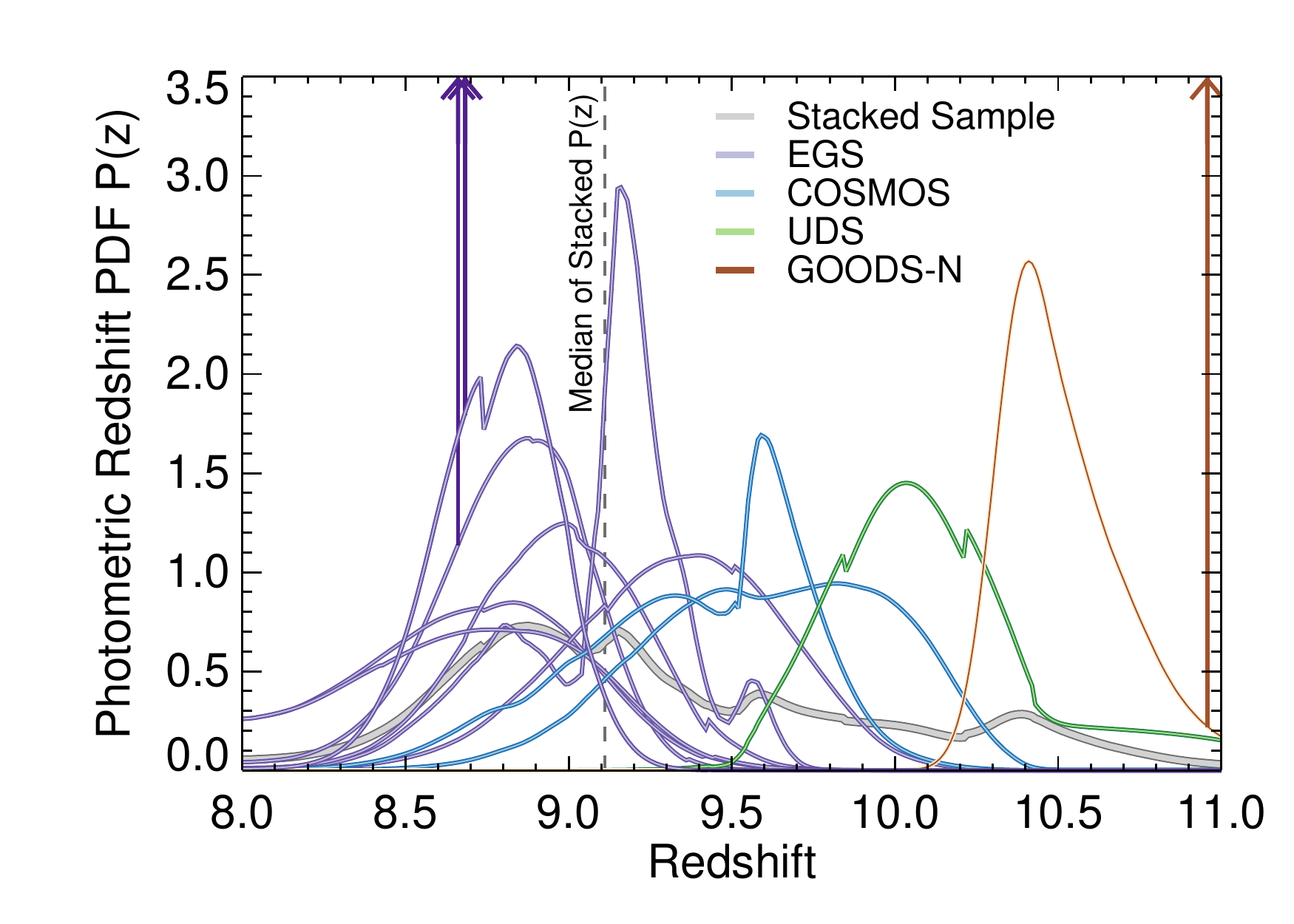}
\vspace{-2mm}
\caption{Left) The number of $z \approx$ 9 galaxies per 170 arcmin$^2$ CANDELS-sized
  field extracted from a 2 deg$^2$ SAM mock catalog.  Our observed
  value of seven such galaxies in the EGS field is more than 37 of the
  39 extracted fields, implying that we are observing a true overdense
  structure at 95\% confidence (encompassing both Poisson and cosmic
  variance uncertainties).  Right) The photometric redshift
  probability distribution functions from our 11 candidates, with the
  EGS, COSMOS, UDS and GOODS-N objects shown in purple, blue, green
  and brown, respectively.  We indicate the known spectroscopic
  redshift of EGS\_z910\_6811 ($z =$ 8.683), EGS\_z910\_44164 ($z =$
  8.665) and GOODSN\_z910\_35589 ($z =$ 10.957) with arrows.  Five of
  the seven EGS candidates have photometric redshift PDFs which peak
  near to this spectroscopic redshift (with the other two EGS
  candidates still consistent with this redshift).  This implies the
  large number of $z \sim$ 9 candidates in the EGS field are part of a
  single overdense structure.  The gray line in the background shows
  the normalized stacked PDF of the entire sample, highlighting the
  median redshift is 9.1.}
\label{fig:overdens}
\end{figure*}

Theoretical predictions in this epoch are rapidly maturing given the
imminent launch of {\it JWST}.  However, many of the above physical
properties require subgrid assumptions, thus by comparing these
predictions to our observations we can begin to bound the assumptions
in these models.  In the left-hand panel of Figure~\ref{fig:lf_theory}
we compare our observations and those from the literature to the
predicted luminosity functions from simulations.  This comparison
includes hydrodynamic simulations (BlueTIDES, \citealt{wilkins17};
IllustrisTNG, \citealt{vogelsberger20}; CROC, \citealt{gnedin16};
FIRE, \citealt{ma18}; FLARES, \citealt{vijayan20}), analytical
calculations \citep{mason15b}, empirical models (UniverseMachine,
\citealt{behroozi20}), and semi-analytic models (Santa Cruz
SAM, \citealt{yung19a}).  As our sample spans $z \sim$ 9--10 we plot
both epochs, though the $z =$ 9 lines are more relevant for comparison
as the median of the stacked $\powerset(z)$ for our entire sample is 9.1
(Figure~\ref{fig:overdens}).

Starting at our faintest observed end, we note that some models
over-predict our observed number densities at $z =$ 9.  These same
models at $z =$ 10 are fairly consistent with our observations, but as
our median redshift is closer to $z =$ 9 it does appear that our
observations imply a somewhat lower number density than some models.
This tension is most noticeable in the FIRE and analytic models
shown, though even those models are consistent within
$\sim$2$\sigma$ confidence.  There are more interesting differences
between the predictions and our observations in our brighter probed luminosities.
Here, the aforementioned three models are very consistent with our
data, but both the CROC and UniverseMachine models significantly
under-predict the observed number density.

This implies that some of the various physical assumptions (e.g., dust
attenuation, star-formation recipes) in these
models need revising to match observations, unsurprising as observations in this epoch are nascent with large
uncertainties.  Nonetheless, in the right panel of
Figure~\ref{fig:lf_theory} we show how the \citet{yung19a} SAM predictions
change if we alter two specific properties likely to significantly
change the bright end.  The red shaded region shows the fiducial
luminosity function from the SAM (where the shading denotes the
evolution from $z =$ 9 to 10), which does a
reasonable job at our faintest observed luminosities, though it does
under-predict our brightest observations.

In blue we show the SAM predictions when no dust attenuation is
applied, which matches the brightest luminosities we probed fairly well, though somewhat
over-predicts the fainter luminosities.  This implies that a modification to
the dust attenuation which results in a decreased attenuation for
brighter galaxies relative to fainter galaxies could improve the
agreement of these predictions with our observations.  However,
\citet{finkelstein12a} found that bright galaxies from $z =$ 4--8 have
seemingly uniform rest-UV colors implying the presence of some dust,
thus this scenario seems less likely.  The colors of our sample of
galaxies are further explored by \citet{tacchella21}, are are
consistent with a similar amount of dust attenuation as similarly
massive galaxies at $z =$ 4--8.  This implies that differences in dust
attenuation at the bright end are not a favored explanation by
observations for the lack of bright-end evolution; we refer the reader
to \citet{tacchella21} for further discussion.

\edit1{The fiducial model of \citet{yung19a} adopts a broken power law
relation between molecular gas surface density and star formation rate
density, with a slope of unity below a critical $H_2$ surface density
($70 M_\odot pc^{-2}$) and a slope of 2 above that critical
density. In green, we show the SAM predictions for an alternative
model in which a classical Kennicutt-Schmidt (KS) relation
\citep{schmidt59,kennicutt89} is adopted, where $\Sigma_{SFR} \propto
\Sigma_{\rm gas}^N$ with $N=1.5$ (where here $\Sigma_{gas}$ includes
both molecular and atomic gas). This model does equally well as the
fiducial one at reproducing the properties of lower redshift galaxies
\citep{somerville15b}, but one can see that the predictions for the
number density of high redshift galaxies are starkly different --- a
shallower SFR law index lowers the entire predicted luminosity
function, with the difference most notable at the bright end. Even
given the sizable observational uncertainties, and the modeling
uncertainties especially with regard to the amount of dust extinction
in these galaxies, our measured luminosity function is strongly
inconsistent with a SF law with a slope of $N=1.5$ assuming the
model's parameterization of dust attenuation (e.g., the  $N=1.5$ SF
law with no dust [dotted green lines] also does well to represent the
data, but as discussed above a dust-free nature is extremely unlikely,
especially for bright galaxies). We note that the
SF law adopted in the fiducial model of \citet{yung19a} is consistent
with the observationally derived SF law in nearby and intermediate
redshift ($z\sim$ 1--2) galaxies
\citep{narayanan12,kennicutt21}. However, uncertainties in the
conversion factor between CO and molecular gas and how it depends on
galaxy properties make it unclear whether the slope of the SF law
steepens at high gas density, or whether instead starburst galaxies
have a SF law with a higher normalization
\citep{kennicutt21}. Regardless, it is clear that these distant
galaxies have much shorter gas depletion times, implying higher SF
efficiencies, than nearby typical spiral galaxies. The star-formation
law in the early universe can be better characterized with {\it JWST},
which will independently constrain the dust attenuation in individual
galaxies with improved photometry (and, for the first time in this
epoch, spectroscopy).}

\section{EGS Overdensity at $z \sim$ 9}

One of the most striking properties of our sample is that although we
included five (roughly) equivalently sized fields, the distribution of
sources is very non-uniform, with seven of our 11 candidates coming
from a single field, the EGS field.  This was also observed by \citet{bouwens19}, who found that the relative
normalization of the luminosity function in the EGS-field alone was
$\sim$2.5 ($\pm$1.1) $\times$ higher than that derived from all five CANDELS
fields.  While our small sample is dominated by uncertainties (primarily Poisson, but also
cosmic variance), it is worth considering this potential overdensity
in more detail.

We first estimate the significance of this potential $z \sim$ 9 structure by drawing random
samples of galaxies from a 2 deg$^2$ mock catalog
\citep{yang21,somerville21} based on the Santa-Cruz
semi-analytic model of \citet{yung19a}.  We extracted the number of
galaxies in unit redshift bins from $z =$ 8.5--9.5 (encompassing the
likely redshifts of the seven EGS sources) down to
our magnitude limit of $H\!\!=$26.6 in non-overlapping CANDELS-sized
fields.  While the five CANDELS fields are all rectangular, the aspect
ratios range from 4.8:1 (EGS) to 1.6:1 (GOODS-N).  For this exercise
we chose to use a single field geometry with an area of 170 arcmin$^2$
and an aspect ratio of 3.1:1, which is the mean of the five fields.
We were able to extract 39 non-overlapping fields of this size.  The
distribution of $z \sim$ 9 galaxies per CANDELS field is shown in the
left panel of Figure~\ref{fig:overdens}.

We find a median expectation of 2 $\pm$ 2 $z \approx$ 9 galaxies per
CANDELS field, highlighting that our discovery of seven such galaxies
in the EGS field is an outlier.  We find that in only two of these
extracted SAM fields does the number of bright $z \sim$ 9 galaxies
match or exceed our detected value (one field with 7, and one with 8),
with 37/39 fields having $\leq$6 galaxies.  Thus, this simulation
implies that observing a field with 7 bright $z \sim$ 9 galaxies is a
95\% (or 2$\sigma$) outlier, implying that it is fairly likely that we
are witnessing a real structure in formation.

In the right panel of Figure~\ref{fig:overdens} we show the
photometric redshift distribution of galaxies in our sample,
highlighting those in the EGS field in purple.  While galaxies in our
full sample have photometric redshift PDFs which span $z \sim$ 8.5-11,
five of the seven EGS candidates have PDFs which peak at $z \sim$
8.5--9 (with the other two candidates still having significant
probability of a redshift in this range).  Interestingly, two of these
sources are EGS\_z910\_6811 which has a Ly$\alpha$-based
spectroscopic redshift of $z =$ 8.683, and EGS\_z910\_44164 with a
Ly$\alpha$-based redshift of $z =$ 8.665 (Larson et al. 2021).  Detecting Ly$\alpha$ from such distant sources is
suggestive that the source resides in a predominantly ionized bubble \citep[e.g.][]{castellano18,tilvi20},
which are predicted to exist around over-densities throughout the epoch
of reionization \citep[e.g.,][]{malhotra06,dijkstra14}.  This apparent overdensity could be powering such an
ionized bubble, which is considered further in Larson et al.\ (2021). If future spectroscopic followup finds that these sources are part of an
overdense structure at $z \sim$ 8.7, it explains why the EGS
candidates are clustered at similar $J-H$ colors in Figure 8.

\section{Conclusions}

We have presented the results of a systematic search through the five
CANDELS fields for galaxy candidates at $z >$ 8.5.  Due to the limited
depth of many of these fields, we limit our search to $H <$ 26.6,
which corresponds approximately to S/N $\sim$ 7 for the shallowest
regions of the CANDELS fields.  We created new photometric catalogs in
these fields, with an emphasis on color accuracy and minimizing
spurious sources, and a detailed empirical estimation of the
photometric noise.  Each {\it HST} source had its {\it Spitzer}/IRAC
photometry modeled with TPHOT, and we presented a discussion of the
utility of IRAC data for the selection of such distant sources.  We measured photometric redshifts for all sources
in these catalogs with EAZY, justifying the inclusion of additional
blue templates based on the measured colors of known $z >$ 6
galaxies.

We selected candidate $z >$ 8.5 galaxies with a customized set of
selection criteria.
We used source photometric redshift probability
distribution functions to select sources likely at $z >$ 8.5, sorting
them into unity redshift bins based on where they had the majority of
their redshift probability density.  This led to an initial sample
size of 140 candidates, which we thoroughly vetted in multiple ways.

As persistance from previous observations could leave a signature
similar to a $z >$ 8.5, we did a detailed screening for persistence,
examining the pixel positons of our candidates in observations taken
24 hours prior to all of our exposures, and removing sources with high
counts at these positions in previous images.  We did an additional
S/N check using noise measured in a local region to each source,
ensuring that we included no spurious sources due to
noisier-than-average regions.  We then did a comprehensive visual
inspection, exploring both artifacts in {\it HST} imaging, and also
removing candidates with poorly deblended IRAC photometry (unless the
photometric redshift from {\it HST}-only satisfied our selection
criteria).  Finally, we explored the potential for stellar
contamination fitting our candidates to low-mass star and brown dwarf
spectral models.  Our final sample after these screening steps was
composed of 14 candidate galaxies.

We further vetted these candidates by including additional {\it HST}
imaging in F098M and F140W, as well as including all available
ground-based photometry.  This step removed three sources.
COSMOS\_z910\_14822 was the brightest source in our sample, but a
combination of a weak detection in F098M and a red $H-K$ slope makes
$z <$ 2.5 more likely.  After the inclusion of ground-based imaging,
UDS\_z910\_731 was also removed as a likely low-redshift galaxy.
Finally, UDS\_z910\_7815 was removed following a 2$\sigma$
detection in the F098M image; while a high-redshift solution was still
preferred for this galaxy, the difference between the goodness-of-fit
for this solution and the best-fitting lower-redshift solution was
only $\Delta \chi^2 =$ 0.9, so this source was removed.  As one final
check, for our final sample of 11 sources, we remeasured the IRAC
photometry using {\sc Galfit} to deblend and subtract neighbors, finding
that in most cases objects were consistent with $z >$ 8.5 regardless
of which IRAC photometry was used.  The exception was
COSMOS\_z910\_20646, where the {\sc Galfit} photometry preferred a
low-redshift solution.  However given that it is unknown whether {\sc Galfit}
or TPHOT are closer to the truth, we kept this source in our sample,
noting this potential uncertainty on its redshift.

We presented a detailed comparison of our search to those in the
literature.  We find significant agreement amongst those studies which
use photometric redshifts \citep[e.g.,][]{bouwens19}, noting that color-selection only would
miss six of our 11 candidates, as shown in Figure~\ref{fig:colcol} \citep[e.g.,][]{bouwens16}.  Most published
sources not in our sample were simply fainter than our detection or
S/N limits, though broadly our photometric redshifts for these sources
are in agreement with those published.

We used the observed photometry to estimate the UV absolute magnitudes
for our sources, calculating line-of-sight lensing due to any
nearby neighbors in the imaging (finding no significant intermediate or strong lensing magnification).  We then
calculated the rest-frame UV luminosity function, using
simulation-based values of the effective volume to correct for
incompleteness.  We pay particular attention to the sizes of simulated
sources, ensuring that the recovered simulated sources match what
little information we have about true bright $z >$ 8.5 galaxies.  Due
to the small number of galaxies in our sample, we measure this
luminosity function in a single redshift bin from 8.5 $< z <$ 11.  We
developed a novel method for calculating the observed number
densities, minimizing any potential bias from choosing a specific set of
magnitude bins.

Comparing to
results in the literature over these same redshifts, we find general
agreement with previous results within the
large Poisson and cosmic variance uncertainties, though our brightest
measurements are at the higher end of some previous results.  Our observations
support a seeming lack of evolution in the bright end of the UV
luminosity function at $z > 8$, previously proposed by
\citet{bowler20}, where the decline in number density towards
brighter luminosities appears shallower than the typical Schechter
exponential decline.
We compared our observed luminosity function to several recent theoretical models,
finding that some under-predict the observed abundance in our
brightest bin, though we find that a combination of either more
efficient star formation in galaxies at this epoch and/or
reduced dust in bright galaxies in this epoch can explain our
observations.

Finally, we explored the distribution of our 11 candidates across the
five CANDELS fields, noting that 7/11 were in a single field -- the
EGS.  We used a SAM mock catalog to explore the expected distribution
of such bright $z >$ 8.5 galaxies in CANDELS-sized fields, and found
that our detection of seven in a single field is a 2$\sigma$ outlier
(encompassing both Poisson and cosmic variance uncertainties). This
implies that this overdensity is real, and we are witnessing the
formation of one of the earliest cosmic structures.  This is supported
by the detection of Ly$\alpha$ emission from two of these galaxies at $z
=$ 8.683 \citep{zitrin15} and $z =$ 8.665 (Larson et al. 2021),
which is consistent with the presence of an ionized bubble formed by
this overdensity deep in the epoch of reionization.

We caution the reader that in spite of our extreme vetting process,
8/11 galaxies still lack spectroscopic confirmation and remain
candidates, thus our conclusions should be weighed alongside the
confidence in these candidates.  \edit1{Likewise, as discussed in detail in
this paper, the selection of these sources with current telescopes is
highly complex, with significant effort required to obtain a pure
sample, which could lead to additional systematic uncertainties in the
luminosity function calculation.}
Our ability to study this early epoch
will soon receive a tremendous boost with the imminent launch of {\it
  JWST}.  While its larger aperture will allow deeper surveys, the key
advance is its ability to probe 1--5$\mu$m all at {\it HST} (or
better) resolution, removing the hurdle of uncertain IRAC deblending.
All five CANDELS fields will receive significant imaging from approved
Cycle 1 programs including JADES (PI Rieke), PRIMER (PI Dunlop),
COSMOS-Webb (PIs Kartaltepe and Casey) and
CEERS (PI Finkelstein).  Early in Cycle 1 the CEERS
program will observe the EGS field, probing
the potential $z \sim$ 9 overdensity by
spectroscopically observing many of the known sources, and
photometrically searching for fainter companions.  While we and others
have pushed {\it HST} to its limits to give us a first glimpse of the
$z >$ 9 universe, {\it JWST} observations will soon unlock the cosmic
secrets hidden in this era.

\acknowledgements

We acknowledge that the location where most of this work took place, the University of Texas
at Austin, that sits on indigenous land. The Tonkawa lived in central
Texas and the Comanche and Apache moved through this area. We pay our
respects to all the American Indian and Indigenous Peoples and
communities who have been or have become a part of these lands and
territories in Texas, on this piece of Turtle Island.  
We thank Adam Kraus, Karl Gebhardt, Stephen Wilkins, Rebecca Bowler, Charlotte Mason,
Jim Dunlop, and Gabe Brammer for helpful conversations.  We thank the
anonymous referee for their detailed and useful comments.
SLF and MB acknowledge support from NASA through ADAP award
80NSSC18K0954.  Support for {\it Hubble Space Telescope} programs \#15697 and \#15862
was provided by NASA through a grant from the Space Telescope Science
Institute, which is operated by the Associations of Universities for
Research in Astronomy, Incorporated, under NASA contract NAS5- 26555.
S.R.R. acknowledges financial support from the
International Max Planck Research School for Astronomy and Cosmic
Physics at the University of Heidelberg (IMPRS--HD).  This research has benefitted from the SpeX Prism Library (and/or SpeX
Prism Library Analysis Toolkit), maintained by Adam Burgasser at
http://www.browndwarfs.org/spexprism

\software{EAZY \citep{brammer08}, Source~Extractor (v2.19.5;
  \citep{bertin96}), T-PHOT \citep{merlin15,merlin16}, Galfit package
  \citep{peng02}, pypher \citep{boucaud16}}


\begin{thebibliography}{140}
\expandafter\ifx\csname natexlab\endcsname\relax\def\natexlab#1{#1}\fi

\bibitem[{{Ashby} {et~al.}(2015){Ashby}, {Willner}, {Fazio}, {Dunlop}, {Egami},
  {Faber}, {Ferguson}, {Grogin}, {Hora}, {Huang}, {Koekemoer}, {Labb{\'e}}, \&
  {Wang}}]{ashby15}
{Ashby}, M.~L.~N., {Willner}, S.~P., {Fazio}, G.~G., {et~al.} 2015, \apjs, 218,
  33

\bibitem[{{Atek} {et~al.}(2015){Atek}, {Richard}, {Jauzac}, {Kneib},
  {Natarajan}, {Limousin}, {Schaerer}, {Jullo}, {Ebeling}, {Egami}, \&
  {Clement}}]{atek15}
{Atek}, H., {Richard}, J., {Jauzac}, M., {et~al.} 2015, \apj, 814, 69

\bibitem[{{Barro} {et~al.}(2019){Barro}, {P{\'e}rez-Gonz{\'a}lez}, {Cava},
  {Brammer}, {Pandya}, {Eliche Moral}, {Esquej}, {Dom{\'\i}nguez-S{\'a}nchez},
  {Alcalde Pampliega}, {Guo}, {Koekemoer}, {Trump}, {Ashby}, {Cardiel},
  {Castellano}, {Conselice}, {Dickinson}, {Dolch}, {Donley}, {Espino Briones},
  {Faber}, {Fazio}, {Ferguson}, {Finkelstein}, {Fontana}, {Galametz},
  {Gardner}, {Gawiser}, {Giavalisco}, {Grazian}, {Grogin}, {Hathi}, {Hemmati},
  {Hern{\'a}n-Caballero}, {Kocevski}, {Koo}, {Kodra}, {Lee}, {Lin}, {Lucas},
  {Mobasher}, {McGrath}, {Nandra}, {Nayyeri}, {Newman}, {Pforr}, {Peth},
  {Rafelski}, {Rodr{\'\i}guez-Munoz}, {Salvato}, {Stefanon}, {van der Wel},
  {Willner}, {Wiklind}, \& {Wuyts}}]{barro19}
{Barro}, G., {P{\'e}rez-Gonz{\'a}lez}, P.~G., {Cava}, A., {et~al.} 2019, \apjs,
  243, 22

\bibitem[{{Becker} {et~al.}(2015){Becker}, {Bolton}, {Madau}, {Pettini},
  {Ryan-Weber}, \& {Venemans}}]{becker15}
{Becker}, G.~D., {Bolton}, J.~S., {Madau}, P., {et~al.} 2015, \mnras, 447, 3402

\bibitem[{{Becker} {et~al.}(2021){Becker}, {D'Aloisio}, {Christenson}, {Zhu},
  {Worseck}, \& {Bolton}}]{becker21}
{Becker}, G.~D., {D'Aloisio}, A., {Christenson}, H.~M., {et~al.} 2021, arXiv
  e-prints, arXiv:2103.16610

\bibitem[{{Beckwith} {et~al.}(2006){Beckwith}, {Stiavelli}, {Koekemoer},
  {Caldwell}, {Ferguson}, {Hook}, {Lucas}, {Bergeron}, {Corbin}, {Jogee},
  {Panagia}, {Robberto}, {Royle}, {Somerville}, \& {Sosey}}]{beckwith06}
{Beckwith}, S.~V.~W., {Stiavelli}, M., {Koekemoer}, A.~M., {et~al.} 2006, \aj,
  132, 1729

\bibitem[{{Behroozi} {et~al.}(2020){Behroozi}, {Conroy}, {Wechsler}, {Hearin},
  {Williams}, {Moster}, {Yung}, {Somerville}, {Gottl{\"o}ber}, {Yepes}, \&
  {Endsley}}]{behroozi20}
{Behroozi}, P., {Conroy}, C., {Wechsler}, R.~H., {et~al.} 2020, \mnras

\bibitem[{{Behroozi} \& {Silk}(2015)}]{behroozi15}
{Behroozi}, P.~S., \& {Silk}, J. 2015, \apj, 799, 32

\bibitem[{{Bernard} {et~al.}(2016){Bernard}, {Carrasco}, {Trenti}, {Oesch},
  {Wu}, {Bradley}, {Schmidt}, {Bouwens}, {Calvi}, {Mason}, {Stiavelli}, \&
  {Treu}}]{bernard16}
{Bernard}, S.~R., {Carrasco}, D., {Trenti}, M., {et~al.} 2016, \apj, 827, 76

\bibitem[{{Bertin} \& {Arnouts}(1996)}]{bertin96}
{Bertin}, E., \& {Arnouts}, S. 1996, \aaps, 117, 393

\bibitem[{{Bhowmick} {et~al.}(2020){Bhowmick}, {Somerville}, {Di Matteo},
  {Wilkins}, {Feng}, \& {Tenneti}}]{bhowmick20}
{Bhowmick}, A.~K., {Somerville}, R.~S., {Di Matteo}, T., {et~al.} 2020, \mnras,
  496, 754

\bibitem[{{Boucaud} {et~al.}(2016){Boucaud}, {Bocchio}, {Abergel}, {Orieux},
  {Dole}, \& {Amine Hadj-Youcef}}]{boucaud16}
{Boucaud}, A., {Bocchio}, M., {Abergel}, A., {et~al.} 2016, {PyPHER:
  Python-based PSF Homogenization kERnels}

\bibitem[{{Bouwens} {et~al.}(2007){Bouwens}, {Illingworth}, {Franx}, \&
  {Ford}}]{bouwens07}
{Bouwens}, R.~J., {Illingworth}, G.~D., {Franx}, M., \& {Ford}, H. 2007, \apj,
  670, 928

\bibitem[{{Bouwens} {et~al.}(2015{\natexlab{a}}){Bouwens}, {Illingworth},
  {Oesch}, {Caruana}, {Holwerda}, {Smit}, \& {Wilkins}}]{bouwens15b}
{Bouwens}, R.~J., {Illingworth}, G.~D., {Oesch}, P.~A., {et~al.}
  2015{\natexlab{a}}, \apj, 811, 140

\bibitem[{{Bouwens} {et~al.}(2019){Bouwens}, {Stefanon}, {Oesch},
  {Illingworth}, {Nanayakkara}, {Roberts-Borsani}, {Labbe'}, \&
  {Smit}}]{bouwens19}
{Bouwens}, R.~J., {Stefanon}, M., {Oesch}, P.~A., {et~al.} 2019, arXiv
  e-prints, arXiv:1905.05202

\bibitem[{{Bouwens} {et~al.}(2014){Bouwens}, {Illingworth}, {Oesch},
  {Labb{\'e}}, {van Dokkum}, {Trenti}, {Franx}, {Smit}, {Gonzalez}, \&
  {Magee}}]{bouwens14}
{Bouwens}, R.~J., {Illingworth}, G.~D., {Oesch}, P.~A., {et~al.} 2014, \apj,
  793, 115

\bibitem[{{Bouwens} {et~al.}(2015{\natexlab{b}}){Bouwens}, {Illingworth},
  {Oesch}, {Trenti}, {Labb{\'e}}, {Bradley}, {Carollo}, {van Dokkum},
  {Gonzalez}, {Holwerda}, {Franx}, {Spitler}, {Smit}, \& {Magee}}]{bouwens15}
---. 2015{\natexlab{b}}, \apj, 803, 34

\bibitem[{{Bouwens} {et~al.}(2016{\natexlab{a}}){Bouwens}, {Oesch},
  {Labb{\'e}}, {Illingworth}, {Fazio}, {Coe}, {Holwerda}, {Smit}, {Stefanon},
  {van Dokkum}, {Trenti}, {Ashby}, {Huang}, {Spitler}, {Straatman}, {Bradley},
  \& {Magee}}]{bouwens16}
{Bouwens}, R.~J., {Oesch}, P.~A., {Labb{\'e}}, I., {et~al.} 2016{\natexlab{a}},
  \apj, 830, 67

\bibitem[{{Bouwens} {et~al.}(2016{\natexlab{b}}){Bouwens}, {Oesch},
  {Labb{\'e}}, {Illingworth}, {Fazio}, {Coe}, {Holwerda}, {Smit}, {Stefanon},
  {van Dokkum}, {Trenti}, {Ashby}, {Huang}, {Spitler}, {Straatman}, {Bradley},
  \& {Magee}}]{bouwens15c}
---. 2016{\natexlab{b}}, \apj, 830, 67

\bibitem[{{Bouwens} {et~al.}(2021){Bouwens}, {Oesch}, {Stefanon},
  {Illingworth}, {Labbe}, {Reddy}, {Atek}, {Montes}, {Naidu}, {Nanayakkara},
  {Nelson}, \& {Wilkins}}]{bouwens21}
{Bouwens}, R.~J., {Oesch}, P.~A., {Stefanon}, M., {et~al.} 2021, arXiv
  e-prints, arXiv:2102.07775

\bibitem[{{Bowler} {et~al.}(2020){Bowler}, {Jarvis}, {Dunlop}, {McLure},
  {McLeod}, {Adams}, {Milvang-Jensen}, \& {McCracken}}]{bowler20}
{Bowler}, R.~A.~A., {Jarvis}, M.~J., {Dunlop}, J.~S., {et~al.} 2020, \mnras,
  493, 2059

\bibitem[{{Bowler} {et~al.}(2012){Bowler}, {Dunlop}, {McLure}, {McCracken},
  {Milvang-Jensen}, {Furusawa}, {Fynbo}, {Le F{\`e}vre}, {Holt}, {Ideue},
  {Ihara}, {Rogers}, \& {Taniguchi}}]{bowler12}
{Bowler}, R.~A.~A., {Dunlop}, J.~S., {McLure}, R.~J., {et~al.} 2012, \mnras,
  426, 2772

\bibitem[{{Bowler} {et~al.}(2014){Bowler}, {Dunlop}, {McLure}, {Rogers},
  {McCracken}, {Milvang-Jensen}, {Furusawa}, {Fynbo}, {Taniguchi}, {Afonso},
  {Bremer}, \& {Le F{\`e}vre}}]{bowler14}
---. 2014, \mnras, 440, 2810

\bibitem[{{Bowler} {et~al.}(2015){Bowler}, {Dunlop}, {McLure}, {McCracken},
  {Furusawa}, {Taniguchi}, {Fynbo}, {Milvang-Jensen}, \& {Le Fevre}}]{bowler15}
---. 2015, ArXiv e-prints

\bibitem[{{Brammer} {et~al.}(2008){Brammer}, {van Dokkum}, \&
  {Coppi}}]{brammer08}
{Brammer}, G.~B., {van Dokkum}, P.~G., \& {Coppi}, P. 2008, \apj, 686, 1503

\bibitem[{{Bridge} {et~al.}(2019){Bridge}, {Holwerda}, {Stefanon}, {Bouwens},
  {Oesch}, {Trenti}, {Bernard}, {Bradley}, {Illingworth}, {Kusmic}, {Magee},
  {Morishita}, {Roberts-Borsani}, {Smit}, \& {Steele}}]{bridge19}
{Bridge}, J.~S., {Holwerda}, B.~W., {Stefanon}, M., {et~al.} 2019, \apj, 882,
  42

\bibitem[{{Bruzual} \& {Charlot}(2003)}]{bruzual03}
{Bruzual}, G., \& {Charlot}, S. 2003, \mnras, 344, 1000

\bibitem[{{Burgasser}(2014)}]{burgasser14}
{Burgasser}, A.~J. 2014, in Astronomical Society of India Conference Series,
  Vol.~11, Astronomical Society of India Conference Series, 7--16

\bibitem[{{Calvi} {et~al.}(2016){Calvi}, {Trenti}, {Stiavelli}, {Oesch},
  {Bradley}, {Schmidt}, {Coe}, {Brammer}, {Bernard}, {Bouwens}, {Carrasco},
  {Carollo}, {Holwerda}, {MacKenty}, {Mason}, {Shull}, \& {Treu}}]{calvi16}
{Calvi}, V., {Trenti}, M., {Stiavelli}, M., {et~al.} 2016, \apj, 817, 120

\bibitem[{{Cardelli} {et~al.}(1989){Cardelli}, {Clayton}, \&
  {Mathis}}]{cardelli89}
{Cardelli}, J.~A., {Clayton}, G.~C., \& {Mathis}, J.~S. 1989, \apj, 345, 245

\bibitem[{{Casey} {et~al.}(2014){Casey}, {Narayanan}, \& {Cooray}}]{casey14}
{Casey}, C.~M., {Narayanan}, D., \& {Cooray}, A. 2014, \physrep, 541, 45

\bibitem[{{Castellano} {et~al.}(2018){Castellano}, {Pentericci}, {Vanzella},
  {Marchi}, {Fontana}, {Dayal}, {Ferrara}, {Hutter}, {Carniani}, {Cristiani},
  {Dickinson}, {Gallerani}, {Giallongo}, {Giavalisco}, {Grazian}, {Maiolino},
  {Merlin}, {Paris}, {Pilo}, \& {Santini}}]{castellano18}
{Castellano}, M., {Pentericci}, L., {Vanzella}, E., {et~al.} 2018, \apjl, 863,
  L3

\bibitem[{{Conroy} \& {Gunn}(2010)}]{conroy10}
{Conroy}, C., \& {Gunn}, J.~E. 2010, {FSPS: Flexible Stellar Population
  Synthesis}

\bibitem[{{DeSantis} {et~al.}(2017){DeSantis}, {Thurman}, {Hix}, \&
  {Ogden}}]{desantis07}
{DeSantis}, Z.~J., {Thurman}, S.~T., {Hix}, T.~T., \& {Ogden}, C.~E. 2017, in
  Advanced Maui Optical and Space Surveillance (AMOS) Technologies Conference,
  ed. S.~{Ryan}, 26

\bibitem[{{Dijkstra}(2014)}]{dijkstra14}
{Dijkstra}, M. 2014, PASA, 31, 40

\bibitem[{{Ellis} {et~al.}(2013){Ellis}, {McLure}, {Dunlop}, {Robertson},
  {Ono}, {Schenker}, {Koekemoer}, {Bowler}, {Ouchi}, {Rogers}, {Curtis-Lake},
  {Schneider}, {Charlot}, {Stark}, {Furlanetto}, \& {Cirasuolo}}]{ellis13}
{Ellis}, R.~S., {McLure}, R.~J., {Dunlop}, J.~S., {et~al.} 2013, \apjl, 763, L7

\bibitem[{{Elvis} {et~al.}(2009){Elvis}, {Civano}, {Vignali}, {Puccetti},
  {Fiore}, {Cappelluti}, {Aldcroft}, {Fruscione}, {Zamorani}, {Comastri},
  {Brusa}, {Gilli}, {Miyaji}, {Damiani}, {Koekemoer}, {Finoguenov}, {Brunner},
  {Urry}, {Silverman}, {Mainieri}, {Hasinger}, {Griffiths}, {Carollo}, {Hao},
  {Guzzo}, {Blain}, {Calzetti}, {Carilli}, {Capak}, {Ettori}, {Fabbiano},
  {Impey}, {Lilly}, {Mobasher}, {Rich}, {Salvato}, {Sanders}, {Schinnerer},
  {Scoville}, {Shopbell}, {Taylor}, {Taniguchi}, \& {Volonteri}}]{elvis09}
{Elvis}, M., {Civano}, F., {Vignali}, C., {et~al.} 2009, \apjs, 184, 158

\bibitem[{{Erb} {et~al.}(2010){Erb}, {Pettini}, {Shapley}, {Steidel}, {Law}, \&
  {Reddy}}]{erb10}
{Erb}, D.~K., {Pettini}, M., {Shapley}, A.~E., {et~al.} 2010, \apj, 719, 1168

\bibitem[{{Fan} {et~al.}(2006){Fan}, {Strauss}, {Becker}, {White}, {Gunn},
  {Knapp}, {Richards}, {Schneider}, {Brinkmann}, \& {Fukugita}}]{fan06}
{Fan}, X., {Strauss}, M.~A., {Becker}, R.~H., {et~al.} 2006, \aj, 132, 117

\bibitem[{{Finkelstein} {et~al.}(2015{\natexlab{a}}){Finkelstein},
  {Finkelstein}, {Tilvi}, {Malhotra}, {Rhoads}, {Grogin}, {Pirzkal}, {Dey},
  {Jannuzi}, {Mobasher}, {Pakzad}, {Salmon}, \& {Wang}}]{kfinkelstein15}
{Finkelstein}, K.~D., {Finkelstein}, S.~L., {Tilvi}, V., {et~al.}
  2015{\natexlab{a}}, \apj, 813, 78

\bibitem[{{Finkelstein} {et~al.}(2017){Finkelstein}, {Dickinson}, {Ferguson},
  {Grazian}, {Grogin}, {Kartaltepe}, {Kewley}, {Kocevski}, {Koekemoer}, {Lotz},
  {Papovich}, {Pentericci}, {Perez-Gonzalez}, {Pirzkal}, {Ravindranath},
  {Somerville}, {Trump}, \& {Wilkins}}]{finkelstein17}
{Finkelstein}, S., {Dickinson}, M., {Ferguson}, H., {et~al.} 2017, {The Cosmic
  Evolution Early Release Science (CEERS) Survey}, JWST Proposal ID 1345. Cycle
  0 Early Release Scienc

\bibitem[{{Finkelstein}(2016)}]{finkelstein16}
{Finkelstein}, S.~L. 2016, \pasa, 33, e037

\bibitem[{{Finkelstein} {et~al.}(2010){Finkelstein}, {Papovich}, {Giavalisco},
  {Reddy}, {Ferguson}, {Koekemoer}, \& {Dickinson}}]{finkelstein10}
{Finkelstein}, S.~L., {Papovich}, C., {Giavalisco}, M., {et~al.} 2010, \apj,
  719, 1250

\bibitem[{{Finkelstein} {et~al.}(2012{\natexlab{a}}){Finkelstein}, {Papovich},
  {Ryan}, {Pawlik}, {Dickinson}, {Ferguson}, {Finlator}, {Koekemoer},
  {Giavalisco}, {Cooray}, {Dunlop}, {Faber}, {Grogin}, {Kocevski}, \&
  {Newman}}]{finkelstein12b}
{Finkelstein}, S.~L., {Papovich}, C., {Ryan}, R.~E., {et~al.}
  2012{\natexlab{a}}, \apj, 758, 93

\bibitem[{{Finkelstein} {et~al.}(2012{\natexlab{b}}){Finkelstein}, {Papovich},
  {Salmon}, {Finlator}, {Dickinson}, {Ferguson}, {Giavalisco}, {Koekemoer},
  {Reddy}, {Bassett}, {Conselice}, {Dunlop}, {Faber}, {Grogin}, {Hathi},
  {Kocevski}, {Lai}, {Lee}, {McLure}, {Mobasher}, \& {Newman}}]{finkelstein12a}
{Finkelstein}, S.~L., {Papovich}, C., {Salmon}, B., {et~al.}
  2012{\natexlab{b}}, \apj, 756, 164

\bibitem[{{Finkelstein} {et~al.}(2015{\natexlab{b}}){Finkelstein}, {Song},
  {Behroozi}, {Somerville}, {Papovich}, {Milosavljevi{\'c}}, {Dekel},
  {Narayanan}, {Ashby}, {Cooray}, {Fazio}, {Ferguson}, {Koekemoer}, {Salmon},
  \& {Willner}}]{finkelstein15b}
{Finkelstein}, S.~L., {Song}, M., {Behroozi}, P., {et~al.} 2015{\natexlab{b}},
  \apj, 814, 95

\bibitem[{{Finkelstein} {et~al.}(2015{\natexlab{c}}){Finkelstein}, {Ryan},
  {Papovich}, {Dickinson}, {Song}, {Somerville}, {Ferguson}, {Salmon},
  {Giavalisco}, {Koekemoer}, {Ashby}, {Behroozi}, {Castellano}, {Dunlop},
  {Faber}, {Fazio}, {Fontana}, {Grogin}, {Hathi}, {Jaacks}, {Kocevski},
  {Livermore}, {McLure}, {Merlin}, {Mobasher}, {Newman}, {Rafelski}, {Tilvi},
  \& {Willner}}]{finkelstein15}
{Finkelstein}, S.~L., {Ryan}, Jr., R.~E., {Papovich}, C., {et~al.}
  2015{\natexlab{c}}, \apj, 810, 71

\bibitem[{{Finkelstein} {et~al.}(2019){Finkelstein}, {D'Aloisio},
  {Paardekooper}, {Ryan}, {Behroozi}, {Finlator}, {Livermore}, {Upton
  Sanderbeck}, {Dalla Vecchia}, \& {Khochfar}}]{finkelstein19}
{Finkelstein}, S.~L., {D'Aloisio}, A., {Paardekooper}, J.-P., {et~al.} 2019,
  \apj, 879, 36

\bibitem[{{Fontana} {et~al.}(2014){Fontana}, {Dunlop}, {Paris}, {Targett},
  {Boutsia}, {Castellano}, {Galametz}, {Grazian}, {McLure}, {Merlin},
  {Pentericci}, {Wuyts}, {Almaini}, {Caputi}, {Chary}, {Cirasuolo},
  {Conselice}, {Cooray}, {Daddi}, {Dickinson}, {Faber}, {Fazio}, {Ferguson},
  {Giallongo}, {Giavalisco}, {Grogin}, {Hathi}, {Koekemoer}, {Koo}, {Lucas},
  {Nonino}, {Rix}, {Renzini}, {Rosario}, {Santini}, {Scarlata}, {Sommariva},
  {Stark}, {van der Wel}, {Vanzella}, {Wild}, {Yan}, \& {Zibetti}}]{fontana14}
{Fontana}, A., {Dunlop}, J.~S., {Paris}, D., {et~al.} 2014, \aap, 570, A11

\bibitem[{{Galametz} {et~al.}(2013){Galametz}, {Grazian}, {Fontana},
  {Ferguson}, {Ashby}, {Barro}, {Castellano}, {Dahlen}, {Donley}, {Faber},
  {Grogin}, {Guo}, {Huang}, {Kocevski}, {Koekemoer}, {Lee}, {McGrath}, {Peth},
  {Willner}, {Almaini}, {Cooper}, {Cooray}, {Conselice}, {Dickinson}, {Dunlop},
  {Fazio}, {Foucaud}, {Gardner}, {Giavalisco}, {Hathi}, {Hartley}, {Koo},
  {Lai}, {de Mello}, {McLure}, {Lucas}, {Paris}, {Pentericci}, {Santini},
  {Simpson}, {Sommariva}, {Targett}, {Weiner}, {Wuyts}, \& {the CANDELS
  Team}}]{galametz13}
{Galametz}, A., {Grazian}, A., {Fontana}, A., {et~al.} 2013, \apjs, 206, 10

\bibitem[{{Giavalisco} {et~al.}(2004{\natexlab{a}}){Giavalisco}, {Ferguson},
  {Koekemoer}, {Dickinson}, {Alexander}, {Bauer}, {Bergeron}, {Biagetti},
  {Brandt}, {Casertano}, {Cesarsky}, {Chatzichristou}, {Conselice},
  {Cristiani}, {Da Costa}, {Dahlen}, {de Mello}, {Eisenhardt}, {Erben}, {Fall},
  {Fassnacht}, {Fosbury}, {Fruchter}, {Gardner}, {Grogin}, {Hook},
  {Hornschemeier}, {Idzi}, {Jogee}, {Kretchmer}, {Laidler}, {Lee}, {Livio},
  {Lucas}, {Madau}, {Mobasher}, {Moustakas}, {Nonino}, {Padovani}, {Papovich},
  {Park}, {Ravindranath}, {Renzini}, {Richardson}, {Riess}, {Rosati},
  {Schirmer}, {Schreier}, {Somerville}, {Spinrad}, {Stern}, {Stiavelli},
  {Strolger}, {Urry}, {Vandame}, {Williams}, \& {Wolf}}]{giavalisco04b}
{Giavalisco}, M., {Ferguson}, H.~C., {Koekemoer}, A.~M., {et~al.}
  2004{\natexlab{a}}, \apjl, 600, L93

\bibitem[{{Giavalisco} {et~al.}(2004{\natexlab{b}}){Giavalisco}, {Dickinson},
  {Ferguson}, {Ravindranath}, {Kretchmer}, {Moustakas}, {Madau}, {Fall},
  {Gardner}, {Livio}, {Papovich}, {Renzini}, {Spinrad}, {Stern}, \&
  {Riess}}]{giavalisco04}
{Giavalisco}, M., {Dickinson}, M., {Ferguson}, H.~C., {et~al.}
  2004{\natexlab{b}}, \apjl, 600, L103

\bibitem[{{Gnedin}(2016)}]{gnedin16}
{Gnedin}, N.~Y. 2016, \apjl, 825, L17

\bibitem[{{Graham} {et~al.}(2005){Graham}, {Driver}, {Petrosian}, {Conselice},
  {Bershady}, {Crawford}, \& {Goto}}]{graham05}
{Graham}, A.~W., {Driver}, S.~P., {Petrosian}, V., {et~al.} 2005, \aj, 130,
  1535

\bibitem[{{Grogin} {et~al.}(2011){Grogin}, {Kocevski}, {Faber}, {Ferguson},
  {Koekemoer}, {Riess}, {Acquaviva}, {Alexander}, {Almaini}, {Ashby}, {Barden},
  {Bell}, {Bournaud}, {Brown}, {Caputi}, {Casertano}, {Cassata}, {Castellano},
  {Challis}, {Chary}, {Cheung}, {Cirasuolo}, {Conselice}, {Roshan Cooray},
  {Croton}, {Daddi}, {Dahlen}, {Dav{\'e}}, {de Mello}, {Dekel}, {Dickinson},
  {Dolch}, {Donley}, {Dunlop}, {Dutton}, {Elbaz}, {Fazio}, {Filippenko},
  {Finkelstein}, {Fontana}, {Gardner}, {Garnavich}, {Gawiser}, {Giavalisco},
  {Grazian}, {Guo}, {Hathi}, {H{\"a}ussler}, {Hopkins}, {Huang}, {Huang},
  {Jha}, {Kartaltepe}, {Kirshner}, {Koo}, {Lai}, {Lee}, {Li}, {Lotz}, {Lucas},
  {Madau}, {McCarthy}, {McGrath}, {McIntosh}, {McLure}, {Mobasher},
  {Moustakas}, {Mozena}, {Nandra}, {Newman}, {Niemi}, {Noeske}, {Papovich},
  {Pentericci}, {Pope}, {Primack}, {Rajan}, {Ravindranath}, {Reddy}, {Renzini},
  {Rix}, {Robaina}, {Rodney}, {Rosario}, {Rosati}, {Salimbeni}, {Scarlata},
  {Siana}, {Simard}, {Smidt}, {Somerville}, {Spinrad}, {Straughn}, {Strolger},
  {Telford}, {Teplitz}, {Trump}, {van der Wel}, {Villforth}, {Wechsler},
  {Weiner}, {Wiklind}, {Wild}, {Wilson}, {Wuyts}, {Yan}, \& {Yun}}]{grogin11}
{Grogin}, N.~A., {Kocevski}, D.~D., {Faber}, S.~M., {et~al.} 2011, \apjs, 197,
  35

\bibitem[{{Guo} {et~al.}(2013){Guo}, {Ferguson}, {Giavalisco}, {Barro},
  {Willner}, {Ashby}, {Dahlen}, {Donley}, {Faber}, {Fontana}, {Galametz},
  {Grazian}, {Huang}, {Kocevski}, {Koekemoer}, {Koo}, {McGrath}, {Peth},
  {Salvato}, {Wuyts}, {Castellano}, {Cooray}, {Dickinson}, {Dunlop}, {Fazio},
  {Gardner}, {Gawiser}, {Grogin}, {Hathi}, {Hsu}, {Lee}, {Lucas}, {Mobasher},
  {Nandra}, {Newman}, \& {van der Wel}}]{guo13}
{Guo}, Y., {Ferguson}, H.~C., {Giavalisco}, M., {et~al.} 2013, \apjs, 207, 24

\bibitem[{{Ishigaki} {et~al.}(2018){Ishigaki}, {Kawamata}, {Ouchi}, {Oguri},
  {Shimasaku}, \& {Ono}}]{ishigaki18}
{Ishigaki}, M., {Kawamata}, R., {Ouchi}, M., {et~al.} 2018, \apj, 854, 73

\bibitem[{{Jiang} {et~al.}(2021){Jiang}, {Kashikawa}, {Wang}, {Walth}, {Ho},
  {Cai}, {Egami}, {Fan}, {Ito}, {Liang}, {Schaerer}, \& {Stark}}]{jiang21}
{Jiang}, L., {Kashikawa}, N., {Wang}, S., {et~al.} 2021, Nature Astronomy, 5,
  256

\bibitem[{{Kawamata} {et~al.}(2018){Kawamata}, {Ishigaki}, {Shimasaku},
  {Oguri}, {Ouchi}, \& {Tanigawa}}]{kawamata18}
{Kawamata}, R., {Ishigaki}, M., {Shimasaku}, K., {et~al.} 2018, \apj, 855, 4

\bibitem[{{Kennicutt}(1989)}]{kennicutt89}
{Kennicutt}, Robert~C., J. 1989, \apj, 344, 685

\bibitem[{{Kennicutt} \& {De Los Reyes}(2021)}]{kennicutt21}
{Kennicutt}, Robert~C., J., \& {De Los Reyes}, M. A.~C. 2021, \apj, 908, 61

\bibitem[{{Kocevski} {et~al.}(2018){Kocevski}, {Hasinger}, {Brightman},
  {Nandra}, {Georgakakis}, {Cappelluti}, {Civano}, {Li}, {Li}, {Aird},
  {Alexander}, {Almaini}, {Brusa}, {Buchner}, {Comastri}, {Conselice},
  {Dickinson}, {Finoguenov}, {Gilli}, {Koekemoer}, {Miyaji}, {Mullaney},
  {Papovich}, {Rosario}, {Salvato}, {Silverman}, {Somerville}, \&
  {Ueda}}]{kocevski18}
{Kocevski}, D.~D., {Hasinger}, G., {Brightman}, M., {et~al.} 2018, \apjs, 236,
  48

\bibitem[{{Koekemoer} {et~al.}(2011){Koekemoer}, {Faber}, {Ferguson}, {Grogin},
  {Kocevski}, {Koo}, {Lai}, {Lotz}, {Lucas}, {McGrath}, {Ogaz}, {Rajan},
  {Riess}, {Rodney}, {Strolger}, {Casertano}, {Castellano}, {Dahlen},
  {Dickinson}, {Dolch}, {Fontana}, {Giavalisco}, {Grazian}, {Guo}, {Hathi},
  {Huang}, {van der Wel}, {Yan}, {Acquaviva}, {Alexander}, {Almaini}, {Ashby},
  {Barden}, {Bell}, {Bournaud}, {Brown}, {Caputi}, {Cassata}, {Challis},
  {Chary}, {Cheung}, {Cirasuolo}, {Conselice}, {Roshan Cooray}, {Croton},
  {Daddi}, {Dav{\'e}}, {de Mello}, {de Ravel}, {Dekel}, {Donley}, {Dunlop},
  {Dutton}, {Elbaz}, {Fazio}, {Filippenko}, {Finkelstein}, {Frazer}, {Gardner},
  {Garnavich}, {Gawiser}, {Gruetzbauch}, {Hartley}, {H{\"a}ussler},
  {Herrington}, {Hopkins}, {Huang}, {Jha}, {Johnson}, {Kartaltepe},
  {Khostovan}, {Kirshner}, {Lani}, {Lee}, {Li}, {Madau}, {McCarthy},
  {McIntosh}, {McLure}, {McPartland}, {Mobasher}, {Moreira}, {Mortlock},
  {Moustakas}, {Mozena}, {Nandra}, {Newman}, {Nielsen}, {Niemi}, {Noeske},
  {Papovich}, {Pentericci}, {Pope}, {Primack}, {Ravindranath}, {Reddy},
  {Renzini}, {Rix}, {Robaina}, {Rosario}, {Rosati}, {Salimbeni}, {Scarlata},
  {Siana}, {Simard}, {Smidt}, {Snyder}, {Somerville}, {Spinrad}, {Straughn},
  {Telford}, {Teplitz}, {Trump}, {Vargas}, {Villforth}, {Wagner}, {Wandro},
  {Wechsler}, {Weiner}, {Wiklind}, {Wild}, {Wilson}, {Wuyts}, \&
  {Yun}}]{koekemoer11}
{Koekemoer}, A.~M., {Faber}, S.~M., {Ferguson}, H.~C., {et~al.} 2011, \apjs,
  197, 36

\bibitem[{{Kulkarni} {et~al.}(2018){Kulkarni}, {Keating}, {Haehnelt}, {Bosman},
  {Puchwein}, {Chardin}, \& {Aubert}}]{kulkarni18}
{Kulkarni}, G., {Keating}, L.~C., {Haehnelt}, M.~G., {et~al.} 2018, arXiv
  e-prints

\bibitem[{{Labb{\'e}} {et~al.}(2007){Labb{\'e}}, {Franx}, {Rudnick},
  {Schreiber}, {van Dokkum}, {Moorwood}, {Rix}, {R{\"o}ttgering}, {Trujillo},
  \& {van der Werf}}]{labbe03}
{Labb{\'e}}, I., {Franx}, M., {Rudnick}, G., {et~al.} 2007, \apj, 665, 944

\bibitem[{{Laporte} {et~al.}(2021){Laporte}, {Meyer}, {Ellis}, {Robertson},
  {Chisholm}, \& {Roberts-Borsani}}]{laporte21}
{Laporte}, N., {Meyer}, R.~A., {Ellis}, R.~S., {et~al.} 2021, arXiv e-prints,
  arXiv:2104.08168

\bibitem[{{Larson} {et~al.}(2021){Larson}, {Finkelstein}, {Finkelstein},
  {Finkelstein}, {Finkelstein}, \& {Finkelstein}}]{larson21}
{Larson}, R., {Finkelstein}, S.~L., {Hutchison}, T., {et~al.} 2021, ApJ
  Submitted


\bibitem[{{Lawrence} {et~al.}(2007){Lawrence}, {Warren}, {Almaini}, {Edge},
  {Hambly}, {Jameson}, {Lucas}, {Casali}, {Adamson}, {Dye}, {Emerson},
  {Foucaud}, {Hewett}, {Hirst}, {Hodgkin}, {Irwin}, {Lodieu}, {McMahon},
  {Simpson}, {Smail}, {Mortlock}, \& {Folger}}]{lawrence07}
{Lawrence}, A., {Warren}, S.~J., {Almaini}, O., {et~al.} 2007, \mnras, 379,
  1599

\bibitem[{{Livermore} {et~al.}(2017){Livermore}, {Finkelstein}, \&
  {Lotz}}]{livermore17}
{Livermore}, R.~C., {Finkelstein}, S.~L., \& {Lotz}, J.~M. 2017, \apj, 835, 113

\bibitem[{{Lotz} {et~al.}(2017){Lotz}, {Koekemoer}, {Coe}, {Grogin}, {Capak},
  {Mack}, {Anderson}, {Avila}, {Barker}, {Borncamp}, {Brammer}, {Durbin},
  {Gunning}, {Hilbert}, {Jenkner}, {Khandrika}, {Levay}, {Lucas}, {MacKenty},
  {Ogaz}, {Porterfield}, {Reid}, {Robberto}, {Royle}, {Smith},
  {Storrie-Lombardi}, {Sunnquist}, {Surace}, {Taylor}, {Williams}, {Bullock},
  {Dickinson}, {Finkelstein}, {Natarajan}, {Richard}, {Robertson}, {Tumlinson},
  {Zitrin}, {Flanagan}, {Sembach}, {Soifer}, \& {Mountain}}]{lotz17}
{Lotz}, J.~M., {Koekemoer}, A., {Coe}, D., {et~al.} 2017, \apj, 837, 97

\bibitem[{{Ma} {et~al.}(2018){Ma}, {Hopkins}, {Garrison-Kimmel},
  {Faucher-Gigu{\`e}re}, {Quataert}, {Boylan-Kolchin}, {Hayward}, {Feldmann},
  \& {Kere{\v{s}}}}]{ma18}
{Ma}, X., {Hopkins}, P.~F., {Garrison-Kimmel}, S., {et~al.} 2018, \mnras, 478,
  1694

\bibitem[{{Mainali} {et~al.}(2018){Mainali}, {Zitrin}, {Stark}, {Ellis},
  {Richard}, {Tang}, {Laporte}, {Oesch}, \& {McGreer}}]{mainali18}
{Mainali}, R., {Zitrin}, A., {Stark}, D.~P., {et~al.} 2018, \mnras, 479, 1180

\bibitem[{{Malhotra} \& {Rhoads}(2006)}]{malhotra06}
{Malhotra}, S., \& {Rhoads}, J.~E. 2006, \apjl, 647, L95

\bibitem[{{Mason} {et~al.}(2015{\natexlab{a}}){Mason}, {Trenti}, \&
  {Treu}}]{mason15b}
{Mason}, C.~A., {Trenti}, M., \& {Treu}, T. 2015{\natexlab{a}}, \apj, 813, 21

\bibitem[{{Mason} {et~al.}(2015{\natexlab{b}}){Mason}, {Treu}, {Schmidt},
  {Collett}, {Trenti}, {Marshall}, {Barone-Nugent}, {Bradley}, {Stiavelli}, \&
  {Wyithe}}]{mason15}
{Mason}, C.~A., {Treu}, T., {Schmidt}, K.~B., {et~al.} 2015{\natexlab{b}},
  ArXiv e-prints

\bibitem[{{McCracken} {et~al.}(2012){McCracken}, {Milvang-Jensen}, {Dunlop},
  {Franx}, {Fynbo}, {Le F{\`e}vre}, {Holt}, {Caputi}, {Goranova}, {Buitrago},
  {Emerson}, {Freudling}, {Hudelot}, {L{\'o}pez-Sanjuan}, {Magnard}, {Mellier},
  {M{\o}ller}, {Nilsson}, {Sutherland}, {Tasca}, \& {Zabl}}]{mccracken12}
{McCracken}, H.~J., {Milvang-Jensen}, B., {Dunlop}, J., {et~al.} 2012, \aap,
  544, A156

\bibitem[{{McLeod} {et~al.}(2016){McLeod}, {McLure}, \& {Dunlop}}]{mcleod16}
{McLeod}, D.~J., {McLure}, R.~J., \& {Dunlop}, J.~S. 2016, ArXiv e-prints

\bibitem[{{McLeod} {et~al.}(2015){McLeod}, {McLure}, {Dunlop}, {Robertson},
  {Ellis}, \& {Targett}}]{mcleod15}
{McLeod}, D.~J., {McLure}, R.~J., {Dunlop}, J.~S., {et~al.} 2015, \mnras, 450,
  3032

\bibitem[{{McLure} {et~al.}(2009){McLure}, {Cirasuolo}, {Dunlop}, {Foucaud}, \&
  {Almaini}}]{mclure09}
{McLure}, R.~J., {Cirasuolo}, M., {Dunlop}, J.~S., {Foucaud}, S., \& {Almaini},
  O. 2009, \mnras, 395, 2196

\bibitem[{{McLure} {et~al.}(2010){McLure}, {Dunlop}, {Cirasuolo}, {Koekemoer},
  {Sabbi}, {Stark}, {Targett}, \& {Ellis}}]{mclure10}
{McLure}, R.~J., {Dunlop}, J.~S., {Cirasuolo}, M., {et~al.} 2010, \mnras, 403,
  960

\bibitem[{{McLure} {et~al.}(2013){McLure}, {Dunlop}, {Bowler}, {Curtis-Lake},
  {Schenker}, {Ellis}, {Robertson}, {Koekemoer}, {Rogers}, {Ono}, {Ouchi},
  {Charlot}, {Wild}, {Stark}, {Furlanetto}, {Cirasuolo}, \&
  {Targett}}]{mclure13}
{McLure}, R.~J., {Dunlop}, J.~S., {Bowler}, R.~A.~A., {et~al.} 2013, \mnras,
  432, 2696

\bibitem[{{Merlin} {et~al.}(2015){Merlin}, {Fontana}, {Ferguson}, {Dunlop},
  {Elbaz}, {Bourne}, {Bruce}, {Buitrago}, {Castellano}, {Schreiber}, {Grazian},
  {McLure}, {Okumura}, {Shu}, {Wang}, {Amorin}, {Boutsia}, {Cappelluti},
  {Comastri}, {Derriere}, {Faber}, \& {Santini}}]{merlin15}
{Merlin}, E., {Fontana}, A., {Ferguson}, H.~C., {et~al.} 2015, ArXiv e-prints

\bibitem[{{Merlin} {et~al.}(2016){Merlin}, {Bourne}, {Castellano}, {Ferguson},
  {Wang}, {Derriere}, {Dunlop}, {Elbaz}, \& {Fontana}}]{merlin16}
{Merlin}, E., {Bourne}, N., {Castellano}, M., {et~al.} 2016, \aap, 595, A97

\bibitem[{{Miralda-Escud{\'e}} \& {Rees}(1998)}]{miralda-escude98}
{Miralda-Escud{\'e}}, J., \& {Rees}, M.~J. 1998, \apj, 497, 21

\bibitem[{{Momcheva} {et~al.}(2016){Momcheva}, {Brammer}, {van Dokkum},
  {Skelton}, {Whitaker}, {Nelson}, {Fumagalli}, {Maseda}, {Leja}, {Franx},
  {Rix}, {Bezanson}, {Da Cunha}, {Dickey}, {F{\"o}rster Schreiber},
  {Illingworth}, {Kriek}, {Labb{\'e}}, {Ulf Lange}, {Lundgren}, {Magee},
  {Marchesini}, {Oesch}, {Pacifici}, {Patel}, {Price}, {Tal}, {Wake}, {van der
  Wel}, \& {Wuyts}}]{momcheva16}
{Momcheva}, I.~G., {Brammer}, G.~B., {van Dokkum}, P.~G., {et~al.} 2016, \apjs,
  225, 27

\bibitem[{{Morishita} {et~al.}(2018){Morishita}, {Trenti}, {Stiavelli},
  {Bradley}, {Coe}, {Oesch}, {Mason}, {Bridge}, {Holwerda}, {Livermore},
  {Salmon}, {Schmidt}, {Shull}, \& {Treu}}]{morishita18}
{Morishita}, T., {Trenti}, M., {Stiavelli}, M., {et~al.} 2018, \apj, 867, 150

\bibitem[{{Naidu} {et~al.}(2020){Naidu}, {Tacchella}, {Mason}, {Bose}, {Oesch},
  \& {Conroy}}]{naidu20}
{Naidu}, R.~P., {Tacchella}, S., {Mason}, C.~A., {et~al.} 2020, \apj, 892, 109

\bibitem[{{Nandra} {et~al.}(2015){Nandra}, {Laird}, {Aird}, {Salvato},
  {Georgakakis}, {Barro}, {Perez-Gonzalez}, {Barmby}, {Chary}, {Coil},
  {Cooper}, {Davis}, {Dickinson}, {Faber}, {Fazio}, {Guhathakurta}, {Gwyn},
  {Hsu}, {Huang}, {Ivison}, {Koo}, {Newman}, {Rangel}, {Yamada}, \&
  {Willmer}}]{nandra15}
{Nandra}, K., {Laird}, E.~S., {Aird}, J.~A., {et~al.} 2015, \apjs, 220, 10

\bibitem[{{Narayanan} {et~al.}(2012){Narayanan}, {Krumholz}, {Ostriker}, \&
  {Hernquist}}]{narayanan12}
{Narayanan}, D., {Krumholz}, M.~R., {Ostriker}, E.~C., \& {Hernquist}, L. 2012,
  \mnras, 421, 3127

\bibitem[{{Nayyeri} {et~al.}(2017){Nayyeri}, {Hemmati}, {Mobasher}, {Ferguson},
  {Cooray}, {Barro}, {Faber}, {Dickinson}, {Koekemoer}, {Peth}, {Salvato},
  {Ashby}, {Darvish}, {Donley}, {Durbin}, {Finkelstein}, {Fontana}, {Grogin},
  {Gruetzbauch}, {Huang}, {Khostovan}, {Kocevski}, {Kodra}, {Lee}, {Newman},
  {Pacifici}, {Pforr}, {Stefanon}, {Wiklind}, {Willner}, {Wuyts}, {Castellano},
  {Conselice}, {Dolch}, {Dunlop}, {Galametz}, {Hathi}, {Lucas}, \&
  {Yan}}]{nayyeri17}
{Nayyeri}, H., {Hemmati}, S., {Mobasher}, B., {et~al.} 2017, \apjs, 228, 7

\bibitem[{{Oesch} {et~al.}(2018){Oesch}, {Bouwens}, {Illingworth}, {Labb{\'e}},
  \& {Stefanon}}]{oesch18}
{Oesch}, P.~A., {Bouwens}, R.~J., {Illingworth}, G.~D., {Labb{\'e}}, I., \&
  {Stefanon}, M. 2018, \apj, 855, 105

\bibitem[{{Oesch} {et~al.}(2013){Oesch}, {Bouwens}, {Illingworth}, {Labb{\'e}},
  {Franx}, {van Dokkum}, {Trenti}, {Stiavelli}, {Gonzalez}, \&
  {Magee}}]{oesch13}
{Oesch}, P.~A., {Bouwens}, R.~J., {Illingworth}, G.~D., {et~al.} 2013, \apj,
  773, 75

\bibitem[{{Oesch} {et~al.}(2014){Oesch}, {Bouwens}, {Illingworth}, {Labb{\'e}},
  {Smit}, {Franx}, {van Dokkum}, {Momcheva}, {Ashby}, {Fazio}, {Huang},
  {Willner}, {Gonzalez}, {Magee}, {Trenti}, {Brammer}, {Skelton}, \&
  {Spitler}}]{oesch14}
---. 2014, \apj, 786, 108

\bibitem[{{Oesch} {et~al.}(2016){Oesch}, {Brammer}, {van Dokkum},
  {Illingworth}, {Bouwens}, {Labb{\'e}}, {Franx}, {Momcheva}, {Ashby}, {Fazio},
  {Gonzalez}, {Holden}, {Magee}, {Skelton}, {Smit}, {Spitler}, {Trenti}, \&
  {Willner}}]{oesch16}
{Oesch}, P.~A., {Brammer}, G., {van Dokkum}, P.~G., {et~al.} 2016, \apj, 819,
  129

\bibitem[{{Paardekooper} {et~al.}(2015){Paardekooper}, {Khochfar}, \& {Dalla
  Vecchia}}]{paardekooper15}
{Paardekooper}, J.-P., {Khochfar}, S., \& {Dalla Vecchia}, C. 2015, \mnras,
  451, 2544

\bibitem[{{Papovich} {et~al.}(2016){Papovich}, {Shipley}, {Mehrtens}, {Lanham},
  {Lacy}, {Ciardullo}, {Finkelstein}, {Bassett}, {Behroozi}, {Blanc}, {de
  Jong}, {DePoy}, {Drory}, {Gawiser}, {Gebhardt}, {Gronwall}, {Hill}, {Hopp},
  {Jogee}, {Kawinwanichakij}, {Marshall}, {McLinden}, {Mentuch Cooper},
  {Somerville}, {Steinmetz}, {Tran}, {Tuttle}, {Viero}, {Wechsler}, \&
  {Zeimann}}]{papovich16}
{Papovich}, C., {Shipley}, H.~V., {Mehrtens}, N., {et~al.} 2016, \apjs, 224, 28

\bibitem[{{Patten} {et~al.}(2006){Patten}, {Stauffer}, {Burrows}, {Marengo},
  {Hora}, {Luhman}, {Sonnett}, {Henry}, {Raghavan}, \& {Megeath}}]{patten06}
{Patten}, B.~M., {Stauffer}, J.~R., {Burrows}, A., {et~al.} 2006, \apj, 651,
  502

\bibitem[{{Peng} {et~al.}(2002){Peng}, {Ho}, {Impey}, \& {Rix}}]{peng02}
{Peng}, C.~Y., {Ho}, L.~C., {Impey}, C.~D., \& {Rix}, H.-W. 2002, \aj, 124, 266

\bibitem[{{Peng} {et~al.}(2010){Peng}, {Lilly}, {Kova{\v{c}}}, {Bolzonella},
  {Pozzetti}, {Renzini}, {Zamorani}, {Ilbert}, {Knobel}, {Iovino}, {Maier},
  {Cucciati}, {Tasca}, {Carollo}, {Silverman}, {Kampczyk}, {de Ravel},
  {Sanders}, {Scoville}, {Contini}, {Mainieri}, {Scodeggio}, {Kneib}, {Le
  F{\`e}vre}, {Bardelli}, {Bongiorno}, {Caputi}, {Coppa}, {de la Torre},
  {Franzetti}, {Garilli}, {Lamareille}, {Le Borgne}, {Le Brun}, {Mignoli},
  {Perez Montero}, {Pello}, {Ricciardelli}, {Tanaka}, {Tresse}, {Vergani},
  {Welikala}, {Zucca}, {Oesch}, {Abbas}, {Barnes}, {Bordoloi}, {Bottini},
  {Cappi}, {Cassata}, {Cimatti}, {Fumana}, {Hasinger}, {Koekemoer},
  {Leauthaud}, {Maccagni}, {Marinoni}, {McCracken}, {Memeo}, {Meneux}, {Nair},
  {Porciani}, {Presotto}, \& {Scaramella}}]{ypeng10}
{Peng}, Y.-j., {Lilly}, S.~J., {Kova{\v{c}}}, K., {et~al.} 2010, \apj, 721, 193

\bibitem[{{Pentericci} {et~al.}(2018){Pentericci}, {Vanzella}, {Castellano},
  {Fontana}, {De Barros}, {Grazian}, {Marchi}, {Bradac}, {Conselice},
  {Cristiani}, {Dickinson}, {Finkelstein}, {Giallongo}, {Guaita}, {Koekemoer},
  {Maiolino}, {Santini}, \& {Tilvi}}]{pentericci18}
{Pentericci}, L., {Vanzella}, E., {Castellano}, M., {et~al.} 2018, ArXiv
  e-prints

\bibitem[{{Planck Collaboration} {et~al.}(2016){Planck Collaboration},
  {Aghanim}, {Ashdown}, {Aumont}, {Baccigalupi}, {Ballardini}, {Banday},
  {Barreiro}, {Bartolo}, {Basak}, {Battye}, {Benabed}, {Bernard}, {Bersanelli},
  {Bielewicz}, {Bock}, {Bonaldi}, {Bonavera}, {Bond}, {Borrill}, {Bouchet},
  {Boulanger}, {Bucher}, {Burigana}, {Butler}, {Calabrese}, {Cardoso},
  {Carron}, {Challinor}, {Chiang}, {Colombo}, {Combet}, {Comis}, {Coulais},
  {Crill}, {Curto}, {Cuttaia}, {Davis}, {de Bernardis}, {de Rosa}, {de Zotti},
  {Delabrouille}, {Delouis}, {Di Valentino}, {Dickinson}, {Diego}, {Dor{\'e}},
  {Douspis}, {Ducout}, {Dupac}, {Efstathiou}, {Elsner}, {En{\ss}lin},
  {Eriksen}, {Falgarone}, {Fantaye}, {Finelli}, {Forastieri}, {Frailis},
  {Fraisse}, {Franceschi}, {Frolov}, {Galeotta}, {Galli}, {Ganga},
  {G{\'e}nova-Santos}, {Gerbino}, {Ghosh}, {Gonz{\'a}lez-Nuevo}, {G{\'o}rski},
  {Gratton}, {Gruppuso}, {Gudmundsson}, {Hansen}, {Helou},
  {Henrot-Versill{\'e}}, {Herranz}, {Hivon}, {Huang}, {Ili{\'c}}, {Jaffe},
  {Jones}, {Keih{\"a}nen}, {Keskitalo}, {Kisner}, {Knox}, {Krachmalnicoff},
  {Kunz}, {Kurki-Suonio}, {Lagache}, {Lamarre}, {Langer}, {Lasenby},
  {Lattanzi}, {Lawrence}, {Le Jeune}, {Leahy}, {Levrier}, {Liguori}, {Lilje},
  {L{\'o}pez-Caniego}, {Ma}, {Mac{\'{\i}}as-P{\'e}rez}, {Maggio}, {Mangilli},
  {Maris}, {Martin}, {Mart{\'{\i}}nez-Gonz{\'a}lez}, {Matarrese}, {Mauri},
  {McEwen}, {Meinhold}, {Melchiorri}, {Mennella}, {Migliaccio},
  {Miville-Desch{\^e}nes}, {Molinari}, {Moneti}, {Montier}, {Morgante}, {Moss},
  {Mottet}, {Naselsky}, {Natoli}, {Oxborrow}, {Pagano}, {Paoletti},
  {Partridge}, {Patanchon}, {Patrizii}, {Perdereau}, {Perotto}, {Pettorino},
  {Piacentini}, {Plaszczynski}, {Polastri}, {Polenta}, {Puget}, {Rachen},
  {Racine}, {Reinecke}, {Remazeilles}, {Renzi}, {Rocha}, {Rossetti}, {Roudier},
  {Rubi{\~n}o-Mart{\'{\i}}n}, {Ruiz-Granados}, {Salvati}, {Sandri},
  {Savelainen}, {Scott}, {Sirri}, {Sunyaev}, {Suur-Uski}, {Tauber}, {Tenti},
  {Toffolatti}, {Tomasi}, {Tristram}, {Trombetti}, {Valiviita}, {Van Tent},
  {Vibert}, {Vielva}, {Villa}, {Vittorio}, {Wandelt}, {Watson}, {Wehus},
  {White}, {Zacchei}, \& {Zonca}}]{planck16}
{Planck Collaboration}, {Aghanim}, N., {Ashdown}, M., {et~al.} 2016, \aap, 596,
  A107

\bibitem[{{Planck Collaboration} {et~al.}(2020){Planck Collaboration},
  {Aghanim}, {Akrami}, {Ashdown}, {Aumont}, {Baccigalupi}, {Ballardini},
  {Banday}, {Barreiro}, {Bartolo}, {Basak}, {Battye}, {Benabed}, {Bernard},
  {Bersanelli}, {Bielewicz}, {Bock}, {Bond}, {Borrill}, {Bouchet}, {Boulanger},
  {Bucher}, {Burigana}, {Butler}, {Calabrese}, {Cardoso}, {Carron},
  {Challinor}, {Chiang}, {Chluba}, {Colombo}, {Combet}, {Contreras}, {Crill},
  {Cuttaia}, {de Bernardis}, {de Zotti}, {Delabrouille}, {Delouis}, {Di
  Valentino}, {Diego}, {Dor{\'e}}, {Douspis}, {Ducout}, {Dupac}, {Dusini},
  {Efstathiou}, {Elsner}, {En{\ss}lin}, {Eriksen}, {Fantaye}, {Farhang},
  {Fergusson}, {Fernandez-Cobos}, {Finelli}, {Forastieri}, {Frailis},
  {Fraisse}, {Franceschi}, {Frolov}, {Galeotta}, {Galli}, {Ganga},
  {G{\'e}nova-Santos}, {Gerbino}, {Ghosh}, {Gonz{\'a}lez-Nuevo}, {G{\'o}rski},
  {Gratton}, {Gruppuso}, {Gudmundsson}, {Hamann}, {Handley}, {Hansen},
  {Herranz}, {Hildebrandt}, {Hivon}, {Huang}, {Jaffe}, {Jones}, {Karakci},
  {Keih{\"a}nen}, {Keskitalo}, {Kiiveri}, {Kim}, {Kisner}, {Knox},
  {Krachmalnicoff}, {Kunz}, {Kurki-Suonio}, {Lagache}, {Lamarre}, {Lasenby},
  {Lattanzi}, {Lawrence}, {Le Jeune}, {Lemos}, {Lesgourgues}, {Levrier},
  {Lewis}, {Liguori}, {Lilje}, {Lilley}, {Lindholm}, {L{\'o}pez-Caniego},
  {Lubin}, {Ma}, {Mac{\'\i}as-P{\'e}rez}, {Maggio}, {Maino}, {Mandolesi},
  {Mangilli}, {Marcos-Caballero}, {Maris}, {Martin}, {Martinelli},
  {Mart{\'\i}nez-Gonz{\'a}lez}, {Matarrese}, {Mauri}, {McEwen}, {Meinhold},
  {Melchiorri}, {Mennella}, {Migliaccio}, {Millea}, {Mitra},
  {Miville-Desch{\^e}nes}, {Molinari}, {Montier}, {Morgante}, {Moss}, {Natoli},
  {N{\o}rgaard-Nielsen}, {Pagano}, {Paoletti}, {Partridge}, {Patanchon},
  {Peiris}, {Perrotta}, {Pettorino}, {Piacentini}, {Polastri}, {Polenta},
  {Puget}, {Rachen}, {Reinecke}, {Remazeilles}, {Renzi}, {Rocha}, {Rosset},
  {Roudier}, {Rubi{\~n}o-Mart{\'\i}n}, {Ruiz-Granados}, {Salvati}, {Sandri},
  {Savelainen}, {Scott}, {Shellard}, {Sirignano}, {Sirri}, {Spencer},
  {Sunyaev}, {Suur-Uski}, {Tauber}, {Tavagnacco}, {Tenti}, {Toffolatti},
  {Tomasi}, {Trombetti}, {Valenziano}, {Valiviita}, {Van Tent}, {Vibert},
  {Vielva}, {Villa}, {Vittorio}, {Wandelt}, {Wehus}, {White}, {White},
  {Zacchei}, \& {Zonca}}]{planck20}
{Planck Collaboration}, {Aghanim}, N., {Akrami}, Y., {et~al.} 2020, \aap, 641,
  A6

\bibitem[{{Roberts-Borsani} {et~al.}(2021){Roberts-Borsani}, {Morishita},
  {Treu}, {Leethochawalit}, \& {Trenti}}]{roberts-borsani21}
{Roberts-Borsani}, G., {Morishita}, T., {Treu}, T., {Leethochawalit}, N., \&
  {Trenti}, M. 2021, arXiv e-prints, arXiv:2106.06544

\bibitem[{{Roberts-Borsani} {et~al.}(2020){Roberts-Borsani}, {Ellis}, \&
  {Laporte}}]{roberts-borsani20}
{Roberts-Borsani}, G.~W., {Ellis}, R.~S., \& {Laporte}, N. 2020, \mnras, 497,
  3440

\bibitem[{{Roberts-Borsani} {et~al.}(2016){Roberts-Borsani}, {Bouwens},
  {Oesch}, {Labbe}, {Smit}, {Illingworth}, {van Dokkum}, {Holden}, {Gonzalez},
  {Stefanon}, {Holwerda}, \& {Wilkins}}]{roberts-borsani16}
{Roberts-Borsani}, G.~W., {Bouwens}, R.~J., {Oesch}, P.~A., {et~al.} 2016,
  \apj, 823, 143

\bibitem[{{Robertson} {et~al.}(2015){Robertson}, {Ellis}, {Furlanetto}, \&
  {Dunlop}}]{robertson15}
{Robertson}, B.~E., {Ellis}, R.~S., {Furlanetto}, S.~R., \& {Dunlop}, J.~S.
  2015, \apjl, 802, L19

\bibitem[{{Robertson} {et~al.}(2013){Robertson}, {Furlanetto}, {Schneider},
  {Charlot}, {Ellis}, {Stark}, {McLure}, {Dunlop}, {Koekemoer}, {Schenker},
  {Ouchi}, {Ono}, {Curtis-Lake}, {Rogers}, {Bowler}, \&
  {Cirasuolo}}]{robertson13}
{Robertson}, B.~E., {Furlanetto}, S.~R., {Schneider}, E., {et~al.} 2013, \apj,
  768, 71

\bibitem[{{Rojas-Ruiz} {et~al.}(2020){Rojas-Ruiz}, {Finkelstein}, {Bagley},
  {Stevans}, {Finkelstein}, {Larson}, {Mechtley}, \& {Diekmann}}]{rojasruiz20}
{Rojas-Ruiz}, S., {Finkelstein}, S.~L., {Bagley}, M.~B., {et~al.} 2020, \apj,
  891, 146

\bibitem[{{Ryan} \& {Reid}(2016)}]{ryan16}
{Ryan}, R.~E., J., \& {Reid}, I.~N. 2016, \aj, 151, 92

\bibitem[{{Schmidt} {et~al.}(2014){Schmidt}, {Treu}, {Trenti}, {Bradley},
  {Kelly}, {Oesch}, {Holwerda}, {Shull}, \& {Stiavelli}}]{schmidt14}
{Schmidt}, K.~B., {Treu}, T., {Trenti}, M., {et~al.} 2014, \apj, 786, 57

\bibitem[{{Schmidt}(1959)}]{schmidt59}
{Schmidt}, M. 1959, \apj, 129, 243

\bibitem[{{Shibuya} {et~al.}(2016){Shibuya}, {Ouchi}, {Kubo}, \&
  {Harikane}}]{shibuya16}
{Shibuya}, T., {Ouchi}, M., {Kubo}, M., \& {Harikane}, Y. 2016, \apj, 821, 72

\bibitem[{{Somerville} {et~al.}(2015){Somerville}, {Popping}, \&
  {Trager}}]{somerville15b}
{Somerville}, R.~S., {Popping}, G., \& {Trager}, S.~C. 2015, \mnras, 453, 4337

\bibitem[{{Somerville} {et~al.}(2021){Somerville}, {Olsen}, {Yung}, {Pacifici},
  {Ferguson}, {Behroozi}, {Osborne}, {Wechsler}, {Pandya}, {Faber}, {Primack},
  \& {Dekel}}]{somerville21}
{Somerville}, R.~S., {Olsen}, C., {Yung}, L.~Y.~A., {et~al.} 2021, \mnras, 502,
  4858

\bibitem[{{Song} {et~al.}(2016){Song}, {Finkelstein}, {Ashby}, {Grazian}, {Lu},
  {Papovich}, {Salmon}, {Somerville}, {Dickinson}, {Duncan}, {Faber}, {Fazio},
  {Ferguson}, {Fontana}, {Guo}, {Hathi}, {Lee}, {Merlin}, \&
  {Willner}}]{song16a}
{Song}, M., {Finkelstein}, S.~L., {Ashby}, M.~L.~N., {et~al.} 2016, \apj, 825,
  5

\bibitem[{{Spitler} {et~al.}(2012){Spitler}, {Labb{\'e}}, {Glazebrook},
  {Persson}, {Monson}, {Papovich}, {Tran}, {Poole}, {Quadri}, \& {van
  Dokkum}}]{spitler12}
{Spitler}, L.~R., {Labb{\'e}}, I., {Glazebrook}, K., {et~al.} 2012, \apj, 748,
  L21

\bibitem[{{Stefanon} {et~al.}(2017{\natexlab{a}}){Stefanon}, {Bouwens},
  {Labb{\'e}}, {Muzzin}, {Marchesini}, {Oesch}, \& {Gonzalez}}]{stefanon17b}
{Stefanon}, M., {Bouwens}, R.~J., {Labb{\'e}}, I., {et~al.} 2017{\natexlab{a}},
  \apj, 843, 36

\bibitem[{{Stefanon} {et~al.}(2017{\natexlab{b}}){Stefanon}, {Yan}, {Mobasher},
  {Barro}, {Donley}, {Fontana}, {Hemmati}, {Koekemoer}, {Lee}, {Lee},
  {Nayyeri}, {Peth}, {Pforr}, {Salvato}, {Wiklind}, {Wuyts}, {Ashby},
  {Castellano}, {Conselice}, {Cooper}, {Cooray}, {Dolch}, {Ferguson},
  {Galametz}, {Giavalisco}, {Guo}, {Willner}, {Dickinson}, {Faber}, {Fazio},
  {Gardner}, {Gawiser}, {Grazian}, {Grogin}, {Kocevski}, {Koo}, {Lee}, {Lucas},
  {McGrath}, {Nandra}, {Newman}, \& {van der Wel}}]{stefanon17}
{Stefanon}, M., {Yan}, H., {Mobasher}, B., {et~al.} 2017{\natexlab{b}}, \apjs,
  229, 32

\bibitem[{{Stefanon} {et~al.}(2019){Stefanon}, {Labb{\'e}}, {Bouwens}, {Oesch},
  {Ashby}, {Caputi}, {Franx}, {Fynbo}, {Illingworth}, {Le F{\`e}vre},
  {Marchesini}, {McCracken}, {Milvang-Jensen}, {Muzzin}, \& {van
  Dokkum}}]{stefanon19}
{Stefanon}, M., {Labb{\'e}}, I., {Bouwens}, R.~J., {et~al.} 2019, \apj, 883, 99

\bibitem[{{Steidel} \& {Hamilton}(1993)}]{steidel93}
{Steidel}, C.~C., \& {Hamilton}, D. 1993, \aj, 105, 2017

\bibitem[{{Stevans} {et~al.}(2021){Stevans}, {Finkelstein}, {Kawinwanichakij},
  {Wold}, {Papovich}, {Somerville}, {Yung}, {Sherman}, {Ciardullo}, {Dave},
  {Florez}, {Gronwall}, \& {Jogee}}]{stevans21}
{Stevans}, M.~L., {Finkelstein}, S.~L., {Kawinwanichakij}, L., {et~al.} 2021,
  arXiv e-prints, arXiv:2103.14690

\bibitem[{{Straatman} {et~al.}(2014){Straatman}, {Labb{\'e}}, {Spitler},
  {Allen}, {Altieri}, {Brammer}, {Dickinson}, {van Dokkum}, {Inami},
  {Glazebrook}, {Kacprzak}, {Kawinwanichakij}, {Kelson}, {McCarthy},
  {Mehrtens}, {Monson}, {Murphy}, {Papovich}, {Persson}, {Quadri}, {Rees},
  {Tomczak}, {Tran}, \& {Tilvi}}]{straatman14}
{Straatman}, C. M.~S., {Labb{\'e}}, I., {Spitler}, L.~R., {et~al.} 2014, \apjl,
  783, L14

\bibitem[{{Tacchella} {et~al.}(2018){Tacchella}, {Bose}, {Conroy},
  {Eisenstein}, \& {Johnson}}]{tacchella18}
{Tacchella}, S., {Bose}, S., {Conroy}, C., {Eisenstein}, D.~J., \& {Johnson},
  B.~D. 2018, \apj, 868, 92

\bibitem[{{Tacchella} {et~al.}(2021){Tacchella}, {Finkelstein}, {Finkelstein},
  {Finkelstein}, {Finkelstein}, \& {Finkelstein}}]{tacchella21}
{Tacchella}, S., {Finkelstein}, S.~L., {Finkelstein}, S.~L., {et~al.} 2021, ApJ
  Submitted

\bibitem[{{Tilvi} {et~al.}(2020){Tilvi}, {Malhotra}, {Rhoads}, {Coughlin},
  {Zheng}, {Finkelstein}, {Veilleux}, {Mobasher}, {Wang}, {Probst}, {Swaters},
  {Hibon}, {Joshi}, {Zabl}, {Jiang}, {Pharo}, \& {Yang}}]{tilvi20}
{Tilvi}, V., {Malhotra}, S., {Rhoads}, J.~E., {et~al.} 2020, \apjl, 891, L10

\bibitem[{{Trenti} {et~al.}(2011){Trenti}, {Bradley}, {Stiavelli}, {Oesch},
  {Treu}, {Bouwens}, {Shull}, {MacKenty}, {Carollo}, \&
  {Illingworth}}]{trenti11}
{Trenti}, M., {Bradley}, L.~D., {Stiavelli}, M., {et~al.} 2011, \apjl, 727, L39

\bibitem[{{Treu} {et~al.}(2017){Treu}, {Abramson}, {Bradac}, {Brammer},
  {Fontana}, {Henry}, {Hoag}, {Huang}, {Mason}, {Morishita}, {Pentericci}, \&
  {Wang}}]{treu17}
{Treu}, T., {Abramson}, L., {Bradac}, M., {et~al.} 2017, {Through the Looking
  GLASS: A JWST Exploration of Galaxy Formation and Evolution from Cosmic Dawn
  to Present Day}, JWST Proposal ID 1324. Cycle 0 Early Release Scienc

\bibitem[{{Vijayan} {et~al.}(2020){Vijayan}, {Lovell}, {Wilkins}, {Thomas},
  {Barnes}, {Irodotou}, {Kuusisto}, \& {Roper}}]{vijayan20}
{Vijayan}, A.~P., {Lovell}, C.~C., {Wilkins}, S.~M., {et~al.} 2020, \mnras

\bibitem[{{Vogelsberger} {et~al.}(2020){Vogelsberger}, {Nelson}, {Pillepich},
  {Shen}, {Marinacci}, {Springel}, {Pakmor}, {Tacchella}, {Weinberger},
  {Torrey}, \& {Hernquist}}]{vogelsberger20}
{Vogelsberger}, M., {Nelson}, D., {Pillepich}, A., {et~al.} 2020, \mnras, 492,
  5167

\bibitem[{{Whitaker} {et~al.}(2011){Whitaker}, {Labb{\'e}}, {van Dokkum},
  {Brammer}, {Kriek}, {Marchesini}, {Quadri}, {Franx}, {Muzzin}, {Williams},
  {Bezanson}, {Illingworth}, {Lee}, {Lundgren}, {Nelson}, {Rudnick}, {Tal}, \&
  {Wake}}]{whitaker11}
{Whitaker}, K.~E., {Labb{\'e}}, I., {van Dokkum}, P.~G., {et~al.} 2011, \apj,
  735, 86

\bibitem[{{Wilkins} {et~al.}(2016){Wilkins}, {Bouwens}, {Oesch}, {Labb{\'e}},
  {Sargent}, {Caruana}, {Wardlow}, \& {Clay}}]{wilkins15}
{Wilkins}, S.~M., {Bouwens}, R.~J., {Oesch}, P.~A., {et~al.} 2016, \mnras, 455,
  659

\bibitem[{{Wilkins} {et~al.}(2017){Wilkins}, {Feng}, {Di Matteo}, {Croft},
  {Lovell}, \& {Waters}}]{wilkins17}
{Wilkins}, S.~M., {Feng}, Y., {Di Matteo}, T., {et~al.} 2017, \mnras, 469, 2517

\bibitem[{{Williams} {et~al.}(2018){Williams}, {Curtis-Lake}, {Hainline},
  {Chevallard}, {Robertson}, {Charlot}, {Endsley}, {Stark}, {Willmer},
  {Alberts}, {Amorin}, {Arribas}, {Baum}, {Bunker}, {Carniani}, {Crand all},
  {Egami}, {Eisenstein}, {Ferruit}, {Husemann}, {Maseda}, {Maiolino}, {Rawle},
  {Rieke}, {Smit}, {Tacchella}, \& {Willott}}]{williams18}
{Williams}, C.~C., {Curtis-Lake}, E., {Hainline}, K.~N., {et~al.} 2018, \apjs,
  236, 33

\bibitem[{{Windhorst} {et~al.}(2011){Windhorst}, {Cohen}, {Hathi}, {McCarthy},
  {Ryan}, {Yan}, {Baldry}, {Driver}, {Frogel}, {Hill}, {Kelvin}, {Koekemoer},
  {Mechtley}, {O'Connell}, {Robotham}, {Rutkowski}, {Seibert}, {Straughn},
  {Tuffs}, {Balick}, {Bond}, {Bushouse}, {Calzetti}, {Crockett}, {Disney},
  {Dopita}, {Hall}, {Holtzman}, {Kaviraj}, {Kimble}, {MacKenty}, {Mutchler},
  {Paresce}, {Saha}, {Silk}, {Trauger}, {Walker}, {Whitmore}, \&
  {Young}}]{windhorst11}
{Windhorst}, R.~A., {Cohen}, S.~H., {Hathi}, N.~P., {et~al.} 2011, \apjs, 193,
  27

\bibitem[{{Wold} {et~al.}(2019){Wold}, {Kawinwanichakij}, {Stevans},
  {Finkelstein}, {Papovich}, {Devarakonda}, {Ciardullo}, {Feldmeier}, {Florez},
  {Gawiser}, {Gronwall}, {Jogee}, {Marshall}, {Sherman}, {Shipley},
  {Somerville}, {Valdes}, \& {Zeimann}}]{wold19}
{Wold}, I. G.~B., {Kawinwanichakij}, L., {Stevans}, M.~L., {et~al.} 2019,
  \apjs, 240, 5

\bibitem[{{Xu} {et~al.}(2016){Xu}, {Wise}, {Norman}, {Ahn}, \& {O'Shea}}]{xu16}
{Xu}, H., {Wise}, J.~H., {Norman}, M.~L., {Ahn}, K., \& {O'Shea}, B.~W. 2016,
  \apj, 833, 84

\bibitem[{{Xue} {et~al.}(2016){Xue}, {Luo}, {Brandt}, {Alexander}, {Bauer},
  {Lehmer}, \& {Yang}}]{xue16}
{Xue}, Y.~Q., {Luo}, B., {Brandt}, W.~N., {et~al.} 2016, \apjs, 224, 15

\bibitem[{{Yang} {et~al.}(2021){Yang}, {Somerville}, {Pullen}, {Popping},
  {Breysse}, \& {Maniyar}}]{yang21}
{Yang}, S., {Somerville}, R.~S., {Pullen}, A.~R., {et~al.} 2021, \apj, 911, 132

\bibitem[{{Yung} {et~al.}(2019){Yung}, {Somerville}, {Finkelstein}, {Popping},
  \& {Dav{\'e}}}]{yung19a}
{Yung}, L.~Y.~A., {Somerville}, R.~S., {Finkelstein}, S.~L., {Popping}, G., \&
  {Dav{\'e}}, R. 2019, \mnras, 483, 2983

\bibitem[{{Yung} {et~al.}(2020){Yung}, {Somerville}, {Finkelstein}, {Popping},
  {Dav{\'e}}, {Venkatesan}, {Behroozi}, \& {Ferguson}}]{yung20b}
{Yung}, L.~Y.~A., {Somerville}, R.~S., {Finkelstein}, S.~L., {et~al.} 2020,
  \mnras, 496, 4574

\bibitem[{{Zitrin} {et~al.}(2015){Zitrin}, {Labb{\'e}}, {Belli}, {Bouwens},
  {Ellis}, {Roberts-Borsani}, {Stark}, {Oesch}, \& {Smit}}]{zitrin15}
{Zitrin}, A., {Labb{\'e}}, I., {Belli}, S., {et~al.} 2015, \apjl, 810, L12

\end{thebibliography}

\appendix

\section{Visually Removed Sources}

\begin{figure*}
\epsscale{1.0}
\plotone{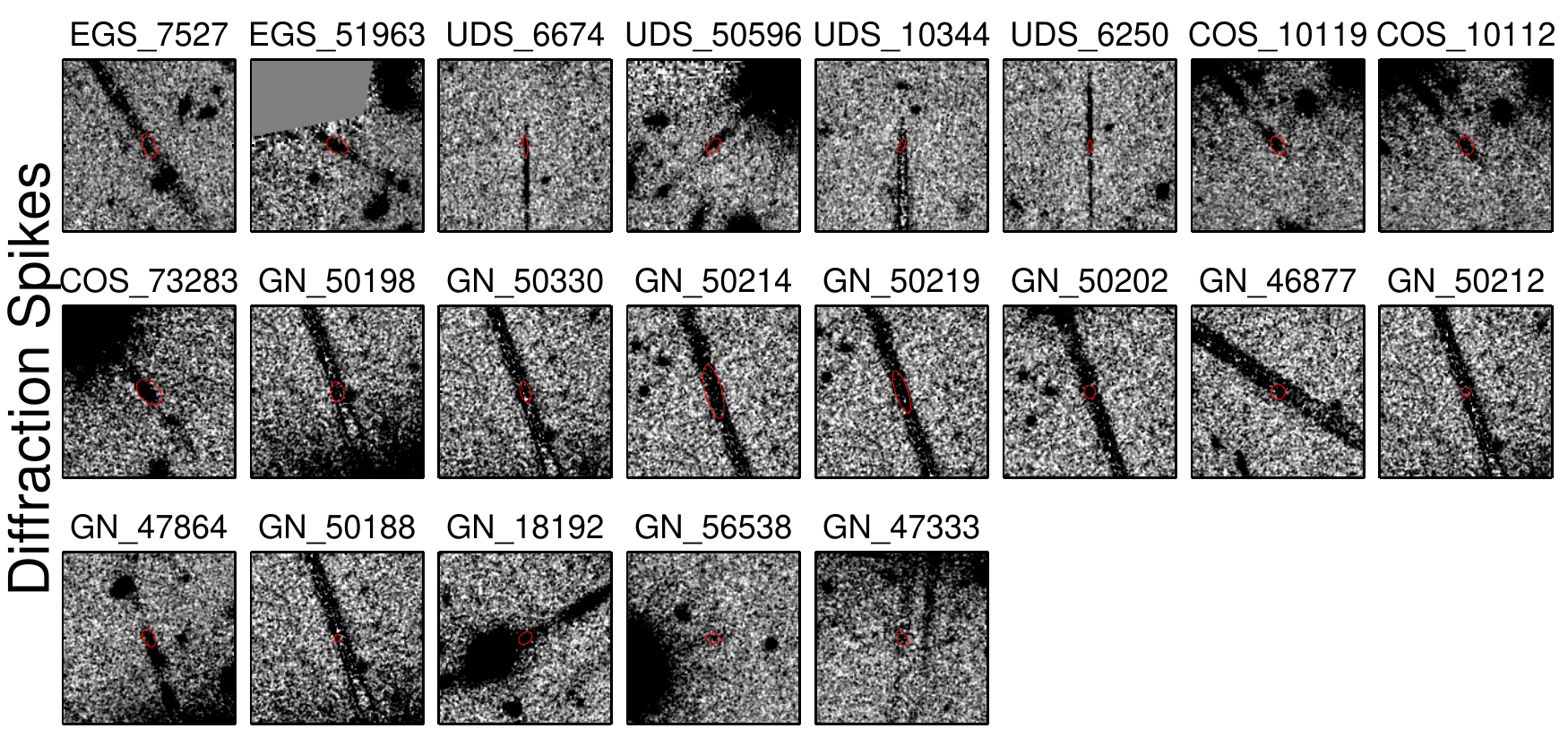}
\plotone{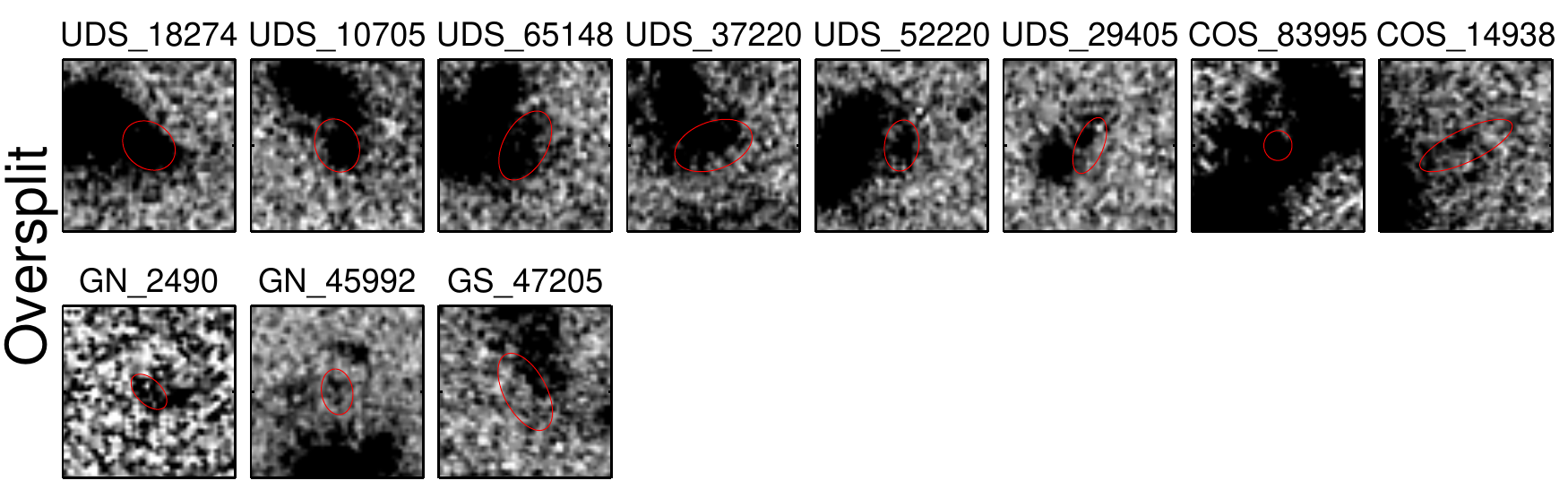}
\plotone{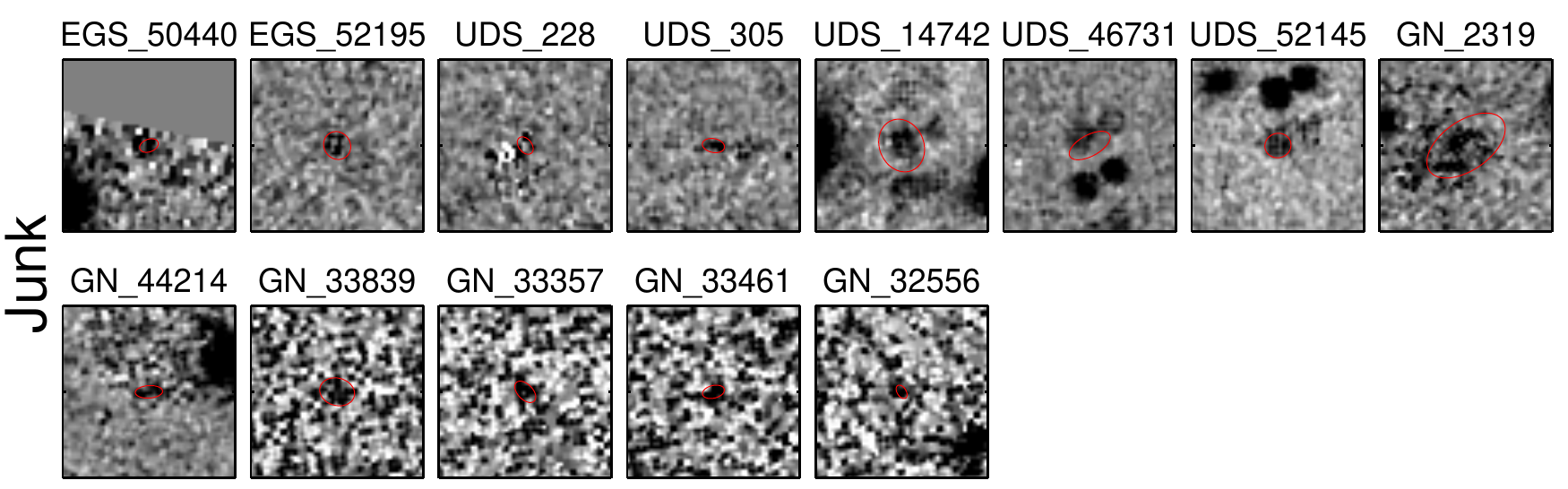}
\caption{Objects removed from our initial sample following visual
  inspection.  In the top section, images are 9\arcs\ in length, while
  they are 3\arcs\ in the remaining two sections.  In all cutout images the
  red ellipses denotes the small Kron aperture identified for the
  source.  The titles for each image show the field it was detected
  in, and its identification number in our photometric catalog.  The top section shows objects identified as being a
  part of a diffraction spike.  The middle section shows objects
  identified as being an oversplit region of a nearby bright galaxy.
  The bottom section shows other objects determined to be non-real
  sources, either due to residing near an image edge or other region
  of enhanced noise, missed persistence (in the case of the UDS objects), or having a morphology inconsistent with a real
  galaxy, indicative of an image defect.}
\label{fig:spurious}
\end{figure*}

\begin{figure*}
\epsscale{1.0}
\plotone{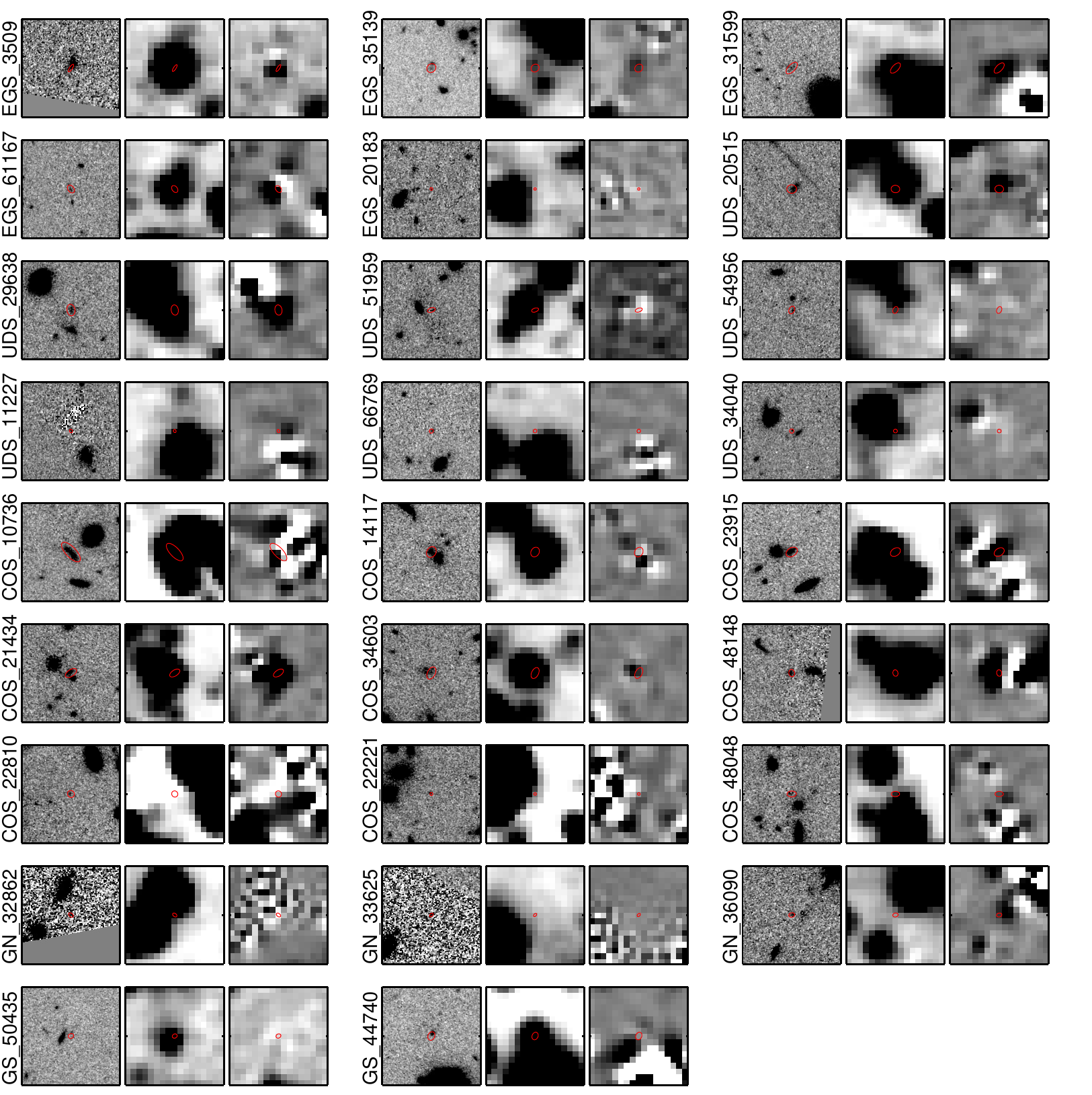}
\caption{Cutout images of the 26 objects removed from our sample when visual inspection
  determined that the TPHOT deblended IRAC photometry was unreliable,
  and the photometric redshift probability distribution function
  without the IRAC photometry did not satisfy our sample selection
  criteria.  From left-to-right, for each object we show the $H_{160}$-band
  image, the 3.6$\mu$m image, and the 3.6$\mu$m residual image, where
  the source of interest was not subtracted (the {\it HST} and IRAC
  cutouts are 9\arcs\ and 10\arcs, respectively).}
\label{fig:baddeblend}
\end{figure*}

In this section of the appendix, we show cutout images of any sources
removed from our sample following any kind of subjective visual
inspection.  Figure~\ref{fig:spurious} shows sources visually
identified as diffraction spikes, oversplit portions of nearby
galaxies, or ``Junk'', which denotes any other kind of non-real
source, typically due to increased noise in the region around a given
object.   This process is discussed in \S 3.3.
Figure~\ref{fig:baddeblend} shows objects removed after visual
inspection determined that the TPHOT IRAC deblending was unreliable,
and the sources did not satisfy our photometric redshift selection
criteria when excluding the IRAC photometry.  Some of these may yet be
real $z >$ 8.5 sources, but for our conservative analysis, we remove
them.  These sources are listed in Table~\ref{tab:appendix}.

\begin{deluxetable}{cccccc}
\vspace{2mm}
\tabletypesize{\small}
\tablecaption{Sources Removed Due to Poor IRAC Deblending}
\tablewidth{\textwidth}
\tablehead{
\colhead{Field} & \colhead{ID} & \colhead{RA} & \colhead{Dec} & \colhead{$\int_8^{15}$ $\powerset(z)$} & \colhead{$\int_8^{15}$ $\powerset(z)$}\\
\colhead{$ $} & \colhead{Limiting$ $} & \colhead{J2000} &
\colhead{J2000} & \colhead{with IRAC} & \colhead{without IRAC}}
\startdata
EGS&3509&214.790502&52.707862&0.77&0.03\\
EGS&35139&215.191622&53.071285&0.77&0.25\\
EGS&31599&214.809988&52.809716&0.68&0.43\\
EGS&61167&215.079527&53.047774&0.98&0.20\\
EGS&20183&215.252461&53.078172&0.99&0.37\\
UDS&20515&34.382023&-5.170840&1.00&0.25\\
UDS&29638&34.371719&-5.192273&0.97&0.17\\
UDS&51959&34.406923&-5.128847&0.93&0.36\\
UDS&54956&34.479757&-5.136519&0.76&0.06\\
UDS&11227&34.490247&-5.251716&1.00&0.24\\
UDS&66769&34.400905&-5.162690&0.98&0.02\\
UDS&34040&34.329161&-5.200860&0.99&0.31\\
COSMOS&10736&150.074713&2.216529&1.00&0.00\\
COSMOS&14117&150.182930&2.230231&0.96&0.04\\
COSMOS&23915&150.068259&2.279747&1.00&0.13\\
COSMOS&21434&150.060263&2.267318&0.99&0.28\\
COSMOS&34603&150.153627&2.327952&1.00&0.30\\
COSMOS&48148&150.054342&2.389392&0.90&0.02\\
COSMOS&22810&150.063056&2.274242&1.00&0.25\\
COSMOS&22221&150.055009&2.271304&1.00&0.11\\
COSMOS&48048&150.075835&2.388783&0.98&0.30\\
GOODSN&32862&189.466464&62.236057&1.00&0.43\\
GOODSN&33625&189.464511&62.237533&0.89&0.00\\
GOODSN&36090&189.323657&62.377288&1.00&0.43\\
GOODSS&50435&53.171530&$-$27.717057&1.00&0.38\\
GOODSS&44740$^{\dagger}$&53.094458&$-$27.739336&0.00&0.90
\enddata
\tablecomments{$^{\dagger}$This source formally satisfies the sample
  selection criteria without IRAC.  However, it is obvious in the
  image that this source is extremely bright in IRAC (even with the
  poor deblending), thus we conclude it is very likely to reside at
  low redshift.}
\label{tab:appendix}
\end{deluxetable}

\end{document}